 \documentclass[sigconf,nonacm]{acmart}

\usepackage[utf8]{inputenc}
\usepackage{xspace}

\usepackage[table]{xcolor}

\definecolor{codegreen}{rgb}{0,0.6,0}
\definecolor{codegray}{rgb}{0.5,0.5,0.5}
\definecolor{codepurple}{rgb}{0.58,0,0.82}
\definecolor{backcolor}{rgb}{0.95,0.95,0.92}

\definecolor{lightgreen}{HTML}{E6F4EA}
\definecolor{lightgrey}{HTML}{F1F3F4}
\definecolor{lightorange}{HTML}{FEF7E0}
\definecolor{lightpink}{HTML}{FCE8E6}
\definecolor{cellgreen}{HTML}{E2EFDA}  
\definecolor{cellyellow}{HTML}{FFF2CC} 
\definecolor{cellgrey}{HTML}{D9D9D9}   
\definecolor{cellred}{HTML}{FCE4D6}    

\usepackage{graphicx}
\usepackage{subcaption}
\usepackage{tikz}
\usetikzlibrary{positioning, shapes.geometric, arrows.meta, calc}

\usepackage{booktabs}
\usepackage{multirow}
\usepackage{makecell}
\usepackage{tabularx}

\usepackage{enumitem}
\usepackage{algorithm}
\usepackage{algpseudocode}

\usepackage{listings}
\lstdefinestyle{mysqlstyle}{
    otherkeywords={<=>, \$},
    commentstyle=\color{codegreen},
    keywordstyle=\color{blue},
    numberstyle=\tiny\color{codegray},
    stringstyle=\color{codepurple},
    basicstyle=\ttfamily\footnotesize,
    breakatwhitespace=false,         
    breaklines=true,                 
    captionpos=b,                    
    keepspaces=true,                 
    numbers=left,                    
    numbersep=5pt,                  
    showspaces=false,                
    showstringspaces=false,
    showtabs=false,                  
    tabsize=2,
    language=SQL
}
\usepackage{mdframed}
\mdfdefinestyle{examplestyle}{
  topline=true,
  bottomline=true,
  rightline=false,
  leftline=false,
  linewidth=0.5pt,
  backgroundcolor=lightgrey,
  leftmargin=-4pt,
  rightmargin=-4pt,
  innerleftmargin=4pt,
  innerrightmargin=4pt,
  innertopmargin=4pt,
  innerbottommargin=4pt,
  skipabove=\topsep,
  skipbelow=\topsep
}
\surroundwithmdframed[style=examplestyle]{example}

\newcommand{\zi}[1]{\textcolor{red}{ZI: #1}}
\newcommand{\jl}[1]{\textcolor{purple}{JL: #1}}
\newcommand{\sys}{\textsc{DASE}\xspace}
\newcommand{\sysb}{\textsc{DASE}\xspace} 
\newcommand{\idx}{\textsc{SemJI}\xspace}

\AtBeginDocument{%
  }

\begin{document}

\title{Efficiently Linking Unstructured Data for Multi-step Reasoning}

\author{Jiaming Liang}
\email{liangjm@seas.upenn.edu}
\orcid{0009-0009-2675-3901}
\affiliation{%
  \institution{University of Pennsylvania}
  \city{Philadelphia}
  \state{PA}
  \country{USA}
}

\author{Haydn Jones}
\email{haydnj@seas.upenn.edu}
\orcid{0000-0002-1006-4126}
\affiliation{%
  \institution{University of Pennsylvania}
  \city{Philadelphia}
  \state{PA}
  \country{USA}
}

\author{Jacob R. Gardner}
\email{jacobrg@cis.upenn.edu}
\orcid{0000-0003-1897-8384}
\affiliation{%
  \institution{University of Pennsylvania}
  \city{Philadelphia}
  \state{PA}
  \country{USA}
}

\author{Mark Yatskar}
\email{myatskar@cis.upenn.edu}
\orcid{0009-0008-8129-0471}
\affiliation{%
  \institution{University of Pennsylvania}
  \city{Philadelphia}
  \state{PA}
  \country{USA}
}

\author{Zachary Ives}
\email{zives@cis.upenn.edu}
\orcid{0000-0001-7527-2957}
\affiliation{%
  \institution{University of Pennsylvania}
  \city{Philadelphia}
  \state{PA}
  \country{USA}
}

\begin{abstract}
Modern LLMs and AI agents increasingly support data engineering workflows that integrate evidence from unstructured sources. Such pipelines typically do data retrieval, integration, and ranking before proceeding to more complex agentic reasoning or actions, e.g., for scientific discovery.  The \textbf{core retrieval problem} in these workflows jointly executes multi-attribute filtering, multi-vector search, exact relational joins, and thresholded embedding-similarity joins. Given a planned query and monotone scoring function, our \sys query engine constructs and ranks candidate evidence tuples. It comprises (i) a multi-step reasoning query model over structured predicates, multiple vectors, and relational links; (ii) \idx, a sparse materialized embedding-similarity join index for rare near-neighbor pairs; and (iii) a co-designed execution layer that combines predicate-aware ANN traversal, batched access, and threshold-based score aggregation.

On scientific-discovery workloads, DASE retrieves candidate evidence for multi-step reasoning queries 6× to 46× faster than strong RDBMS, rerank, and vector-database baselines at comparable recall; and for tasks that require semantic-operator post-processing, DASE acts as a high-recall prefilter that makes downstream LLM evaluation both cheaper and more accurate—e.g., on SemBench E-Commerce it improves BigQuery quality from 0.67 to 0.80 while cutting cost from \$2.42 to \$0.54.
\end{abstract}

\maketitle

\section{Introduction}

Novel database and AI systems have emerged that construct answers by integrating information from multiple unstructured sources, leveraging LLMs to reason about semantic content and relationships.
Examples of such \emph{agentic data engineering systems} include ``deep research''~\cite{zheng2025deepresearcher,guo_deepseek-r1_2025,yao_react_2022} and ``wide search'' agents~\cite{widesearch,lan2026table,lan2025deepwidesearch,huang2026wideseek,xu2026wideseekr1exploringwidthscaling,chen2026mapreduce}, systems for scientific discovery and literature synthesis~\cite{ghareeb2026multi,gottweis2026accelerating,li2026multi,asai2026synthesizing}, LLM-based information extraction tools~\cite{shankar2025docetl,anderson_design_2024}, and unstructured-to-structured query systems based on \emph{semantic operators}~\cite{patel_semantic_2025,liu2025palimpzest,russo_abacus_2025,trummer_implementing_2025}.
Despite fundamental differences in architecture and scope, these systems generally share a core operation: given a user query, they decompose it into plans, where an initial \emph{multi-step reasoning} retrieval primitive finds and assembles entity combinations that best satisfy multiple search criteria. These integrated results are then fed into pipelines containing rich agentic reasoning steps, which may perform aggregate reasoning or initiate other actions. These agent-generated retrieval and integration workloads motivate rethinking query execution in data systems, in line with the broader call for \emph{agent-first} architectures~\cite{liu2025supporting}. Our paper targets the initial data-retrieval task and layer: given a planned query and scoring function, we construct and rank candidate evidence tuples. Our LLM-based planner, which decomposes natural language tasks into retrieval and agentic steps using a combination of chain-of-thought reasoning~\cite{wei_chain--thought_2022}, tool-calling~\cite{yang_gpt4tools_2023}, and ReAct-based reasoning~\cite{yao_react_2022}, is beyond the scope of the paper.

Our team, in collaboration with chemists and life scientists, is building an agentic data engineering tool to accelerate scientific discovery. Two initial classes of applications involve discovering candidate molecules with pharmaceutical benefits and engineering RNA sequences with therapeutic properties. Both require reasoning over mostly natural-language data that has been preprocessed by (1) indexing document paragraphs with embeddings, as in retrieval-augmented generation (RAG)~\cite{gao_retrieval-augmented_2023}, and (2) extracting knowledge-graph-like annotations of those paragraphs~\cite{peng_graph_2024,saleh_sg-rag_2024} and indexing annotation labels and values by their embeddings. We use molecular discovery as the running example; the RNA workload has a similar retrieval structure.
A molecule's structure can be encoded as a string (SMILES)~\cite{weininger1988smiles}, a bit vector (ECFP)~\cite{rogers_extended-connectivity_2010}, or as a specialized embedding (ChemBERTa)~\cite{chithrananda_chemberta_2020} depending on the downstream reasoning steps required. Accompanying textual passages, as well as knowledge graph properties such as outcomes or experimental settings, might also be embedded using standard text or multimodal embeddings.

\begin{example}[Agentic data engineering for molecular discovery]
\label{ex:motivating}
The process of discovering and optimizing molecules or RNA starts with an information gathering stage, i.e., a query that assembles candidate results.  Since evidence is often fragmented across documents and their extracted annotations, a query about a molecule might (1) approximately match against a ChemBERTa embedding, extracted from the text into a knowledge graph, but linked back to the paragraph in which it appears; (2) match on an extracted \texttt{organism}; (3) match against embeddings of experimental conclusions.  All of these matches must be joined to ensure they are within a common paper; and ranked using a function that \textbf{combines all of the individual matches}.
Finally, a ranked list of promising candidate molecules from this query stage is often fed into additional tools, e.g., to optimize their chemistry, or to generate new molecules using fine-tuned LLMs.
\end{example}

\smallskip

Such queries may also need to link information across documents based on approximate matches in their content:





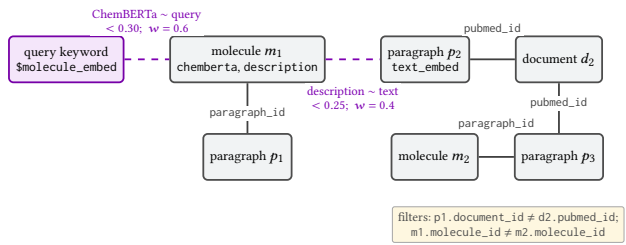
\begin{figure}[t]
\centering
\resizebox{0.97\columnwidth}{!}{%
\begin{tikzpicture}[
    relation/.style={rectangle, rounded corners=2pt, draw=black!70, thick,
        fill=lightgrey, minimum width=1.35cm, minimum height=0.72cm,
        align=center, font=\scriptsize},
    query/.style={rectangle, rounded corners=2pt, draw=codepurple!80!black,
        thick, fill=codepurple!10, minimum width=1.35cm, minimum height=0.72cm,
        align=center, font=\scriptsize},
    exact/.style={thick, black!75},
    semantic/.style={thick, dashed, codepurple!85!black},
    elabel/.style={font=\fontsize{5.6}{6.4}\selectfont, fill=white,
        inner sep=1pt, align=center},
    constraint/.style={font=\fontsize{5.7}{6.5}\selectfont, text=black!70,
        fill=lightorange, draw=black!35, rounded corners=1pt, inner sep=2.5pt,
        align=center}
]
    \node[query] (q) {query keyword\\\texttt{\$molecule\_embed}};
    \node[relation, right=0.70cm of q] (m1) {molecule \(m_1\)\\\texttt{chemberta}, \texttt{description}};
    \node[relation, right=0.85cm of m1] (p2) {paragraph \(p_2\)\\\texttt{text\_embed}};
    \node[relation, right=0.70cm of p2] (d2) {document \(d_2\)};
    \node[relation, below=0.78cm of m1] (p1) {paragraph \(p_1\)};
    \node[relation, below=0.78cm of d2] (p3) {paragraph \(p_3\)};
    \node[relation, left=0.55cm of p3] (m2) {molecule \(m_2\)};

    \draw[semantic] (q) -- node[elabel, above=10pt] {ChemBERTa \(\sim\) query\\\(< 0.30;\ w=0.6\)} (m1);
    \draw[semantic] (m1) -- node[elabel, below=10.5pt] {description \(\sim\) text\\\(< 0.25;\ w=0.4\)} (p2);

    \draw[exact] (m1) -- node[elabel, pos=0.60] {\texttt{paragraph\_id}} (p1);
    \draw[exact] (p2) -- node[elabel, above=10pt] {\texttt{pubmed\_id}} (d2);
    \draw[exact] (d2) -- node[elabel, pos=0.40] {\texttt{pubmed\_id}} (p3);
    \draw[exact] (p3) -- node[elabel, above=12pt, yshift=-1.2pt] {\texttt{paragraph\_id}} (m2);

    \node[constraint, below=0.40cm of m2, xshift=1.15cm] {filters: \texttt{p1.document\_id} \(\neq\) \texttt{d2.pubmed\_id};\\\texttt{m1.molecule\_id} \(\neq\) \texttt{m2.molecule\_id}};
\end{tikzpicture}
}
\vspace{-1mm}
\caption{A graph pattern for the query described in Example~\ref{ex:join}, corresponding to agentic planner-generated SQL (Listing~\ref{lst:join-query}). The leftmost keyword-input leaf, \texttt{\$molecule\_embed}, matches molecule variable \(m_1\); dashed purple edges are thresholded embedding-similarity predicates whose distances form the ranking score, and solid edges are exact joins.}
\label{fig:query_graph}
\vspace{-3mm}
\end{figure}

\begin{example} \label{ex:join}
Building upon Example~\ref{ex:motivating}, a scientific-discovery query might be based on the observation that small molecules may have similar characteristics even if they are not particularly similar in ChemBERTa embedding space. Thus we might seek to \emph{find related molecules}, by matching on a first document, then \textbf{linking and joining via text description embedding} with \emph{other} documents, then traversing (joining) to their extracted molecular descriptions.  Figure~\ref{fig:query_graph} visualizes this. We will formalize and revisit this query in subsequent examples.
\end{example}

\smallskip


The retrieval task is not just to retrieve relevant records step by step. A candidate must be evaluated jointly against the query's criteria, even when the evidence for those criteria is scattered across sources. This requires linking the relevant records to the same candidate and ranking the resulting combinations as a whole. As collections grow, the system must search a large space of candidate--evidence combinations while returning only a short list for downstream reasoning.

Concretely, a query can combine structured predicates, exact joins, and thresholded embedding-similarity joins. It returns the top-\(k\) evidence tuples according to a scoring function that combines multiple similarity signals.
Conventional vector DBMSs such as Pinecone or Weaviate~\cite{pinecone,weaviate} do not natively plan multi-table joins and joint multi-signal ranking, so an external query processor must stitch such operations together. Milvus~\cite{wang_milvus_2021}, designed for hybrid search, supports logical predicates and vector retrieval but does not provide relational join planning. We therefore study how to implement multi-step reasoning queries in a relational DBMS with approximate-nearest-neighbor (ANN) vector-index support.

As we demonstrate in Section~\ref{sec:experiments}, standard RDBMS execution strategies for combining predicates with approximate-nearest-neighbor search~\cite{gollapudi_filtered-diskann_2023,lu2026depth} perform poorly on join-heavy multi-step reasoning queries and other semantic-query tasks~\cite{lao_sembench_2025}. Similarly, information-retrieval-style reranking~\cite{nogueira_passage_2020,cao_feedback-driven_2010} can incur low recall. Multi-step reasoning queries therefore introduce challenges in query indexing, optimization, and execution, which we address with the following contributions:

\begin{itemize}[leftmargin=*]
    \item \textbf{Query formalism for agentic systems.} We formalize ranked multi-step reasoning queries that support structured filters, vector scores, and cross-table joins.
    \item \textbf{Indexing strategy.} We develop \idx, a materialized embedding-similarity join index that represents near-neighbor pairs as a relational access path for top-\(k\) multi-step reasoning queries.
    \item \textbf{Algorithm/system co-design.} We extend PostgreSQL's ANN-index support with batched access and develop \sys, which combines predicate-aware ANN traversal, filtered score streams, and threshold-based aggregation to jointly optimize attribute predicates and vector scoring across relational joins.
    \item \textbf{Benchmark and evaluation.} We contribute a benchmark over IMDB and MOLECULE with eight query classes covering structured filtering, multi-signal scoring, and cross-table joins, separately and in combination. Fixed query semantics and exhaustive top-\(k\) reference results enable comparisons of retrieval quality and throughput across systems. We also evaluate prefiltering for downstream semantic queries, scaling with join depth, and robustness to embedding anisotropy and shared context.
\end{itemize}

\noindent
To ensure reproducibility, our code, data, and full technical report are available at \url{https://anonymous.4open.science/r/DASE-2C61/}.

\section{Query Model and Index Strategy}
\label{sec:prob_def}

This paper targets query answering tasks over unstructured data that has been indexed and annotated within a local relational DBMS, which may include both approximate nearest-neighbor search over embeddings, as well as logical predicates over relationships.

\paragraph{Data model}
Our underlying data model is based on object-rel\-a\-tion\-al tables, representing raw document and file content. Tables are linked through foreign key relationships.  Using a combination of user-defined functions and semantic operators~\cite{patel_semantic_2025,shankar2025docetl}, indexed data may be parsed, mapped to an embedding, and then analyzed using LLM operators such as \emph{semantic map} to extract structured information from the parsed content; \emph{semantic filter} to test tuples' content against a prompt, etc. The results of semantic maps may be saved as labeled \emph{annotations} in a structured relational table, in a controlled vocabulary, with explicit provenance links to the source content, and with embeddings over their values.



\begin{figure}[htbp]
\centering
\vspace{-2mm}
\includegraphics[width=0.95\linewidth]{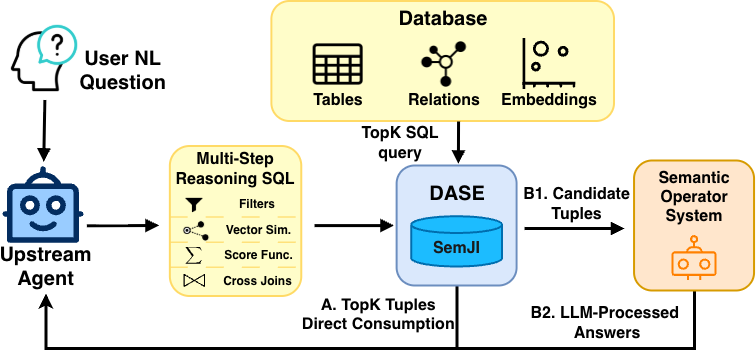}
\vspace{-3mm}
    \caption{\sys interactions with other system components.}
    \label{fig.usecase}
\vspace{-3mm}
\end{figure}

\subsection{Top-\(k\) Multi-Step Reasoning Queries}
\label{sec:multistep_query}
Our target use case involves an upper-layer agentic system, performing deep or wide search, or another form of multi-step reasoning over a user's natural-language question. This creates an agentic pipeline, in which the agent, or a planning frontend, creates an initial retrieval and integration task involving a structured query over a local dataset. This \emph{multi-step reasoning query} combines multi-attribute filtering, multi-vector similarity scoring, and cross-table joins, together with a scoring function that aggregates these signals into a unified ranking objective. Our \sys engine executes this query and returns the top-\(k\) results. We assume the scoring function is supplied by the upstream agent. \sys is agnostic to its origin; in our own research agent, the weights are trained using logistic regression over a \emph{calibration set}~\cite{patel_semantic_2025} of labeled samples. These assumptions are consistent with Fagin's monotonicity assumption~\cite{fagin_optimal_2003} and thus admit efficient threshold-based top-\(k\) evaluation.  \sys then outputs a \emph{ranked list}, which the upstream agent consumes in one of two modes:

\paragraph{1. Direct consumption.} When the agent's question reduces to finding the best-matching combinations of evidence---a typical pattern in wide search and multi-objective discovery---the ranked output is the answer. Cross-encoders~\cite{nogueira_passage_2020} are a common way for the agent to map per-source signals into a unified ranking.

\paragraph{2. Prefilter for semantic operators.} For tasks that require a final LLM judgment---for example, evidence validation, entity resolution, or answer synthesis---\sys returns a high-recall candidate set to a downstream semantic-operator system~\cite{russo_abacus_2025,trummer_implementing_2025,patel_semantic_2025}. This realizes the standard calibrated prefilter--LLM pattern: cheap proxy signals prune candidates before expensive semantic evaluation~\cite{patel_semantic_2025}. In our setting, the prefilter is a ranked, thresholded select--join query over embedding distances and structured attributes.

In either mode, \sys executes the agent-supplied top-\(k\) scoring function over structured attributes, multiple embedding spaces, and cross-table joins. We now formalize this query.

\subsubsection{Query Model.}
\label{para:query_model} 
A multi-step reasoning query \(Q\) is formally defined by the triple \((\mathcal{P}, \mathcal{S}, \mathcal{J})\), which characterizes the feasibility space, ranking objectives, and relational dependencies.

\textbf{Predicates (\(\mathcal{P}\)):} \(\mathcal{P} = \{p_1, \dots, p_{|\mathcal{P}|}\}\) is a set of heterogeneous predicates over 1 or more \textbf{structured components} (which may be the results of a join or Cartesian product). This set may be empty (\(\mathcal{P} = \emptyset\)), indicating an unconstrained search space. Regardless of their specific semantic types (such as range constraints, set memberships, or probabilistic filters) each \(p_i \in \mathcal{P}\) can be conceptually modeled as a Boolean indicator function \(p_i: \mathcal{R} \to \{0, 1\}\). For a given query subexpression \(q\) and any resulting row instance \(r \in q(\cdot)\), \(p_i(r)\) logically evaluates to \(1\) if \(r\) satisfies the constraint, and \(0\) otherwise.
    
\textbf{Similarity Join Conditions (\(\mathcal{J}\)):} \(\mathcal{J} = \{(e_{1,i}, e_{2,i}, dis_i, \tau_i)\}_{i=1}^{|\mathcal{J}|}\) is an optional set of relational constraints (\(\mathcal{J} = \emptyset\) denotes a standard single-table query). Each condition specifies a thresholded embedding-similarity join between \textbf{unstructured components} across tables: a link is established when the embedding distance satisfies \(dis_i(e_{1,i}, e_{2,i}) \le \tau_i\). Thus, distances are lower-is-better feasibility predicates, not ranking scores.

\textbf{Scoring Model (\(\mathcal{S}\)):} \(\mathcal{S} = \text{score}_f(s_1, \dots, s_{|\mathcal{S}|})\) defines the unified ranking objective. To guarantee a definable top-\(k\) retrieval, the query must specify at least one scoring signal (\(|\mathcal{S}| \ge 1\)). Every \(s_i\) is a \emph{higher-is-better affinity}: an embedding distance \(d\) is converted by a calibrated non-increasing transform \(\phi_i\), e.g., \(s_i=\phi_i(d)=-d\); and a scalar attribute is mapped to a desirability score by an analogous transform. Components therefore include (1) query-to-object affinity \(\phi_i(dis(\mathbf{q}_v,e_i))\), (2) relational-link affinity \(\phi_i(dis(e_{1,i},e_{2,i}))\), and (3) structural-attribute desirability. The aggregation function \(f\) is coordinatewise non-decreasing (for example, a nonnegative weighted sum), so higher component affinities cannot lower the final score. The goal is to retrieve the top-\(k\) tuples that satisfy \(\mathcal{P}\) and \(\mathcal{J}\) while maximizing \(\mathcal{S}\).


As discussed above, the scoring signals and weights are provided by an upper-layer agentic system (e.g., through calibration or learned rankings~\cite{chen_learning_2016,li_learning_2011,yan_actively_2013}) or by direct user specification; our methods are agnostic to their origin, provided the resulting component scores obey the higher-is-better, monotone contract above.



Our task extends classic rank aggregation~\cite{fagin_optimal_2003, tziavelis_ranked_2024}. Traditional settings find top answers by aggregating \emph{independent} scores that evaluate the query against individual records. In contrast, our core challenge is computing dynamic, \emph{interdependent} scores to evaluate embedding-similarity join conditions on the fly. Reconciling these highly coupled data-to-data affinities with structured logical predicates, while avoiding exhaustive search and leveraging indexing, is the primary computational hurdle in retrieving the top-\(k\) results.

\begin{example} \label{ex:query}
Listing~\ref{lst:join-query} gives an abbreviated SQL query for
Example~\ref{ex:join} (eliding result-formatting details).  The \texttt{semantic\_score} function computes a higher-is-better result, by negating the calibrated weighted combination of the lookup and join distances.
\end{example}

\begin{figure}
\begin{lstlisting}[language=SQL, caption={Abbreviated molecular-relationship query pattern for \texttt{\$molecule\_embed}, for Example~\ref{ex:join}.},label={lst:join-query}]
SELECT m1.molecule_id, m2.molecule_id, d2.pubmed_id,
       semantic_score(m1, p2, $molecule_embed) AS semantic_score
FROM molecule m1
    JOIN paragraph p1 ON m1.paragraph_id = p1.paragraph_id
    JOIN paragraph p2 ON p2.text_embed <=> m1.description_embed < 0.25
    JOIN document d2 ON p2.pubmed_id = d2.pubmed_id
    JOIN paragraph p3 ON d2.pubmed_id = p3.pubmed_id
    JOIN molecule m2 ON p3.paragraph_id = m2.paragraph_id
WHERE  m1.chemberta_embed <=> $molecule_embed < 0.3
    AND p1.document_id <> d2.pubmed_id
    AND m1.molecule_id <> m2.molecule_id
ORDER BY semantic_score DESC LIMIT 100;
\end{lstlisting}
\vspace{-6mm}
\end{figure}

\medskip
Standard ANN indices fall short here for two reasons. First, as they are implemented in engines such as PostgreSQL, they are designed for 1-to-N point lookups, making them \textbf{unable to execute M-to-N embedding-similarity joins following a block nested loops pattern}. Instead, they essentially only support tuple nested loops joins. Second, because they \textbf{cannot jointly evaluate the multi-signal scoring function}, the true top results for the overall query might be buried arbitrarily deep within the ANN's ranked stream, precluding early termination.

The M-to-N embedding-similarity pattern is a critical part of many object-linking tasks, such as record linking~\cite{christen_survey_2011} or textual similarity detection.  In many cases, we will have a block of \(M\) items from the left relation (collection) that are highly similar to \(N\) items in the right relation; the final answer set may be ranked with a holistic combination of score components (the multi-signal scoring function) including the similarities of each similarity step, as well as the intrinsic scores of the original items versus specific query constraints.
We thus develop an alternative approach: \textbf{precompute and materialize} partial answers based on distributional knowledge, then expand the answer frontier as needed. This decouples join discovery from scoring and supports threshold algorithms~\cite{fagin_optimal_2003} for earlier termination.

\subsection{Precomputing Semantic Join Tail-Events}
\label{sec:motivate_materializing_tail_events}

\begin{figure}[htbp]
    \centering
     \includegraphics[width=0.9\linewidth]{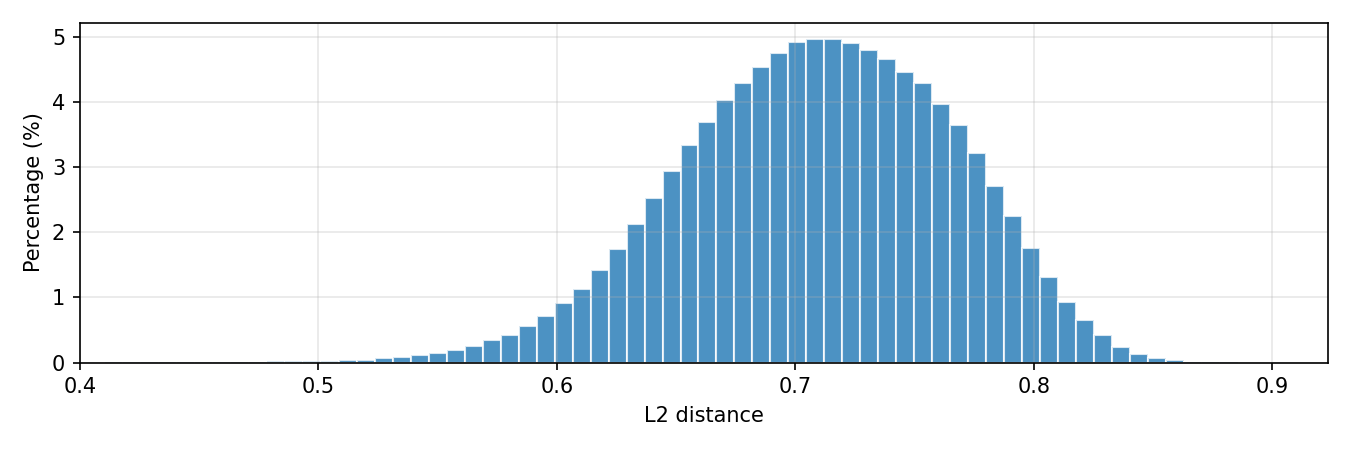}
\vspace{-4mm}
    \caption{\label{fig:pairwise-dist} L2 distance distribution: pairwise among IMDB Actor\_Director embeddings (Gemini-embedding-001)}
\vspace{-4mm}
\end{figure}

Evaluating an embedding-similarity join naively requires distances for a Cartesian product of endpoint embeddings. In high-dimensional, approximately isotropic embedding spaces\footnote{``Approximately isotropic'' means that variance is not concentrated in a small number of embedding directions.}, pairwise distances concentrate in a dense bulk, while unusually close pairs occupy a small left tail~\cite{vershynin2018high}. The join candidates that carry semantic affinity are therefore sparse relative to all pairs. Figure~\ref{fig:pairwise-dist} illustrates this empirical pattern for IMDB actor--director embeddings: most distances lie between \(0.55\) and \(0.85\), with only a small close-pair tail.

This observation motivates materializing the close-pair tail as a \textbf{sparse embedding-similarity join index} and feeding its endpoints to batched retrieval. (Real endpoint populations can be anisotropic and their distance distributions need not be Gaussian, but still exhibit the tail property; Section~\ref{sec:pancaking} discusses mitigations.)

\section{Semantic Join Index}
\label{sec:sji_index}

Based on the intuition from the previous discussion, the \textbf{Semantic Join Index (\idx)} materializes close vector pairs, instantiating the embedding-similarity-join topology \(\mathcal{J}\) of our query model and avoiding repeated join discovery at query time. Here ``semantic'' refers to the embedding-based proxy; it is distinct from a downstream LLM semantic operator.

\subsection{Motivation and Definition}

For materialization radius \(\tau_{cap}\), we materialize relation \(\mathcal{J}_{\tau_{cap}}=\{(A,B)\mid \text{dist}(A,B)\le\tau_{cap}\}\). Each entry \((id_A,tid_A,id_B,tid_B,\text{dist}(A,B))\) records the source fields, tuple pointers, and distance; the distance is converted to the query's higher-is-better affinity only during ranking (e.g., by subtracting from 1). Thus, \(\tau_{cap}\) is an \textbf{index-coverage boundary}.  Standard relational joins chain multiple \idx steps for multi-hop patterns; this composes thresholded links and does not assume transitive embedding similarity. (We discuss correlation in Section~\ref{sec:pancaking}.)

\subsection{Correctness and Coverage}
\label{sec:semji-contract}
An exact \idx construction stores every pair in \(\mathcal{J}_{\tau_{cap}}\). For a query threshold \(\tau_q\le\tau_{cap}\), the index answers its similarity-join predicate exactly. For \(\tau_q>\tau_{cap}\), exact execution additionally enumerates the \emph{outer band} \(\tau_{cap}<\text{dist}(A,B)\le\tau_q\) through the base engine.

An ANN-built \idx is instead approximate: it stores \(\widehat{\mathcal{J}}_{\tau_{cap}}\subseteq\mathcal{J}_{\tau_{cap}}\). Filtering by distance removes false positives among retrieved candidates, but ANN traversal can miss an under-the-cap pair and hence create false negatives. Completing only the outer band cannot repair such a miss. Therefore, a result using an approximate \idx is exact only with respect to its retrieved candidate set. The approximate \idx inherits any guarantee of its underlying ANN method; HNSW has no general completeness guarantee, whereas some alternatives have approximation bounds~\cite{gollapudi_filtered-diskann_2023}.



\subsection{Index Construction}
\label{sec:index_construction}

The \idx construction process determines what guarantees it provides.


\paragraph{Exact construction}
Exact construction enumerates the Cartesian product and retains every pair within \(\tau_{cap}\); it provides the ground truth for approximate-construction recall.

\paragraph{Approximate construction} leverages an ANN structure. For a neighbor relationship, it probes one table against an index over the second one, and exactly refines all returned candidates. This filter-and-refine procedure removes returned candidates outside \(\tau_{cap}\) but does not establish coverage.

\textbf{E2LSH filter-and-refine (alternative strategy).}
For \(L_2\) distance, E2LSH~\cite{datar2004locality} hashes each vector \(x\) using
\(h_{a,b}(x)=\lfloor(a^\top x+b)/w\rfloor\), where
\(a\sim\mathcal{N}(0,I)\) and \(b\sim\mathrm{Uniform}[0,w)\).
The build phase stores each larger-relation vector in one or more hash tables;
the probe phase retrieves larger-relation vectors sharing a hash bucket with
each smaller-relation vector. We then compute exact distances for these
collision candidates and materialize only pairs satisfying
\(\mathrm{dist}(a,b)\le\tau_{cap}\). Thus, refinement eliminates false
positives, whereas missed collisions may yield false negatives. We did not implement this strategy in practice, since a hash-table and bucket-management path is less
well suited to an in-engine PostgreSQL implementation than HNSW traversal.
If enabled, it requires reporting the number of tables, concatenated hashes,
and tuned width (for example, \(w\approx4\tau_{cap}\) is only a high-recall
heuristic).

\textbf{Radius-expanded HNSW (Algorithm~\ref{alg:radius_hnsw}), preferred strategy.}
Standard HNSW is optimized for top-\(k\) retrieval and can prune candidates inside a requested radius. Our \textbf{radius-expanded traversal} relaxes the pruning gate to the maximum of the beam boundary and \(\tau_{cap}\):
    \(\text{gate} = \max(\mathcal{W}.\text{furthest}(), \tau_{cap})\).
This improves empirical range-query recall but does not establish completeness; we measure it against exact construction.

\begin{algorithm}
\caption{Layer Traversal in Radius-Expanded HNSW}
\label{alg:radius_hnsw}
\begin{algorithmic}[1]
\Require Query \(q\), Entry Point \(ep\), Beam \(ef\), Query Radius \(\tau_q\le\tau_{cap}\), Construction Cap \(\tau_{cap}\)
\Ensure Approximate Result Set \(\mathcal{R}\)
\State \(d_{ep} \leftarrow \text{dist}(ep,q)\); \(\mathcal{C} \leftarrow \text{MinHeap}(ep,d_{ep})\); \(\mathcal{W} \leftarrow \text{MaxHeap}(ep,d_{ep})\); \(\mathcal{V} \leftarrow \{ep\}\)
\State \(\mathcal{R} \leftarrow \emptyset\)
\If{\(d_{ep} \leq \tau_q\)} \(\mathcal{R}.\text{add}(ep)\) \EndIf
\While{\(\mathcal{C}\) is not empty}
    \State \(c \leftarrow \mathcal{C}.\text{pop}()\); \(w \leftarrow \mathcal{W}.\text{peek}()\)
    \If{\(c.dist > w.dist\) \textbf{and} \(c.dist > \tau_{cap}\)}
        \State \textbf{break} \Comment{Prune only if outside both beam and radius}
    \EndIf
    \State \(gate \leftarrow \max(\mathcal{W}.\text{peek}().dist, \tau_{cap})\)
    \For{\(e \in \text{neighbors}(c)\)}
        \If{\(e \notin \mathcal{V}\)}
            \State \(\mathcal{V}.\text{add}(e)\); \(d_e \leftarrow \text{dist}(e, q)\)
            \If{\(d_e \le gate\) \textbf{or} \(|\mathcal{W}| < ef\)}
                \State \(\mathcal{C}.\text{push}(e, d_e)\); \(\mathcal{W}.\text{push}(e, d_e)\)
                \If{\(|\mathcal{W}| > ef\)} \(\mathcal{W}.\text{pop\_furthest}()\) \EndIf
                \If{\(d_e \le \tau_q\)} \(\mathcal{R}.\text{add}(e)\) \EndIf
            \EndIf
        \EndIf
    \EndFor
\EndWhile
\State \Return \(\mathcal{R}\)
\end{algorithmic}
\end{algorithm}

\subsection{Space Cost}
\label{sec:semji-space}
\idx{} is a materialized relation, with \(J=|\mathcal{J}_{\tau_{cap}}|\) retained pairs. Its base representation and any distance/order or endpoint lookup indexes require \(O(J)\) records. Because the \idx is constructed only for frequently used embedding-column pairs whose calibrated tail has a bounded expected degree, the retained-pair fraction \(\rho=J/(|A||B|)\) is small, so storage scales as \(\rho|A||B|\) rather than the full product. Nonetheless, we enforce a maximum space budget per index, adjusting \(\tau_{cap}\) to reach that budget. The configurations in Section~\ref{sec:experiments} show practical settings.



\subsection{Batched Execution Optimization}
\label{sec:batched_execution}
Both exact and approximate \idx construction options batch probes. Exact construction uses blocked, vectorized distance computation and writes only surviving pairs. Approximate construction groups HNSW probes with shared upper-layer traversals and E2LSH probes by hash bucket, improving locality for candidate access.

\subsection{Multi-Step Reasoning}
\label{sec:multi_hop_linkage_def}

\idx supports multi-hop paths (e.g., \(A \xrightarrow{\tau_1} B \xrightarrow{\tau_2} C\)) by treating its materialized connectivity as a relational view. For a chain \(\mathcal{J}_{chain}=\{(\mathcal{J}_1,\tau_1),\ldots,(\mathcal{J}_m,\tau_m)\}\), the engine computes:
\begin{equation}
    \mathcal{R}_{chain} \leftarrow \textit{Fetch}(\mathcal{J}_1, \tau_1) \bowtie \textit{Fetch}(\mathcal{J}_2, \tau_2) \bowtie \dots \bowtie \textit{Fetch}(\mathcal{J}_m, \tau_m)
\end{equation}
where each \textit{Fetch} applies its runtime threshold \(\tau_i\le\tau_{cap}\). Projecting \(\mathcal{R}_{chain}\) onto a target table yields the semantic semijoin predicate \(P_{\mathcal{J}}\), which the Filtered-Score Streamer (Section~\ref{sec:fss}) pushes down before scoring.

\subsection{Capacity-Based Threshold Maintenance}
\label{sec:maintenance}

We can maintain both exact and approximate \idx structures with the same basic mechanism.  
Initially, the \idx has a \(\tau_{cap}\) set based on a small warmup sample and query.  Of course, the workload may introduce queries in the future that require a different threshold.  Thus, the controller grows the index by its \emph{materialized-pair capacity}, which allows us to limit the estimated number of growth steps. Let \(J(\tau)=|\mathcal{J}_{\tau}|\) and let \(J=J(\tau_{cap})\) be the current number of retained pairs. A space budget \(B_{max}\) bounds every \idx{}.

The budget is on retained pairs, not on the numerical radius:
embedding geometry can make a small change in radius admit many pairs, whereas
\(J\) is directly proportional to base \idx{} storage (up to fixed per-pair index
overhead). Even with a large budget, we would expect to materialize a very small percentage
of the possible pairs, as we see in Section~\ref{sec:experiments}.

\emph{Overflow and capacity doubling.}
When a query requests \(\tau_q>\tau_{cap}\), \sys answers it both using the \idx along with base-engine enumeration of the outer band \(\tau_{cap}<\mathrm{dist}(a,b)\leq\tau_q\). Thus, index growth is not on the query's critical path. The controller records the overflow and asynchronously doubles its
retained-pair capacity, targeting \(B'\) pairs at radius \(\tau_{next}\),
\[
 B' = \min\{B_{max},\max\{2J,J_{min}\}\},\quad
 \tau_{next}=\inf\{\tau\geq\tau_{cap}: J(\tau)\geq B'\},
\]
where \(J_{min}\) prevents a near-empty initial index from growing in negligible increments. The controller expands only when \(J<B_{max}\), materializes the new values in the range \((\tau_{cap},\tau_{next}]\), and publishes a new index version. A single unusually permissive query need not make the index cover \(\tau_q\): strict capacity doubling preserves the budget, while that query still uses its fallback. Ties or a sharp density increase can make \(J(\tau_{next})\) exceed \(B'\); the controller rejects that expansion if it would exceed \(B_{max}\).

This policy is independent of the distribution of pairs under the threshold, although anisotropy can change the radius needed to admit \(B'\) pairs. We additionally monitor endpoint-degree skew, since a modest total \(J\) concentrated at a few endpoints can still make downstream joins expensive.

\emph{Amortization and data evolution.}
For a static relation, capacity doubling yields at most \(\lceil\log_2(B_{max}/J_{min})\rceil\) expansions. If an expansion discovers, writes, and indexes only new distance-bands, the total pair-materialization work through final size \(J_f\) is \(O(J_f-J_0)\); even rebuilding the materialized relation at every doubling has a geometric-series bound below \(2J_f\) for pair writing. This is an output-space amortization guarantee, where the cost of the rescan is dependent on the radius-expanded HNSW of Algorithm~\ref{alg:radius_hnsw}.

\subsection{Robustness to Correlated Embeddings}
\label{sec:pancaking}

Our \idx{} materialization-radius heuristic relies on the pairwise-distance population being sufficiently isotropic for a tail threshold such as \(\tau_{cap}=\mu-3\sigma\) to be informative. This can fail when endpoint embeddings are anisotropic or share dominant directions, because they were extracted from overlapping source context. Such geometry can distort the apparent distance distribution, so a statistically chosen materialization radius may omit useful close pairs~\cite{ethayarajh2019contextual,mu_allbutthetop_2018,su_whitening_2021}.
We can detect anisotropy by computing the covariance of the centered embeddings from a query; if the variance is concentrated in a few directions, this indicates that a raw-distance tail estimate may be error-prone once predicates or other neighborhood relationships are factored in.


In lieu of the baseline strategy of calibrating thresholds by \(\mu-3\sigma\) over a sample, \idx{} can adopt mitigation strategies from the literature for anisotropic data. \textbf{PCA whitening} mean-centers and rescales principal axes~\cite{su_whitening_2021}; \textbf{all-but-the-top (ABTT)} removes a configured number of dominant directions~\cite{mu_allbutthetop_2018}. Because these transformations define different retrieval metrics, \idx{} recalibrates \(\tau_{cap}\) and builds a matching transformed index; it never queries transformed vectors against the raw index.

Even when anisotropy makes the choice of the index-selection threshold unreliable, it need not make the admitted relation dense. \idx{} materializes pairs with \(\mathrm{dist}(a,b)\leq\tau_{cap}\); equivalently, \(\tau_{cap}\) retains pairs whose distance-derived affinity is at least the calibrated threshold. As Section~\ref{sec:pancaking-experiments} shows, these near-neighbor edges remain a tail of the real endpoint-pair products (0.03--1.09\% under the evaluated policies), even for the anisotropic corpora.

\begin{figure}[tb]
    \centering
    \includegraphics[width=1\linewidth]{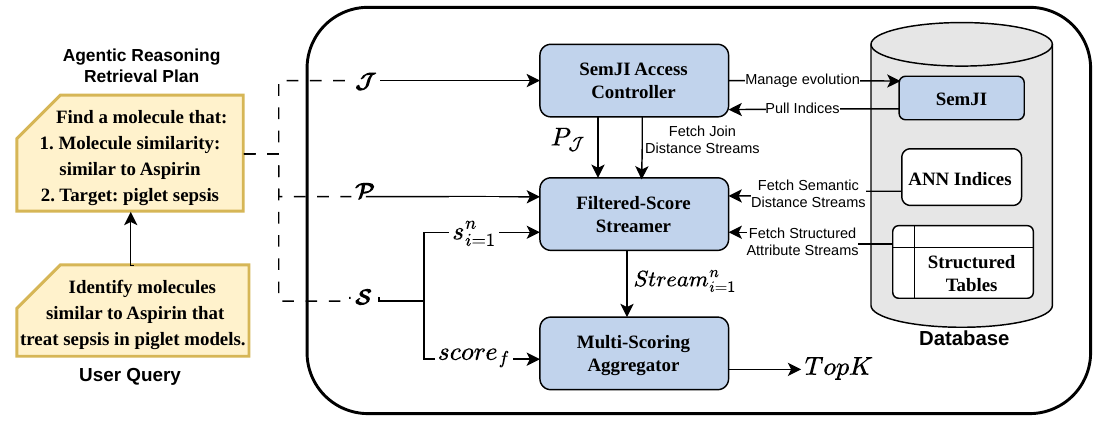}  
    \vspace{-8mm}
    \caption{\label{fig:system_arch}\sysb architecture}
    \vspace{-3mm}
\end{figure}

\section{\sysb Execution Strategy}
\label{sec:execution}

The execution layer of \sysb consists of three core components that interoperate to tackle multi-step reasoning with both similarity and Boolean predicates: the \textbf{Multi-Scoring Aggregator} (MSA), multiple \textbf{Filtered-Score Streamers} (FSSs), and the \textbf{\idx Access Controller}, which exposes materialized connectivity. Predicate-aware ANN traversal is one access mode within FSS, alongside predicate-first and vector-first execution. Our approach requires a coordinatewise non-decreasing scoring function over higher-is-better component scores~\cite{fagin_optimal_2003}.

\subsection{Predicate-aware ANN Index Traversal}
\label{sec:predicate_ann}

Most hybrid search systems split their strategies into two camps~\cite{gollapudi_filtered-diskann_2023,lu2026depth}: \textit{vector-first} (post-filtering) and \textit{filter-first} (pre-filtering). This creates a performance gap: vector-first works well when most items match the filter, while filter-first is better when the filter eliminates almost everything. Between these two extremes lies a ``middle ground'' where neither approach is efficient. To handle this, we implemented a \textbf{Predicate-aware ANN Index Traversal} that checks filters while walking the index graph. \textbf{Often one of the filter conditions is whether the ANN node endpoints lie within a batch of path endpoints retrieved from the \idx index.}

\textbf{Simulating an Oracle Graph.} Our method is inspired by the \textit{Oracle Graph} idea from ACORN~\cite{patel2024acorn}. Conceptually, if we had a custom index built only from the items that pass the filter, search would be optimal. Since we cannot build a new index for every query, we simulate this experience using the global HNSW index.

\textbf{Transitive neighbor expansion.} The main problem with walking a global ANN graph under constraints is that a path to a good candidate may need to traverse a node that fails the filter. To mitigate this, we modify the greedy search: When the immediate neighbors of a node are all filtered out, we look at their neighbors (2 hops away). By using these invalid nodes as bridges rather than dead ends, we can jump over gaps in the graph and find the true nearest neighbors that satisfy the query.

\textbf{Bloom filter optimization.} As noted above, \sys will frequently combine \idx results with an ANN traversal --- doing a similarity join over a similarity-based lookup. Checking the filter condition for every node during a graph traversal is too slow if it requires fetching data from the main table. To speed this up, we compile the query's node filter set into a \textbf{Bloom filter} 
~\cite{bloom_spacetime_1970} before the search begins. This allows the traversal engine to check if a node is valid using a fast bitwise operation.
For a target false-positive rate \(\epsilon\), the Bloom filter uses \(O(N\log(1/\epsilon))\) bits and takes \(O(N)\) time to build. If the dataset is too large, the system falls back to standard disk-based strategies to avoid exhausting memory.


\subsection{Filtered-Score Streamer}
\label{sec:fss}
The \textbf{Filtered-Score Streamer (FSS)} generates a sorted stream for \textbf{one} higher-is-better scoring signal under constraints. Formally, its input is (1) a set of predicates \(\mathcal{P}\) and (2) a scoring signal \(s\), which is one of: (a) query-to-object affinity, (b) relational-link affinity, or (c) structural-attribute desirability, according to the Query Model in Section~\ref{sec:prob_def}. Distance-derived signals are first transformed by the query's non-increasing affinity function. The output is an iterable stream of entries that pass the filters in descending score order: \(FSS(\mathcal{P},s)\rightarrow Stream((id_i,score_i))\).
The FSS adapts its execution path based on the data source and predicate selectivity \(\sigma\).

 \textbf{(a) Semantic Path}: For a query-to-object affinity derived from \(\text{dist}(\mathbf{q},e_i)\), the scorer selects between attribute-first filtering, vector-first ANN traversal, and the predicate-aware ANN traversal based on \(\sigma\) to minimize distance-computation overhead. When \(\sigma<\sigma_1\), attribute-first filtering leaves a short candidate list for direct distance computation; when \(\sigma>\sigma_2\), vector-first search with expanded retrieval and post-filtering is preferable; and when \(\sigma_1<\sigma<\sigma_2\), the system uses predicate-aware traversal.
        
       \textbf{\textit{Selection of \(\sigma\) thresholds.}} Offline profiling identifies crossover selectivities \(\sigma_1,\sigma_2\) at a fixed recall target: attribute-first below \(\sigma_1\), vector-first above \(\sigma_2\), and predicate-aware ANN between them.
        
         \textbf{(b) Relational Path}: For a relational-link affinity derived from \(\text{dist}(A,B)\), the scorer scans the \textbf{join index} in ascending distance order (equivalently, descending affinity order), while applying \(\mathcal{P}\) to filtering attributes stored in the join index or entity metadata. This can be accelerated by traditional B-tree indices on the join index.
         
        \textbf{(c) Attribute Path}: For a structured attribute scoring signal, we perform filtering with B-tree indices over the relational data.

\subsection{Multi-Scoring Aggregator}
\label{sec:msa}
The \textbf{Multi-Scoring Aggregator (MSA)} combines the sorted streams \(\{\mathcal{S}_i\}_{i=1}^n\) produced by FSS under a coordinatewise non-decreasing, higher-is-better aggregation function \(score_f\), returning \linebreak
\(\text{MSA}(\{\mathcal{S}_i\}_{i=1}^n,score_f,K)\). It is a direct adaptation of Fagin's Threshold Algorithm (TA)~\cite{fagin_optimal_2003}, with Rank-Join-style random access~\cite{li_ranksql_2005} to reconstruct relationally valid tuples. Thus, MSA does not replace TA's ranking logic: it retains sorted access, random access, a top-\(k\) buffer, and TA's threshold stopping test.

On observing an entry \(e\) from any stream, MSA uses \idx lookups and sibling attributes to materialize every valid tuple containing \(e\), scores each newly discovered tuple, and maintains the best \(K\). The only DASE-specific changes are that random access can traverse a multi-hop \idx predicate and that the next stream need not be selected round-robin. If \(L[i]\) is the most recently seen score of stream \(i\), every unmaterialized tuple lies below every current cursor and is bounded by \(\tau=score_f(L[1],\ldots,L[n])\). MSA stops when its current \(K\)-th score is at least \(\tau\).

\paragraph{Exact answers.} This is the usual TA guarantee: the result is exact when all streams are complete and globally sorted, random access reconstructs every valid tuple containing a seen entry, and component scores are exact. If we leverage ANN traversal or an approximate \idx, the stopping rule provides exact answers only with respect to the set of retrieved candidates.

\paragraph{Distribution-Aware Stream Selection}
\label{sec:adaptive_selection}
TA normally advances streams in round-robin order. MSA instead chooses the stream expected to reduce the termination gap \(G=\tau-\theta_K\) most, where \(\theta_K\) is the current \(K\)-th score. For a linear \(score_f=\sum_i w_i s_i\), profiling supplies each stream's local score density \(f_i\) (an empirical density or a fitted Gaussian). The expected threshold reduction from advancing stream \(i\) is approximately \(w_i/(Nf_i(L[i]))\). MSA also estimates the expected increase in \(\theta_K\) from the newly materialized tuples using the other streams' cursor-truncated score distributions, and advances
\[
i^*=\arg\max_i\{\mathbb{E}[\Delta\tau_i]+\mathbb{E}[\Delta\theta_{K,i}]\}.
\]
Density and truncated-expectation values are tabulated during profiling. This is a greedy scheduling policy layered on TA, without changing its stopping condition or exactness guarantees. Its candidate-elevation estimate uses marginal stream distributions, and is therefore heuristic when ranking signals are correlated.

\subsection{\idx Access Controller}
\label{sec:controller}
The \textbf{\idx Access Controller} functions as the storage abstraction layer that exposes the materialized semantic connectivity of \idx to the query pipeline. It serves two distinct query-time roles depending on whether the relational data is treated as a hard constraint or a ranking signal. Additionally, it acts as the runtime monitor that drives the structural evolution of the index.

\textbf{Materializing join predicates.}
When the query includes hard relational constraints (\(\mathcal{J}\)), the controller's objective is to define the valid search space. It retrieves the set of materialized pairs for each join condition and, in the case of multi-hop dependencies, joins these sets to synthesize a unified \textbf{candidate set} of valid tuples. This process effectively converts topological constraints into a flat filter predicate \(P_{\mathcal{J}}\), which is pushed down to the \textit{Filtered-Score Streamer} to prune the search space. Symbolically,
\(P_{\mathcal{J}} \leftarrow \bowtie_{i=1}^{|\mathcal{J}|} \text{SemJIFetch}(\mathcal{J}_i)\).


\textbf{Streaming relational scores.}
When the query utilizes relational proximity as a ranking objective (i.e., a scoring component \(s \in \mathcal{S}\) is defined by the semantic link distance), the controller acts as a streaming source. It scans the underlying index to yield pairs \((id, dis)\) sorted by their precomputed semantic distance. This stream is fed directly into the \textit{Multi-Scoring Aggregator} as a standard ranking signal: \(\mathcal{S}_{rel} \leftarrow \text{SemJIFetch}(\mathcal{J}_S)\).


\textbf{Workload monitoring and evolution trigger.}
As described in Section~\ref{sec:maintenance}, the controller performs workload monitoring. During query execution, it tracks the \emph{base-engine completion rate}: the fraction of queries whose materialized \idx coverage is exhausted, i.e., that overflow to \(\tau_q > \tau_{cap}\). If this rate consistently exceeds its tolerance, the controller asynchronously triggers a capacity-doubling expansion, growing \(\tau_{cap} \rightarrow \tau_{next}\) within the budget \(B_{max}\) via Radius-Expanded HNSW. This allows the index to evolve without bottlenecking real-time execution.

\section{Experiments}
\label{sec:experiments}

\begin{table*}[t]
\centering
\caption{\label{tab:exp-setup} Main experimental data and configuration.  \idx{} counts are the materialized-pair records in the released configurations; a workload selects the radius appropriate to its join threshold.}
\vspace{-3mm}
\small
\setlength{\tabcolsep}{4pt}
\begin{tabular}{p{0.08\textwidth}p{0.29\textwidth}p{0.30\textwidth}p{0.25\textwidth}}
\toprule
\textbf{Scenario} & \textbf{Base relations} & \textbf{Candidate Links / \idx{} pairs} & \textbf{Workload (queries/class, top-\(20\))} \\
\midrule
\textbf{IMDB} & \texttt{imdb\_t1}: 42,378 rows; \texttt{imdb\_t2}: 43,237 rows &
  \(\tau_{cap}=0.5\): 2.10M; \(0.6\): 58.41M pairs &
  W1--W8; 84--100 \\
\textbf{MOLECULE} & \texttt{facts\_50k}: 50,000 rows; \texttt{paper}: 62,857 rows &
  \(\tau_{cap}=0.5\): 0.032M; \(0.6\): 1.38M; \(0.7\): 63.48M pairs &
  W1--W8; 93--100 \\
\textbf{SemBench} & Movie, E-Commerce, Wildlife, MMQA, and Cars scenarios &
  Calibrated embedding candidate links for semantic joins; no shared deep-research \idx{} &
  F/L/J/M/C/R tasks; \sys prefilter and \sys + BigQuery \\
\textbf{NFCorpus} & 2,868 papers; title embeddings &
  \(\tau=0.35\): 21,466 intervention--condition; 254 study--outcome pairs &
  depth 2--8; 90 instances (3 \(\times\) 30) \\
\midrule
\bottomrule
\end{tabular}
\vspace{-4mm}
\end{table*}

\begin{table}[tb]
\caption{\label{tab:workloads} Characterization of query workload classes.}
\vspace{-4mm}
\footnotesize
\begin{tabular}{@{}ccccp{1.8cm}p{3cm}@{}}
\toprule
\textbf{WL} & $\mathcal{P}$ & \textbf{Multi-}$\mathcal{S}$ & $\mathcal{J}$ & \textbf{Pattern} & \textbf{Expression} \\
\midrule
\multicolumn{6}{@{}l}{\textit{Intra-Entity Retrieval}} \\
\midrule
$W_1$ & $\times$ & $\times$ & $\times$
      & Semantic retrieval
      & $\mathit{TopK}\{\mathit{FSS}(\emptyset,\, s)\}$ \\[2pt]
$W_2$ & \checkmark & $\times$ & $\times$
      & Filtered retrieval
      & $\mathit{TopK}\{\mathit{FSS}(\mathcal{P},\, s)\}$ \\[2pt]
$W_3$ & $\times$ & \checkmark & $\times$
      & \makecell[tl]{Multi-scoring \\ retrieval}
      & $\mathit{MSA}(\{\mathit{FSS}(\emptyset,\, s_i)\}_{i=1}^{n},\;$ $ \mathit{score}_f,\, K)$ \\[2pt]
$W_4$ & \checkmark & \checkmark & $\times$
      & \makecell[tl]{Filtered \\ multi-scoring}
      & $\mathit{MSA}(\{\mathit{FSS}(\mathcal{P},\, s_i)\}_{i=1}^{n},\; $ $\mathit{score}_f,\, K)$ \\
\midrule
\multicolumn{6}{@{}l}{\textit{Inter-Entity Synthesis}} \\
\midrule
$W_5$ & $\times$ & $\times$ & \checkmark
      & Semantic join
      & $\mathit{TopK}\{\mathit{FSS}(P_{\mathcal{J}},\, s)\}$ \\[2pt]
$W_6$ & \checkmark & $\times$ & \checkmark
      & Filtered join
      & $\mathit{TopK}\{\mathit{FSS}(\mathcal{P} \cup P_{\mathcal{J}},\, s)\}$ \\[2pt]
$W_7$ & $\times$ & \checkmark & \checkmark
      & Multi-scoring join
      & $\mathit{MSA}(\{\mathit{FSS}(P_{\mathcal{J}},\, s_i)\}_{i=1}^{n},\;$ $ \mathit{score}_f,\, K)$ \\[2pt]
$W_8$ & \checkmark & \checkmark & \checkmark
      & \makecell[tl]{Filtered \\ multi-scoring join}
      & $\begin{aligned}[t]
  &\mathit{MSA}(\{\mathit{FSS}(\mathcal{P} \cup P_{\mathcal{J}}, s_i)\}_{i=1}^{n}, \\
  &\mathit{score}_f, K)
\end{aligned}$ \\
\bottomrule
\end{tabular}
\vspace{-6mm}
\end{table}
\begin{figure*}
    \centering
    \includegraphics[width=1\linewidth]{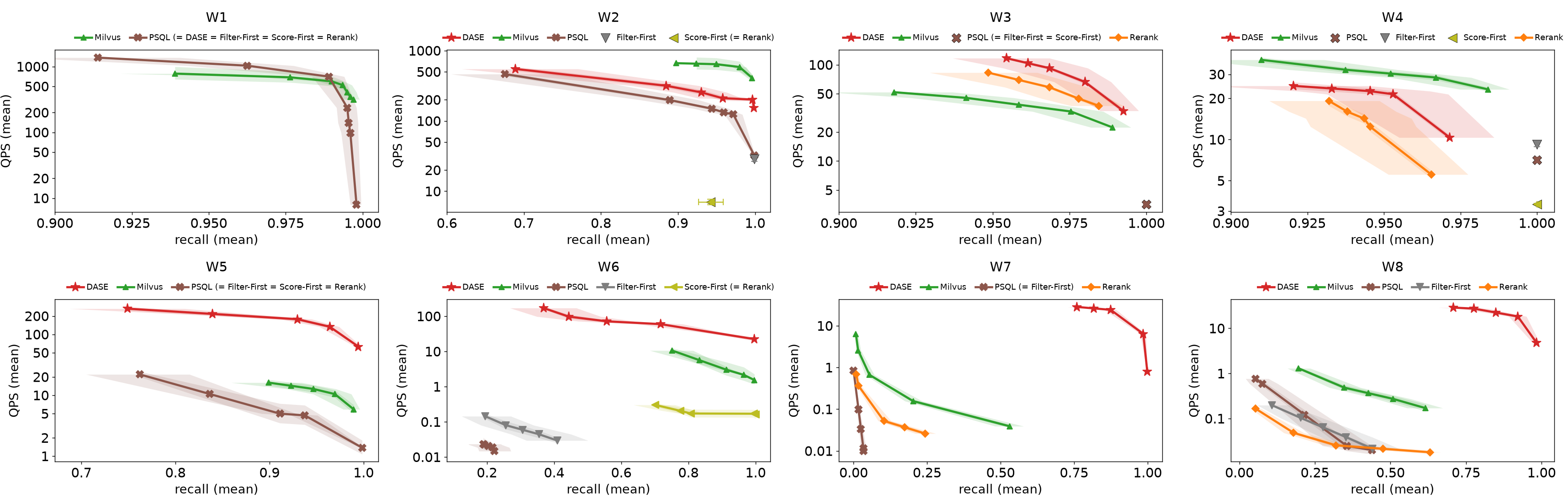}  
    \vspace{-8mm}
    \caption{\label{fig:mrq-imdb} Multi-step Reasoning Top-k: Deep Research workload on IMDB}
    \vspace{-3mm}
\end{figure*}

\begin{figure*}
    \centering
    \includegraphics[width=1\linewidth]{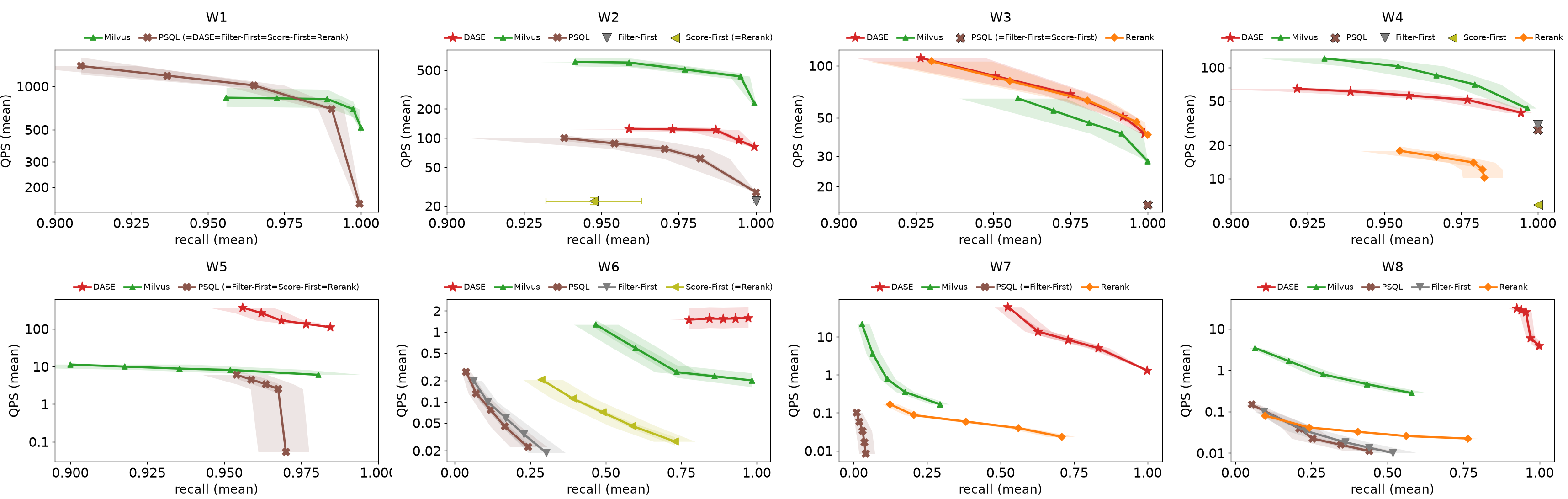}  
    \vspace{-8mm}
    \caption{\label{fig:mrq-molecule} Multi-step Reasoning Top-k: Deep Research workload on MOLECULE}
    \vspace{-3mm}
\end{figure*}

\begin{table*}[tbp]
\caption{\label{tab:sembench} SemBench cost, quality, and latency {\footnotesize (notation: F = filter; L = limit-$k$; J = join; M = semantic map; C = semantic classify; R = semantic rank)}}
\vspace{-4mm}
\centering
\fontsize{8.5pt}{1.5pt}\selectfont
\setlength{\tabcolsep}{0.5pt}
\begin{tabular}{l | ccc | ccc | ccc | ccc | ccc | ccc}
\toprule
& \multicolumn{3}{c|}{\textbf{LOTUS \cite{patel_semantic_2025}}} & \multicolumn{3}{c|}{\textbf{Palimpzest \cite{liu2025palimpzest}}} & \multicolumn{3}{c|}{\textbf{ThalamusDB \cite{jo_thalamusdb_2024}}} & \multicolumn{3}{c|}{\textbf{BigQuery \cite{fernandes2015bigquery}}} & \multicolumn{3}{c|}{\textbf{\sys}} & \multicolumn{3}{c}{\textbf{\sys + BigQuery}} \\
 & \textbf{Cost} & \textbf{Qual.} & \textbf{Lat.} & \textbf{Cost} & \textbf{Qual.} & \textbf{Lat.} & \textbf{Cost} & \textbf{Qual.} & \textbf{Lat.} & \textbf{Cost} & \textbf{Qual.} & \textbf{Lat.} & \textbf{Cost} & \textbf{Qual.} & \textbf{Lat.}& \textbf{Cost} & \textbf{Qual.} & \textbf{Lat.}   \\
\specialrule{1.2pt}{2pt}{2pt}


\multicolumn{1}{l}{} & \multicolumn{9}{l}{\Large \textbf{(a) Movie Scenario}} &  \multicolumn{3}{c}{\emph{DASE Preprocess}} & \cellcolor{cellgreen}\$3$\cdot 10^{-3}$ & \textbf{ } & \cellcolor{cellgreen}27.6 s & \cellcolor{cellgreen}\$3$\cdot 10^{-3}$ & \textbf{ } & \cellcolor{cellgreen}27.6 s \\
\textbf{Q1: F L} & \cellcolor{cellgrey}\$0.09 & \cellcolor{cellgreen}\textbf{1.00} & \cellcolor{cellgrey}33.1 s & \cellcolor{cellyellow}\$1$\cdot 10^{-3}$ & \cellcolor{cellgreen}\textbf{1.00} & \cellcolor{cellyellow}3.8 s & \cellcolor{cellyellow}\$4$\cdot 10^{-4}$ & \cellcolor{cellgreen}0.95 & \cellcolor{cellyellow}4.2 s & \cellcolor{cellgrey}\$0.05 & \cellcolor{cellgreen}\textbf{1.00} & \cellcolor{cellgrey}26.3 s & \cellcolor{cellgreen}\$2.4$\cdot 10^{-6}$ & \cellcolor{cellgreen}\textbf{1.00} & \cellcolor{cellgreen}0.5 s & \cellcolor{cellgreen}\$1$\cdot 10^{-4}$ & \cellcolor{cellgreen}\textbf{1.00} & \cellcolor{cellyellow}3.8 s \\
\textbf{Q2: F L} & \cellcolor{cellgrey}\$0.01 & \cellcolor{cellgreen}\textbf{1.00} & \cellcolor{cellyellow}2.1 s & \cellcolor{cellgrey}\$0.01 & \cellcolor{cellgreen}\textbf{1.00} & \cellcolor{cellgrey}29.7 s & \cellcolor{cellyellow}\$2$\cdot 10^{-3}$ & \cellcolor{cellyellow}0.92 & \cellcolor{cellyellow}1.9 s & \cellcolor{cellyellow}\$3$\cdot 10^{-3}$ & \cellcolor{cellgreen}\textbf{1.00} & \cellcolor{cellgrey}9.5 s & \cellcolor{cellgreen}\$2.4$\cdot 10^{-6}$ & \cellcolor{cellgreen}\textbf{1.00} & \cellcolor{cellgreen}0.5 s & \cellcolor{cellgreen}\$1$\cdot 10^{-4}$ & \cellcolor{cellgreen}\textbf{1.00} & \cellcolor{cellyellow}4.2 s \\
\textbf{Q3: F} & \cellcolor{cellyellow}\$5$\cdot 10^{-3}$ & \cellcolor{cellyellow}0.64 & \cellcolor{cellyellow}2.1 s & \cellcolor{cellgrey}\$0.01 & \cellcolor{cellyellow}0.64 & \cellcolor{cellyellow}4.6 s & \cellcolor{cellyellow}\$2$\cdot 10^{-3}$ & \cellcolor{cellyellow}0.74 & \cellcolor{cellyellow}3.1 s & \cellcolor{cellyellow}\$3$\cdot 10^{-3}$ & \cellcolor{cellyellow}0.64 & \cellcolor{cellgrey}11.0 s & \cellcolor{cellgreen}\$3.6$\cdot 10^{-6}$ & \cellcolor{cellgreen}\textbf{0.82} & \cellcolor{cellgreen}0.5 s & \cellcolor{cellgreen}\$7$\cdot 10^{-4}$ & \cellcolor{cellgreen}\textbf{0.82} & \cellcolor{cellyellow}6.2 s \\
\textbf{Q4: F} & \cellcolor{cellyellow}\$5$\cdot 10^{-3}$ & \cellcolor{cellyellow}0.64 & \cellcolor{cellyellow}2.8 s & \cellcolor{cellgrey}\$0.02 & \cellcolor{cellyellow}0.74 & \cellcolor{cellyellow}4.4 s & \cellcolor{cellyellow}\$2$\cdot 10^{-3}$ & \cellcolor{cellyellow}0.74 & \cellcolor{cellyellow}3.8 s & \cellcolor{cellyellow}\$3$\cdot 10^{-3}$ & \cellcolor{cellyellow}0.64 & \cellcolor{cellgrey}11.4 s & \cellcolor{cellgreen}\$3.6$\cdot 10^{-6}$ & \cellcolor{cellgreen}\textbf{0.82} & \cellcolor{cellgreen}0.5 s & \cellcolor{cellgreen}\$7$\cdot 10^{-4}$ & \cellcolor{cellgreen}\textbf{0.82} & \cellcolor{cellyellow}5.3 s \\
\textbf{Q5: J L} & \cellcolor{cellgrey}\$2.38 & \cellcolor{cellgrey}0.59 & \cellcolor{cellgrey}536.5 s & \cellcolor{cellyellow}\$0.01 & \cellcolor{cellgrey}0.39 & \cellcolor{cellyellow}1.9 s & \cellcolor{cellgreen}\$1$\cdot 10^{-3}$ & \cellcolor{cellgreen}\textbf{1.00} & \cellcolor{cellyellow}2.3 s & \cellcolor{cellgrey}\$1.01 & \cellcolor{cellyellow}0.89 & \cellcolor{cellgrey}54.5 s & \cellcolor{cellgreen}\$3.6$\cdot 10^{-6}$ & \cellcolor{cellgreen}\textbf{1.00} & \cellcolor{cellgreen}0.5 s & \cellcolor{cellgreen}\$6$\cdot 10^{-4}$ & \cellcolor{cellyellow}0.82 & \cellcolor{cellyellow}3.9 s \\
\textbf{Q6: J L} & \cellcolor{cellgrey}\$1.81 & \cellcolor{cellgrey}0.67 & \cellcolor{cellgrey}432.4 s & \cellcolor{cellyellow}\$0.01 & \cellcolor{cellyellow}0.83 & \cellcolor{cellyellow}2.3 s & \cellcolor{cellgreen}\$9$\cdot 10^{-4}$ & \cellcolor{cellyellow}0.84 & \cellcolor{cellyellow}1.7 s & \cellcolor{cellgrey}\$1.00 & \cellcolor{cellgrey}0.69 & \cellcolor{cellgrey}54.5 s & \cellcolor{cellgreen}\$3.6$\cdot 10^{-6}$ & \cellcolor{cellgreen}\textbf{1.00} & \cellcolor{cellgreen}0.5 s & \cellcolor{cellgreen}\$7$\cdot 10^{-4}$ & \cellcolor{cellyellow}0.94 & \cellcolor{cellyellow}3.8 s \\
\textbf{Q7: J} & \cellcolor{cellyellow}\$1.81 & \cellcolor{cellgrey}0.21 & \cellcolor{cellyellow}431.8 s & \cellcolor{cellgrey}\$7.72 & \cellcolor{cellgreen}0.68 & \cellcolor{cellgrey}1056.1 s & \cellcolor{cellgreen}\$0.15 & \cellcolor{cellyellow}0.57 & \cellcolor{cellgrey}636.9 s & \cellcolor{cellgrey}\$3.31 & \cellcolor{cellgreen}0.70 & \cellcolor{cellyellow}198.3 s & \cellcolor{cellgreen}\$3.6$\cdot 10^{-6}$ & \cellcolor{cellgreen}\textbf{0.73} & \cellcolor{cellgreen}0.5 s & \cellcolor{cellgreen}\$9$\cdot 10^{-4}$ & \cellcolor{cellgreen}0.71 & \cellcolor{cellgreen}3.8 s \\
\textbf{Q8: C} & \cellcolor{cellyellow}\$4$\cdot 10^{-3}$ & \cellcolor{cellgreen}\textbf{0.93} & \cellcolor{cellyellow}2.3 s & \cellcolor{cellgrey}\$0.02 & \cellcolor{cellyellow}0.86 & \cellcolor{cellyellow}4.3 s & \cellcolor{cellyellow}\$5$\cdot 10^{-3}$ & \cellcolor{cellyellow}0.83 & \cellcolor{cellyellow}6.8 s & \cellcolor{cellyellow}\$3$\cdot 10^{-3}$ & \cellcolor{cellyellow}0.76 & \cellcolor{cellgrey}10.9 s & \cellcolor{cellgreen}\$3.6$\cdot 10^{-6}$ & \cellcolor{cellgreen}0.89 & \cellcolor{cellgreen}0.5 s & \cellcolor{cellgreen}\$7$\cdot 10^{-4}$ & \cellcolor{cellgreen}0.89 & \cellcolor{cellyellow}4.2 s \\
\textbf{Q9: R} & \cellcolor{cellyellow}\$0.02 & \cellcolor{cellgreen}0.75 & \cellcolor{cellyellow}4.9 s & \cellcolor{cellgrey}\$0.05 & \cellcolor{cellgreen}\textbf{0.78} & \cellcolor{cellyellow}5.7 s & \cellcolor{cellred}\textbf{X} & \cellcolor{cellred}\textbf{X} & \cellcolor{cellred}\textbf{X} & \cellcolor{cellyellow}\$0.02 & \cellcolor{cellgreen}\textbf{0.78} & \cellcolor{cellgrey}13.3 s & \cellcolor{cellgreen}\$4.8$\cdot 10^{-6}$ & \cellcolor{cellyellow}0.66 & \cellcolor{cellgreen}0.5 s & \cellcolor{cellgreen}\$6$\cdot 10^{-3}$ & \cellcolor{cellgreen}0.74 & \cellcolor{cellgrey}12.8 s \\
\textbf{Q10: R} & \cellcolor{cellyellow}\$0.13 & \cellcolor{cellgreen}0.40 & \cellcolor{cellyellow}30.9 s & \cellcolor{cellgrey}\$0.38 & \cellcolor{cellgreen}0.42 & \cellcolor{cellyellow}39.2 s & \cellcolor{cellred}\textbf{X} & \cellcolor{cellred}\textbf{X} & \cellcolor{cellred}\textbf{X} & \cellcolor{cellyellow}\$0.13 & \cellcolor{cellgreen}\textbf{0.44} & \cellcolor{cellyellow}32.1 s & \cellcolor{cellgreen}\$4.8$\cdot 10^{-6}$ & \cellcolor{cellgreen}0.43 & \cellcolor{cellgreen}0.5 s & \cellcolor{cellyellow}\$0.085 & \cellcolor{cellgreen}0.41 & \cellcolor{cellyellow}26.1 s \\
\multicolumn{1}{c|}{\emph{Avg}} & \cellcolor{cellgrey}\$0.626 & \cellcolor{cellyellow}0.68 & \cellcolor{cellgrey}147.9 s & \cellcolor{cellgrey}\$0.823 & \cellcolor{cellyellow}0.73 & \cellcolor{cellyellow}115.2 s & \cellcolor{cellgreen}\$0.02 & \cellcolor{cellgreen}0.82 & \cellcolor{cellyellow}82.6 s & \cellcolor{cellyellow}\$0.553 & \cellcolor{cellyellow}0.75 & \cellcolor{cellyellow}42.2 s & \cellcolor{cellgreen}\$3$\cdot 10^{-4}$ & \cellcolor{cellgreen}\textbf{0.83} & \cellcolor{cellgreen}3.0 s & \cellcolor{cellgreen}\$9$\cdot 10^{-3}$ & \cellcolor{cellgreen}0.81 & \cellcolor{cellgreen}9.2 s \\

\specialrule{1.2pt}{2pt}{2pt}
\multicolumn{1}{l}{} & \multicolumn{9}{l}{\Large \textbf{(b) E-Commerce Scenario}} &  \multicolumn{3}{c}{\emph{DASE Preprocess}} &  \cellcolor{cellgreen}\$0.03 & \textbf{ } & \cellcolor{cellgreen}124.0 s & \cellcolor{cellgreen}\$0.03 & \textbf{ } & \cellcolor{cellgreen}124.0 s \\
\textbf{Q1: F} & \cellcolor{cellyellow}\$0.06 & \cellcolor{cellgreen}\textbf{1.00} & \cellcolor{cellyellow}12.2 s & \cellcolor{cellgrey}\$0.08 & \cellcolor{cellgreen}\textbf{1.00} & \cellcolor{cellyellow}12.2 s & \cellcolor{cellyellow}\$0.03 & \cellcolor{cellgreen}\textbf{1.00} & \cellcolor{cellgrey}34.8 s & \cellcolor{cellyellow}\$0.04 & \cellcolor{cellgrey}0.59 & \cellcolor{cellyellow}21.2 s & \cellcolor{cellgreen}\$5$\cdot 10^{-6}$ & \cellcolor{cellyellow}0.94 & \cellcolor{cellgreen}0.7 s & \cellcolor{cellgreen}\$7$\cdot 10^{-3}$ & \cellcolor{cellgreen}\textbf{1.00} & \cellcolor{cellyellow}8.0 s \\
\textbf{Q2: F} & \cellcolor{cellyellow}\$0.18 & \cellcolor{cellgreen}\textbf{0.87} & \cellcolor{cellgrey}166.1 s & \cellcolor{cellyellow}\$0.38 & \cellcolor{cellgreen}0.83 & \cellcolor{cellyellow}57.5 s & \cellcolor{cellyellow}\$0.31 & \cellcolor{cellyellow}0.67 & \cellcolor{cellyellow}100.8 s & \cellcolor{cellgrey}\$3.96 & \cellcolor{cellgrey}0.21 & \cellcolor{cellyellow}55.7 s & \cellcolor{cellgreen}\$7$\cdot 10^{-6}$ & \cellcolor{cellgrey}0.47 & \cellcolor{cellgreen}0.7 s & \cellcolor{cellgreen}\$0.02 & \cellcolor{cellgrey}0.66 & \cellcolor{cellgreen}5.5 s \\
\textbf{Q3: M} & \cellcolor{cellyellow}\$0.07 & \cellcolor{cellgreen}0.97 & \cellcolor{cellyellow}17.1 s & \cellcolor{cellyellow}\$0.12 & \cellcolor{cellgreen}\textbf{0.98} & \cellcolor{cellyellow}16.5 s & \cellcolor{cellred}\textbf{X} & \cellcolor{cellred}\textbf{X} & \cellcolor{cellred}\textbf{X} & \cellcolor{cellyellow}\$0.12 & \cellcolor{cellgreen}0.97 & \cellcolor{cellyellow}21.2 s & \cellcolor{cellred}\textbf{X} & \cellcolor{cellred}\textbf{X} & \cellcolor{cellred}\textbf{X} & \cellcolor{cellgreen}\$0.04 & \cellcolor{cellgreen}0.93 & \cellcolor{cellyellow}10.5 s \\
\textbf{Q4: M} & \cellcolor{cellyellow}\$0.14 & \cellcolor{cellgrey}0.45 & \cellcolor{cellyellow}156.7 s & \cellcolor{cellyellow}\$0.24 & \cellcolor{cellyellow}0.53 & \cellcolor{cellgrey}335.0 s & \cellcolor{cellred}\textbf{X} & \cellcolor{cellred}\textbf{X} & \cellcolor{cellred}\textbf{X} & \cellcolor{cellyellow}\$0.37 & \cellcolor{cellgreen}\textbf{0.69} & \cellcolor{cellgreen}31.0 s & \cellcolor{cellred}\textbf{X} & \cellcolor{cellred}\textbf{X} & \cellcolor{cellred}\textbf{X} & \cellcolor{cellyellow}\$0.17 & \cellcolor{cellgreen}0.67 & \cellcolor{cellgreen}18.3 s \\
\textbf{Q5: C} & \cellcolor{cellyellow}\$0.04 & \cellcolor{cellgreen}\textbf{0.99} & \cellcolor{cellyellow}7.4 s & \cellcolor{cellyellow}\$0.07 & \cellcolor{cellgreen}0.98 & \cellcolor{cellyellow}6.6 s & \cellcolor{cellred}\textbf{X} & \cellcolor{cellred}\textbf{X} & \cellcolor{cellred}\textbf{X} & \cellcolor{cellgrey}\$0.17 & \cellcolor{cellgreen}0.98 & \cellcolor{cellgrey}25.7 s & \cellcolor{cellgreen}\$4$\cdot 10^{-6}$ & \cellcolor{cellgreen}0.96 & \cellcolor{cellgreen}0.7 s & \cellcolor{cellgreen}\$0.01 & \cellcolor{cellgreen}0.98 & \cellcolor{cellyellow}5.8 s \\
\textbf{Q6: C} & \cellcolor{cellyellow}\$0.12 & \cellcolor{cellyellow}0.89 & \cellcolor{cellgrey}114.8 s & \cellcolor{cellgrey}\$0.43 & \cellcolor{cellyellow}0.89 & \cellcolor{cellgrey}143.7 s & \cellcolor{cellred}\textbf{X} & \cellcolor{cellred}\textbf{X} & \cellcolor{cellred}\textbf{X} & \cellcolor{cellgrey}\$0.35 & \cellcolor{cellyellow}0.88 & \cellcolor{cellyellow}34.9 s & \cellcolor{cellgreen}\$4$\cdot 10^{-6}$ & \cellcolor{cellgrey}0.47 & \cellcolor{cellgreen}0.7 s & \cellcolor{cellyellow}\$0.16 & \cellcolor{cellgreen}\textbf{0.95} & \cellcolor{cellyellow}19.5 s \\
\textbf{Q7: J} & \cellcolor{cellgrey}\$1.33 & \cellcolor{cellyellow}0.75 & \cellcolor{cellgrey}199.4 s & \cellcolor{cellgrey}\$1.79 & \cellcolor{cellgreen}\textbf{0.92} & \cellcolor{cellgrey}287.6 s & \cellcolor{cellgreen}\$0.08 & \cellcolor{cellgrey}0.51 & \cellcolor{cellyellow}97.7 s & \cellcolor{cellyellow}\$0.86 & \cellcolor{cellyellow}0.83 & \cellcolor{cellyellow}45.4 s & \cellcolor{cellgreen}\$1$\cdot 10^{-9}$ & \cellcolor{cellyellow}0.78 & \cellcolor{cellgreen}2$\cdot 10^{-3}$ s & \cellcolor{cellgreen}\$0.09 & \cellcolor{cellyellow}0.81 & \cellcolor{cellgreen}7.4 s \\
\textbf{Q8: J} & \cellcolor{cellgreen}\$4$\cdot 10^{-3}$ & \cellcolor{cellgreen}\textbf{1.00} & \cellcolor{cellyellow}4.1 s & \cellcolor{cellgreen}\$0.01 & \cellcolor{cellgreen}\textbf{1.00} & \cellcolor{cellyellow}3.9 s & \cellcolor{cellgrey}\$0.30 & \cellcolor{cellgrey}0.00 & \cellcolor{cellgrey}713.0 s & \cellcolor{cellgrey}\$18.23 & \cellcolor{cellgrey}0.29 & \cellcolor{cellgrey}126.2 s & \cellcolor{cellgreen}\$1$\cdot 10^{-9}$ & \cellcolor{cellgrey}0.25 & \cellcolor{cellgreen}1$\cdot 10^{-3}$ s & \cellcolor{cellyellow}\$0.10 & \cellcolor{cellgreen}\textbf{1.00} & \cellcolor{cellyellow}3.4 s \\
\textbf{Q9: J} & \cellcolor{cellyellow}\$0.06 & \cellcolor{cellgrey}0.55 & \cellcolor{cellgrey}243.4 s & \cellcolor{cellyellow}\$0.10 & \cellcolor{cellgrey}0.49 & \cellcolor{cellyellow}44.9 s & \cellcolor{cellgrey}\$0.31 & \cellcolor{cellgrey}0.00 & \cellcolor{cellgrey}872.2 s & \cellcolor{cellgrey}\$0.21 & \cellcolor{cellgrey}0.58 & \cellcolor{cellyellow}48.6 s & \cellcolor{cellgreen}\$1$\cdot 10^{-9}$ & \cellcolor{cellgrey}0.41 & \cellcolor{cellgreen}1$\cdot 10^{-3}$ s & \cellcolor{cellyellow}\$0.07 & \cellcolor{cellgreen}\textbf{0.83} & \cellcolor{cellgreen}18.8 s \\
\textbf{Q10: F J} & \cellcolor{cellgreen}\$0.21 & \cellcolor{cellgrey}0.00 & \cellcolor{cellyellow}519.5 s & \cellcolor{cellgrey}\$1.02 & \cellcolor{cellgrey}0.06 & \cellcolor{cellgrey}1192.5 s & \cellcolor{cellred}\textbf{X} & \cellcolor{cellred}\textbf{X} & \cellcolor{cellred}\textbf{X} & \cellcolor{cellred}\textbf{X} & \cellcolor{cellred}\textbf{X} & \cellcolor{cellred}\textbf{X} & \cellcolor{cellgreen}\$1$\cdot 10^{-5}$ & \cellcolor{cellred}\textbf{X} & \cellcolor{cellgreen}0.7 s & \cellcolor{cellyellow}\$0.65 & \cellcolor{cellgreen}\textbf{0.34} & \cellcolor{cellgreen}90.1 s \\
\textbf{Q11: F J} & \cellcolor{cellgreen}\$0.27 & \cellcolor{cellgreen}\textbf{0.78} & \cellcolor{cellyellow}158.7 s & \cellcolor{cellyellow}\$0.61 & \cellcolor{cellgreen}0.73 & \cellcolor{cellyellow}132.4 s & \cellcolor{cellred}\textbf{X} & \cellcolor{cellred}\textbf{X} & \cellcolor{cellred}\textbf{X} & \cellcolor{cellred}\textbf{X} & \cellcolor{cellred}\textbf{X} & \cellcolor{cellred}\textbf{X} & \cellcolor{cellred}\textbf{X} & \cellcolor{cellred}\textbf{X} & \cellcolor{cellred}\textbf{X} & \cellcolor{cellgrey}\$1.68 & \cellcolor{cellgrey}0.40 & \cellcolor{cellyellow}131.3 s \\
\textbf{Q12: F M} & \cellcolor{cellyellow}\$0.10 & \cellcolor{cellgrey}0.60 & \cellcolor{cellyellow}36.4 s & \cellcolor{cellyellow}\$0.14 & \cellcolor{cellgrey}0.00 & \cellcolor{cellyellow}31.9 s & \cellcolor{cellred}\textbf{X} & \cellcolor{cellred}\textbf{X} & \cellcolor{cellred}\textbf{X} & \cellcolor{cellyellow}\$0.10 & \cellcolor{cellgreen}\textbf{0.97} & \cellcolor{cellyellow}31.1 s & \cellcolor{cellgreen}\$6$\cdot 10^{-6}$ & \cellcolor{cellyellow}0.85 & \cellcolor{cellgreen}0.7 s & \cellcolor{cellyellow}\$0.08 & \cellcolor{cellgreen}0.95 & \cellcolor{cellyellow}32.1 s \\
\textbf{Q13: F} & \cellcolor{cellyellow}\$0.24 & \cellcolor{cellgreen}\textbf{0.74} & \cellcolor{cellgrey}274.8 s & \cellcolor{cellyellow}\$0.44 & \cellcolor{cellgreen}\textbf{0.74} & \cellcolor{cellgrey}238.3 s & \cellcolor{cellred}\textbf{X} & \cellcolor{cellred}\textbf{X} & \cellcolor{cellred}\textbf{X} & \cellcolor{cellyellow}\$0.38 & \cellcolor{cellgreen}0.70 & \cellcolor{cellyellow}22.4 s & \cellcolor{cellgreen}\$2$\cdot 10^{-5}$ & \cellcolor{cellyellow}0.58 & \cellcolor{cellgreen}1.0 s & \cellcolor{cellyellow}\$0.13 & \cellcolor{cellgreen}0.72 & \cellcolor{cellyellow}11.6 s \\
\textbf{Q14: F J R} & \cellcolor{cellgreen}\$0.23 & \cellcolor{cellyellow}0.87 & \cellcolor{cellgrey}178.2 s & \cellcolor{cellred}\textbf{X} & \cellcolor{cellred}\textbf{X} & \cellcolor{cellred}\textbf{X} & \cellcolor{cellred}\textbf{X} & \cellcolor{cellred}\textbf{X} & \cellcolor{cellred}\textbf{X} & \cellcolor{cellyellow}\$4.26 & \cellcolor{cellgrey}0.37 & \cellcolor{cellyellow}73.6 s & \cellcolor{cellgreen}\$5$\cdot 10^{-6}$ & \cellcolor{cellgrey}0.00 & \cellcolor{cellgreen}0.7 s & \cellcolor{cellgrey}\$4.83 & \cellcolor{cellgreen}\textbf{1.00} & \cellcolor{cellyellow}87.3 s \\
\multicolumn{1}{c|}{\emph{Avg}} & \cellcolor{cellyellow}\$0.22 & \cellcolor{cellyellow}0.74 & \cellcolor{cellyellow}149.2 s & \cellcolor{cellyellow}\$0.42 & \cellcolor{cellyellow}0.70 & \cellcolor{cellyellow}192.5 s & \cellcolor{cellyellow}\$0.21 & \cellcolor{cellgrey}0.43 & \cellcolor{cellgrey}363.7 s & \cellcolor{cellgrey}\$2.42 & \cellcolor{cellyellow}0.67 & \cellcolor{cellgreen}44.8 s & \cellcolor{cellgreen}\$3$\cdot 10^{-3}$ & \cellcolor{cellgrey}0.57 & \cellcolor{cellgreen}10.8 s & \cellcolor{cellyellow}\$0.54 & \cellcolor{cellgreen}\textbf{0.80} & \cellcolor{cellgreen}38.2 s \\

\midrule

\multicolumn{1}{l}{} & \multicolumn{9}{l}{\Large \textbf{(c) Wildlife Scenario}} &  \multicolumn{3}{c}{\emph{DASE Preprocess}} & \cellcolor{cellgreen}\$0.01 & \textbf{ } & \cellcolor{cellgreen}53.9 s & \cellcolor{cellgreen}\$0.01 & \textbf{ } & \cellcolor{cellgreen}53.9 s \\
\textbf{Q1: F} & \cellcolor{cellyellow}\$0.11 & \cellcolor{cellgreen}\textbf{0.79} & \cellcolor{cellgrey}92.4 s & \cellcolor{cellyellow}\$0.13 & \cellcolor{cellgreen}\textbf{0.79} & \cellcolor{cellyellow}32.8 s & \cellcolor{cellyellow}\$0.11 & \cellcolor{cellgreen}\textbf{0.79} & \cellcolor{cellyellow}19.6 s & \cellcolor{cellyellow}\$0.11 & \cellcolor{cellgreen}\textbf{0.79} & \cellcolor{cellyellow}32.0 s & \cellcolor{cellgreen}\$1$\cdot 10^{-9}$ & \cellcolor{cellgreen}\textbf{0.79} & \cellcolor{cellgreen}5$\cdot 10^{-4}$ s & \cellcolor{cellgreen}\$0.02 & \cellcolor{cellgreen}0.78 & \cellcolor{cellgreen}10.1 s \\
\textbf{Q2: F} & \cellcolor{cellred}\textbf{X} & \cellcolor{cellred}\textbf{X} & \cellcolor{cellred}\textbf{X} & \cellcolor{cellyellow}\$0.01 & \cellcolor{cellgrey}0.17 & \cellcolor{cellyellow}2.8 s & \cellcolor{cellyellow}\$0.01 & \cellcolor{cellgrey}0.14 & \cellcolor{cellyellow}4.5 s & \cellcolor{cellyellow}\$0.01 & \cellcolor{cellgrey}0.19 & \cellcolor{cellgrey}9.4 s & \cellcolor{cellgreen}\$1$\cdot 10^{-9}$ & \cellcolor{cellgreen}\textbf{0.83} & \cellcolor{cellgreen}5$\cdot 10^{-4}$ s & \cellcolor{cellgreen}\$3$\cdot 10^{-3}$ & \cellcolor{cellgrey}0.50 & \cellcolor{cellyellow}6.2 s \\
\textbf{Q3: F L} & \cellcolor{cellyellow}{\$0.11} & \cellcolor{cellgreen}\textbf{1.00} & \cellcolor{cellgrey}99.4 s & \cellcolor{cellyellow}\$0.13 & \cellcolor{cellgrey}0.00 & \cellcolor{cellyellow}22.7 s & \cellcolor{cellgreen}\$0.03 & \cellcolor{cellgrey}0.00 & \cellcolor{cellgreen}4.3 s & \cellcolor{cellyellow}\$0.11 & \cellcolor{cellgrey}0.00 & \cellcolor{cellyellow}25.7 s & \cellcolor{cellgreen}\$1$\cdot 10^{-9}$ & \cellcolor{cellgrey}0.00 & \cellcolor{cellgreen}8$\cdot 10^{-4}$ s & \cellcolor{cellgreen}\$0.02 & \cellcolor{cellgreen}\textbf{1.00} & \cellcolor{cellyellow}9.0 s \\
\textbf{Q4: F L} & \cellcolor{cellred}\textbf{X} & \cellcolor{cellred}\textbf{X} & \cellcolor{cellred}\textbf{X} & \cellcolor{cellgrey}\$0.01 & \cellcolor{cellgrey}0.00 & \cellcolor{cellyellow}2.7 s & \cellcolor{cellgreen}\$1$\cdot 10^{-3}$ & \cellcolor{cellgrey}0.00 & \cellcolor{cellgreen}1.3 s & \cellcolor{cellgrey}\$0.01 & \cellcolor{cellgreen}\textbf{1.00} & \cellcolor{cellgrey}9.4 s & \cellcolor{cellgreen}\$1$\cdot 10^{-9}$ & \cellcolor{cellgreen}\textbf{1.00} & \cellcolor{cellgreen}8$\cdot 10^{-4}$ s & \cellcolor{cellyellow}\$3$\cdot 10^{-3}$ & \cellcolor{cellgreen}\textbf{1.00} & \cellcolor{cellgrey}6.1 s \\
\textbf{Q5: F} & \cellcolor{cellred}\textbf{X} & \cellcolor{cellred}\textbf{X} & \cellcolor{cellred}\textbf{X} & \cellcolor{cellgrey}\$0.13 & \cellcolor{cellgrey}0.75 & \cellcolor{cellyellow}13.5 s & \cellcolor{cellyellow}\$0.01 & \cellcolor{cellgrey}0.75 & \cellcolor{cellgreen}2.3 s & \cellcolor{cellgrey}\$0.12 & \cellcolor{cellgrey}0.75 & \cellcolor{cellgrey}19.2 s & \cellcolor{cellgreen}\$1$\cdot 10^{-9}$ & \cellcolor{cellgreen}\textbf{1.00} & \cellcolor{cellgreen}1$\cdot 10^{-3}$ s & \cellcolor{cellyellow}\$0.02 & \cellcolor{cellyellow}0.85 & \cellcolor{cellyellow}7.0 s \\
\textbf{Q6: F} & \cellcolor{cellred}\textbf{X} & \cellcolor{cellred}\textbf{X} & \cellcolor{cellred}\textbf{X} & \cellcolor{cellgrey}\$0.13 & \cellcolor{cellgrey}0.00 & \cellcolor{cellyellow}19.3 s & \cellcolor{cellyellow}\$0.06 & \cellcolor{cellgreen}\textbf{0.50} & \cellcolor{cellyellow}13.2 s & \cellcolor{cellyellow}\$0.12 & \cellcolor{cellgrey}0.20 & \cellcolor{cellyellow}24.3 s & \cellcolor{cellgreen}\$1$\cdot 10^{-9}$ & \cellcolor{cellgrey}0.00 & \cellcolor{cellgreen}1$\cdot 10^{-3}$ s & \cellcolor{cellyellow}\$0.06 & \cellcolor{cellyellow}0.40 & \cellcolor{cellyellow}13.9 s \\
\textbf{Q7: F} & \cellcolor{cellyellow}\$0.23 & \cellcolor{cellgreen}\textbf{1.00} & \cellcolor{cellgrey}188.3 s & \cellcolor{cellyellow}\$0.13 & \cellcolor{cellgreen}\textbf{1.00} & \cellcolor{cellyellow}43.9 s & \cellcolor{cellyellow}\$0.20 & \cellcolor{cellgreen}\textbf{1.00} & \cellcolor{cellyellow}28.8 s & \cellcolor{cellyellow}\$0.22 & \cellcolor{cellgreen}\textbf{1.00} & \cellcolor{cellyellow}24.6 s & \cellcolor{cellgreen}\$1$\cdot 10^{-9}$ & \cellcolor{cellgrey}0.00 & \cellcolor{cellgreen}1$\cdot 10^{-3}$ s & \cellcolor{cellyellow}\$0.14 & \cellcolor{cellgreen}\textbf{1.00} & \cellcolor{cellyellow}19.9 s \\
\textbf{Q8: F} & \cellcolor{cellred}\textbf{X} & \cellcolor{cellred}\textbf{X} & \cellcolor{cellred}\textbf{X} & \cellcolor{cellyellow}\$0.13 & \cellcolor{cellgreen}\textbf{0.75} & \cellcolor{cellyellow}34.9 s & \cellcolor{cellyellow}\$0.12 & \cellcolor{cellgreen}\textbf{0.75} & \cellcolor{cellyellow}23.8 s & \cellcolor{cellyellow}\$0.23 & \cellcolor{cellgreen}\textbf{0.75} & \cellcolor{cellyellow}35.5 s & \cellcolor{cellgreen}\$1$\cdot 10^{-9}$ & \cellcolor{cellgrey}0.00 & \cellcolor{cellgreen}2$\cdot 10^{-3}$ s & \cellcolor{cellyellow}\$0.16 & \cellcolor{cellgreen}0.74 & \cellcolor{cellyellow}24.8 s \\
\textbf{Q9: F} & \cellcolor{cellred}\textbf{X} & \cellcolor{cellred}\textbf{X} & \cellcolor{cellred}\textbf{X} & \cellcolor{cellyellow}\$0.13 & \cellcolor{cellyellow}0.57 & \cellcolor{cellyellow}19.1 s & \cellcolor{cellyellow}\$0.08 & \cellcolor{cellgreen}\textbf{0.67} & \cellcolor{cellyellow}17.2 s & \cellcolor{cellyellow}\$0.12 & \cellcolor{cellyellow}0.59 & \cellcolor{cellyellow}37.6 s & \cellcolor{cellgreen}\$1$\cdot 10^{-9}$ & \cellcolor{cellgrey}0.00 & \cellcolor{cellgreen}1$\cdot 10^{-3}$ s & \cellcolor{cellyellow}\$0.06 & \cellcolor{cellgrey}0.40 & \cellcolor{cellyellow}19.5 s \\
\textbf{Q10: F L} & \cellcolor{cellyellow}\$0.11 & \cellcolor{cellgreen}\textbf{1.00} & \cellcolor{cellgrey}87.8 s & \cellcolor{cellyellow}\$0.13 & \cellcolor{cellgrey}0.00 & \cellcolor{cellyellow}19.6 s & \cellcolor{cellgreen}\$0.03 & \cellcolor{cellgrey}0.00 & \cellcolor{cellgreen}4.4 s & \cellcolor{cellyellow}\$0.11 & \cellcolor{cellgrey}0.00 & \cellcolor{cellyellow}40.3 s & \cellcolor{cellgreen}\$1$\cdot 10^{-9}$ & \cellcolor{cellgreen}\textbf{1.00} & \cellcolor{cellgreen}8$\cdot 10^{-4}$ s & \cellcolor{cellyellow}\$0.06 & \cellcolor{cellgrey}0.25 & \cellcolor{cellyellow}24.2 s \\
\multicolumn{1}{c|}{\emph{Avg}} & \cellcolor{cellyellow}\$0.14 & \cellcolor{cellgreen}\textbf{0.94} & \cellcolor{cellgrey}117.0 s & \cellcolor{cellyellow}\$0.11 & \cellcolor{cellgrey}0.40 & \cellcolor{cellyellow}21.1 s & \cellcolor{cellyellow}\$0.07 & \cellcolor{cellgrey}0.46 & \cellcolor{cellyellow}11.9 s & \cellcolor{cellyellow}\$0.12 & \cellcolor{cellgrey}0.52 & \cellcolor{cellyellow}25.8 s & \cellcolor{cellgreen}\$9$\cdot 10^{-4}$ & \cellcolor{cellgrey}0.46 & \cellcolor{cellgreen}4.9 s & \cellcolor{cellyellow}\$0.05 & \cellcolor{cellgrey}0.69 & \cellcolor{cellyellow}17.7 s \\

\specialrule{1.2pt}{2pt}{2pt}
\multicolumn{1}{l}{} & \multicolumn{9}{l}{\Large \textbf{(d) MMQA Scenario}} &  \multicolumn{3}{c}{\emph{DASE Preprocess}} & \cellcolor{cellgreen}\$0.01 & \textbf{ } & \cellcolor{cellgreen}50.1 s & \cellcolor{cellgreen}\$0.01 & \textbf{ } & \cellcolor{cellgreen}50.1 s \\
\textbf{Q1: M} & \cellcolor{cellyellow}\$0.02 & \cellcolor{cellgreen}\textbf{1.00} & \cellcolor{cellyellow}7.4 s & \cellcolor{cellgreen}\$3$\cdot 10^{-3}$ & \cellcolor{cellgreen}\textbf{1.00} & \cellcolor{cellgreen}4.5 s & \cellcolor{cellred}\textbf{X} & \cellcolor{cellred}\textbf{X} & \cellcolor{cellred}\textbf{X} & \cellcolor{cellyellow}\$0.01 & \cellcolor{cellgreen}\textbf{1.00} & \cellcolor{cellyellow}14.1 s & \cellcolor{cellred}\textbf{X} & \cellcolor{cellred}\textbf{X} & \cellcolor{cellred}\textbf{X} & \cellcolor{cellyellow}\$0.01 & \cellcolor{cellgreen}\textbf{1.00} & \cellcolor{cellyellow}14.1 s \\
\textbf{Q2a: J} & \cellcolor{cellgrey}\$0.89 & \cellcolor{cellyellow}0.83 & \cellcolor{cellgrey}169.6 s & \cellcolor{cellgrey}\$1.04 & \cellcolor{cellgreen}\textbf{1.00} & \cellcolor{cellgrey}135.9 s & \cellcolor{cellred}\textbf{X} & \cellcolor{cellred}\textbf{X} & \cellcolor{cellred}\textbf{X} & \cellcolor{cellyellow}\$0.08 & \cellcolor{cellgrey}0.00 & \cellcolor{cellyellow}53.8 s & \cellcolor{cellgreen}\$1$\cdot 10^{-5}$ & \cellcolor{cellgrey}0.08 & \cellcolor{cellgreen}1.0 s & \cellcolor{cellyellow}\$0.04 & \cellcolor{cellgrey}0.57 & \cellcolor{cellyellow}28.6 s \\
\textbf{Q2b: J} & \cellcolor{cellgrey}\$0.89 & \cellcolor{cellyellow}0.83 & \cellcolor{cellgrey}166.8 s & \cellcolor{cellgrey}\$1.04 & \cellcolor{cellgreen}\textbf{1.00} & \cellcolor{cellgrey}133.8 s & \cellcolor{cellred}\textbf{X} & \cellcolor{cellred}\textbf{X} & \cellcolor{cellred}\textbf{X} & \cellcolor{cellyellow}\$0.12 & \cellcolor{cellgrey}0.00 & \cellcolor{cellyellow}38.4 s & \cellcolor{cellgreen}\$1$\cdot 10^{-5}$ & \cellcolor{cellgrey}0.08 & \cellcolor{cellgreen}1.0 s & \cellcolor{cellyellow}\$0.06 & \cellcolor{cellgrey}0.57 & \cellcolor{cellyellow}21.4 s \\
\textbf{Q3a: F} & \cellcolor{cellyellow}\$0.01 & \cellcolor{cellgreen}\textbf{0.83} & \cellcolor{cellyellow}4.2 s & \cellcolor{cellyellow}\$0.02 & \cellcolor{cellgreen}0.80 & \cellcolor{cellyellow}4.5 s & \cellcolor{cellyellow}\$0.01 & \cellcolor{cellyellow}0.75 & \cellcolor{cellyellow}11.0 s & \cellcolor{cellyellow}\$0.01 & \cellcolor{cellyellow}0.72 & \cellcolor{cellgrey}19.6 s & \cellcolor{cellgreen}\$5$\cdot 10^{-6}$ & \cellcolor{cellgrey}0.38 & \cellcolor{cellgreen}0.7 s & \cellcolor{cellyellow}\$9$\cdot 10^{-3}$ & \cellcolor{cellyellow}0.72 & \cellcolor{cellgrey}20.5 s \\
\textbf{Q3f: F} & \cellcolor{cellyellow}\$0.01 & \cellcolor{cellgreen}\textbf{1.00} & \cellcolor{cellyellow}4.2 s & \cellcolor{cellyellow}\$0.02 & \cellcolor{cellgreen}\textbf{1.00} & \cellcolor{cellyellow}8.0 s & \cellcolor{cellyellow}\$0.01 & \cellcolor{cellgreen}\textbf{1.00} & \cellcolor{cellyellow}7.0 s & \cellcolor{cellyellow}\$0.01 & \cellcolor{cellgrey}0.67 & \cellcolor{cellgrey}22.3 s & \cellcolor{cellgreen}\$5$\cdot 10^{-6}$ & \cellcolor{cellgrey}0.09 & \cellcolor{cellgreen}0.7 s & \cellcolor{cellyellow}\$0.01 & \cellcolor{cellgrey}0.67 & \cellcolor{cellgrey}22.3 s \\
\textbf{Q4: M} & \cellcolor{cellred}\textbf{X} & \cellcolor{cellred}\textbf{X} & \cellcolor{cellred}\textbf{X} & \cellcolor{cellgrey}\$5$\cdot 10^{-3}$ & \cellcolor{cellyellow}0.54 & \cellcolor{cellgreen}1.2 s & \cellcolor{cellred}\textbf{X} & \cellcolor{cellred}\textbf{X} & \cellcolor{cellred}\textbf{X} & \cellcolor{cellyellow}\$2$\cdot 10^{-3}$ & \cellcolor{cellgreen}0.60 & \cellcolor{cellyellow}9.7 s & \cellcolor{cellred}\textbf{X} & \cellcolor{cellred}\textbf{X} & \cellcolor{cellred}\textbf{X} & \cellcolor{cellyellow}\$2$\cdot 10^{-3}$ & \cellcolor{cellgreen}\textbf{0.61} & \cellcolor{cellyellow}9.5 s \\
\textbf{Q5: M} & \cellcolor{cellyellow}\$1$\cdot 10^{-3}$ & \cellcolor{cellgreen}\textbf{1.00} & \cellcolor{cellyellow}0.48 s & \cellcolor{cellyellow}\$1$\cdot 10^{-3}$ & \cellcolor{cellgreen}\textbf{1.00} & \cellcolor{cellyellow}0.50 s & \cellcolor{cellred}\textbf{X} & \cellcolor{cellred}\textbf{X} & \cellcolor{cellred}\textbf{X} & \cellcolor{cellred}\textbf{X} & \cellcolor{cellred}\textbf{X} & \cellcolor{cellred}\textbf{X} & \cellcolor{cellred}\textbf{X} & \cellcolor{cellred}\textbf{X} & \cellcolor{cellred}\textbf{X} & \cellcolor{cellred}\textbf{X} & \cellcolor{cellred}\textbf{X} & \cellcolor{cellred}\textbf{X} \\
\textbf{Q6a: F} & \cellcolor{cellgrey}\$0.01 & \cellcolor{cellgreen}\textbf{1.00} & \cellcolor{cellyellow}4.4 s & \cellcolor{cellgrey}\$0.02 & \cellcolor{cellgreen}\textbf{1.00} & \cellcolor{cellyellow}4.0 s & \cellcolor{cellyellow}\$2$\cdot 10^{-3}$ & \cellcolor{cellgrey}0.33 & \cellcolor{cellyellow}5.6 s & \cellcolor{cellyellow}\$4$\cdot 10^{-3}$ & \cellcolor{cellgrey}0.03 & \cellcolor{cellgrey}18.9 s & \cellcolor{cellgreen}\$5$\cdot 10^{-6}$ & \cellcolor{cellgrey}0.04 & \cellcolor{cellgreen}0.7 s & \cellcolor{cellyellow}\$3$\cdot 10^{-3}$ & \cellcolor{cellgreen}\textbf{1.00} & \cellcolor{cellgrey}15.8 s \\
\textbf{Q6b: F} & \cellcolor{cellgrey}\$0.01 & \cellcolor{cellgreen}\textbf{1.00} & \cellcolor{cellyellow}3.8 s & \cellcolor{cellgrey}\$0.02 & \cellcolor{cellgreen}\textbf{1.00} & \cellcolor{cellyellow}4.1 s & \cellcolor{cellyellow}\$2$\cdot 10^{-3}$ & \cellcolor{cellgreen}\textbf{1.00} & \cellcolor{cellyellow}5.5 s & \cellcolor{cellyellow}\$4$\cdot 10^{-3}$ & \cellcolor{cellgrey}0.04 & \cellcolor{cellgrey}17.3 s & \cellcolor{cellgreen}\$5$\cdot 10^{-6}$ & \cellcolor{cellgrey}0.01 & \cellcolor{cellgreen}0.7 s & \cellcolor{cellyellow}\$4$\cdot 10^{-3}$ & \cellcolor{cellyellow}0.87 & \cellcolor{cellgrey}20.4 s \\
\textbf{Q6c: F} & \cellcolor{cellgrey}\$0.01 & \cellcolor{cellgreen}\textbf{1.00} & \cellcolor{cellyellow}4.6 s & \cellcolor{cellgrey}\$0.02 & \cellcolor{cellgreen}\textbf{1.00} & \cellcolor{cellyellow}4.7 s & \cellcolor{cellyellow}\$2$\cdot 10^{-3}$ & \cellcolor{cellgrey}0.53 & \cellcolor{cellyellow}13.4 s & \cellcolor{cellyellow}\$4$\cdot 10^{-3}$ & \cellcolor{cellgrey}0.13 & \cellcolor{cellyellow}17.4 s & \cellcolor{cellgreen}\$5$\cdot 10^{-6}$ & \cellcolor{cellgrey}0.10 & \cellcolor{cellgreen}0.7 s & \cellcolor{cellyellow}\$4$\cdot 10^{-3}$ & \cellcolor{cellgrey}0.66 & \cellcolor{cellgrey}19.1 s \\
\textbf{Q7: J} & \cellcolor{cellgrey}\$13.61 & \cellcolor{cellgrey}0.32 & \cellcolor{cellgrey}2311.4 s & \cellcolor{cellgrey}\$15.65 & \cellcolor{cellgrey}0.31 & \cellcolor{cellgrey}2101.6 s & \cellcolor{cellred}\textbf{X} & \cellcolor{cellred}\textbf{X} & \cellcolor{cellred}\textbf{X} & \cellcolor{cellyellow}\$1.18 & \cellcolor{cellgrey}0.00 & \cellcolor{cellyellow}91.7 s & \cellcolor{cellgreen}\$1$\cdot 10^{-5}$ & \cellcolor{cellgrey}0.10 & \cellcolor{cellgreen}0.7 s & \cellcolor{cellgreen}\$0.04 & \cellcolor{cellgreen}\textbf{1.00} & \cellcolor{cellgreen}6.9 s \\
\multicolumn{1}{c|}{\emph{Avg}} & \cellcolor{cellgrey}\$1.55 & \cellcolor{cellgreen}\textbf{0.88} & \cellcolor{cellgrey}267.7 s & \cellcolor{cellgrey}\$1.62 & \cellcolor{cellgreen}0.87 & \cellcolor{cellgrey}218.4 s & \cellcolor{cellgreen}\$5$\cdot 10^{-3}$ & \cellcolor{cellyellow}0.72 & \cellcolor{cellgreen}8.5 s & \cellcolor{cellyellow}\$0.14 & \cellcolor{cellgrey}0.31 & \cellcolor{cellyellow}30.3 s & \cellcolor{cellgreen}\$1$\cdot 10^{-3}$ & \cellcolor{cellgrey}0.11 & \cellcolor{cellgreen}6.3 s & \cellcolor{cellgreen}\$0.02 & \cellcolor{cellyellow}0.76 & \cellcolor{cellyellow}20.8 s \\

\specialrule{1.2pt}{2pt}{2pt}
\multicolumn{1}{l}{} & \multicolumn{9}{l}{\Large \textbf{(e) Cars Scenario}} &  \multicolumn{3}{c}{\emph{DASE Preprocess}} & \cellcolor{cellgreen}\$2.50 & \textbf{ } & \cellcolor{cellgreen}1300 s & \cellcolor{cellgreen}\$2.50 & \textbf{ } & \cellcolor{cellgreen}1300 s \\
\textbf{Q1: F} & \cellcolor{cellyellow}\$1.74 & \cellcolor{cellgreen}\textbf{0.90} & \cellcolor{cellgrey}550.0 s & \cellcolor{cellyellow}\$2.44 & \cellcolor{cellgrey}0.69 & \cellcolor{cellyellow}465.6 s & \cellcolor{cellyellow}\$1.37 & \cellcolor{cellyellow}0.81 & \cellcolor{cellgrey}829.7 s & \cellcolor{cellyellow}\$1.44 & \cellcolor{cellyellow}0.71 & \cellcolor{cellgreen}61.7 s & \cellcolor{cellgreen}\$5$\cdot 10^{-6}$ & \cellcolor{cellgrey}0.66 & \cellcolor{cellgreen}0.9 s & \cellcolor{cellgreen}\$0.31 & \cellcolor{cellyellow}0.78 & \cellcolor{cellgreen}22.0 s \\
\textbf{Q2: F} & \cellcolor{cellred}\textbf{X} & \cellcolor{cellred}\textbf{X} & \cellcolor{cellred}\textbf{X} & \cellcolor{cellyellow}\$4$\cdot 10^{-3}$ & \cellcolor{cellyellow}0.00 & \cellcolor{cellyellow}4.1 s & \cellcolor{cellgrey}\$0.01 & \cellcolor{cellyellow}0.09 & \cellcolor{cellgrey}14.1 s & \cellcolor{cellgrey}\$0.01 & \cellcolor{cellyellow}0.08 & \cellcolor{cellgrey}14.1 s & \cellcolor{cellgreen}\$5$\cdot 10^{-6}$ & \cellcolor{cellyellow}0.00 & \cellcolor{cellgreen}0.7 s & \cellcolor{cellgreen}\$1$\cdot 10^{-3}$ & \cellcolor{cellgreen}\textbf{0.16} & \cellcolor{cellyellow}5.1 s \\
\textbf{Q3: F L} & \cellcolor{cellgrey}\$0.60 & \cellcolor{cellyellow}0.90 & \cellcolor{cellgrey}456.2 s & \cellcolor{cellyellow}\$0.01 & \cellcolor{cellyellow}0.92 & \cellcolor{cellyellow}6.1 s & \cellcolor{cellred}\textbf{X} & \cellcolor{cellred}\textbf{X} & \cellcolor{cellred}\textbf{X} & \cellcolor{cellgrey}\$1.66 & \cellcolor{cellgreen}\textbf{1.00} & \cellcolor{cellgrey}36.4 s & \cellcolor{cellgreen}\$5$\cdot 10^{-6}$ & \cellcolor{cellgreen}\textbf{1.00} & \cellcolor{cellgreen}0.7 s & \cellcolor{cellgreen}\$2$\cdot 10^{-3}$ & \cellcolor{cellgreen}\textbf{1.00} & \cellcolor{cellyellow}4.3 s \\
\textbf{Q4: F} & \cellcolor{cellyellow}\$1.71 & \cellcolor{cellgreen}0.99 & \cellcolor{cellgrey}822.0 s & \cellcolor{cellyellow}\$2.41 & \cellcolor{cellgreen}0.99 & \cellcolor{cellyellow}443.6 s & \cellcolor{cellyellow}\$1.34 & \cellcolor{cellgreen}\textbf{1.00} & \cellcolor{cellgrey}768.8 s & \cellcolor{cellyellow}\$1.41 & \cellcolor{cellgreen}0.99 & \cellcolor{cellgreen}68.7 s & \cellcolor{cellgreen}\$5$\cdot 10^{-6}$ & \cellcolor{cellgreen}0.99 & \cellcolor{cellgreen}0.9 s & \cellcolor{cellgreen}\$0.31 & \cellcolor{cellgreen}0.99 & \cellcolor{cellgreen}23.4 s \\
\textbf{Q5: F} & \cellcolor{cellred}\textbf{X} & \cellcolor{cellred}\textbf{X} & \cellcolor{cellred}\textbf{X} & \cellcolor{cellyellow}\$0.01 & \cellcolor{cellgreen}\textbf{1.00} & \cellcolor{cellyellow}6.3 s & \cellcolor{cellgrey}\$1.61 & \cellcolor{cellgreen}\textbf{1.00} & \cellcolor{cellgrey}483.7 s & \cellcolor{cellgrey}\$1.47 & \cellcolor{cellgreen}\textbf{1.00} & \cellcolor{cellgrey}58.9 s & \cellcolor{cellgreen}\$1$\cdot 10^{-5}$ & \cellcolor{cellgrey}0.50 & \cellcolor{cellgreen}1.1 s & \cellcolor{cellyellow}\$7$\cdot 10^{-3}$ & \cellcolor{cellgreen}\textbf{1.00} & \cellcolor{cellyellow}5.1 s \\
\textbf{Q6: F J} & \cellcolor{cellred}\textbf{X} & \cellcolor{cellred}\textbf{X} & \cellcolor{cellred}\textbf{X} & \cellcolor{cellyellow}\$2.51 & \cellcolor{cellgreen}0.96 & \cellcolor{cellgrey}427.3 s & \cellcolor{cellyellow}\$1.96 & \cellcolor{cellgreen}\textbf{0.97} & \cellcolor{cellgrey}775.4 s & \cellcolor{cellyellow}\$2.00 & \cellcolor{cellgreen}0.96 & \cellcolor{cellyellow}44.3 s & \cellcolor{cellgreen}\$1$\cdot 10^{-5}$ & \cellcolor{cellyellow}0.88 & \cellcolor{cellgreen}1.3 s & \cellcolor{cellgreen}\$0.89 & \cellcolor{cellgreen}0.96 & \cellcolor{cellyellow}41.2 s \\
\textbf{Q7: F} & \cellcolor{cellred}\textbf{X} & \cellcolor{cellred}\textbf{X} & \cellcolor{cellred}\textbf{X} & \cellcolor{cellyellow}\$4.47 & \cellcolor{cellgreen}0.56 & \cellcolor{cellgrey}882.7 s & \cellcolor{cellyellow}\$3.06 & \cellcolor{cellgreen}\textbf{0.58} & \cellcolor{cellgrey}2146.2 s & \cellcolor{cellyellow}\$3.17 & \cellcolor{cellyellow}0.45 & \cellcolor{cellyellow}86.0 s & \cellcolor{cellgreen}\$1$\cdot 10^{-5}$ & \cellcolor{cellyellow}0.47 & \cellcolor{cellgreen}1.3 s & \cellcolor{cellyellow}\$1.55 & \cellcolor{cellyellow}0.52 & \cellcolor{cellyellow}131.5 s \\
\textbf{Q8: F L} & \cellcolor{cellyellow}\$1.78 & \cellcolor{cellgreen}\textbf{0.45} & \cellcolor{cellgrey}1349.8 s & \cellcolor{cellgrey}\$2.02 & \cellcolor{cellyellow}0.29 & \cellcolor{cellgrey}268.7 s & \cellcolor{cellgreen}\$0.21 & \cellcolor{cellgrey}0.20 & \cellcolor{cellyellow}68.0 s & \cellcolor{cellyellow}\$1.69 & \cellcolor{cellgrey}0.24 & \cellcolor{cellyellow}38.2 s & \cellcolor{cellgreen}\$5$\cdot 10^{-6}$ & \cellcolor{cellgrey}0.15 & \cellcolor{cellgreen}0.7 s & \cellcolor{cellgreen}\$0.04 & \cellcolor{cellyellow}0.39 & \cellcolor{cellgreen}6.0 s \\
\textbf{Q9: F} & \cellcolor{cellred}\textbf{X} & \cellcolor{cellred}\textbf{X} & \cellcolor{cellred}\textbf{X} & \cellcolor{cellgrey}\$2.05 & \cellcolor{cellgrey}0.00 & \cellcolor{cellgrey}308.1 s & \cellcolor{cellgrey}\$1.61 & \cellcolor{cellgreen}\textbf{1.00} & \cellcolor{cellgrey}335.8 s & \cellcolor{cellyellow}\$0.03 & \cellcolor{cellgrey}0.00 & \cellcolor{cellyellow}13.6 s & \cellcolor{cellgreen}\$5$\cdot 10^{-6}$ & \cellcolor{cellgreen}\textbf{1.00} & \cellcolor{cellgreen}0.7 s & \cellcolor{cellyellow}\$0.03 & \cellcolor{cellgrey}0.00 & \cellcolor{cellyellow}16.4 s \\
\textbf{Q10: C} & \cellcolor{cellyellow}\$3.09 & \cellcolor{cellyellow}0.41 & \cellcolor{cellgrey}618.1 s & \cellcolor{cellyellow}\$4.69 & \cellcolor{cellyellow}0.51 & \cellcolor{cellgrey}594.9 s & \cellcolor{cellred}\textbf{X} & \cellcolor{cellred}\textbf{X} & \cellcolor{cellred}\textbf{X} & \cellcolor{cellyellow}\$2.70 & \cellcolor{cellgreen}\textbf{0.57} & \cellcolor{cellyellow}62.0 s & \cellcolor{cellgreen}\$2$\cdot 10^{-5}$ & \cellcolor{cellyellow}0.45 & \cellcolor{cellgreen}1.8 s & \cellcolor{cellgreen}\$0.86 & \cellcolor{cellgreen}0.55 & \cellcolor{cellgreen}27.6 s \\
\multicolumn{1}{c|}{\emph{Avg}} & \cellcolor{cellyellow}\$1.78 & \cellcolor{cellgreen}\textbf{0.73} & \cellcolor{cellgrey}759.2 s & \cellcolor{cellyellow}\$2.06 & \cellcolor{cellyellow}0.59 & \cellcolor{cellyellow}340.7 s & \cellcolor{cellyellow}\$1.40 & \cellcolor{cellgreen}0.70 & \cellcolor{cellgrey}677.7 s & \cellcolor{cellyellow}\$1.56 & \cellcolor{cellyellow}0.60 & \cellcolor{cellgreen}48.4 s & \cellcolor{cellgreen}\$0.23 & \cellcolor{cellyellow}0.61 & \cellcolor{cellgreen}119.1 s & \cellcolor{cellgreen}\$0.59 & \cellcolor{cellyellow}0.63 & \cellcolor{cellyellow}143.9 s \\

\bottomrule
\end{tabular}%
\end{table*}

\emph{Datasets.} \label{subsec:dataset}
To evaluate \sys{} on deep research reasoning tasks, we constructed two benchmark datasets.
In \textbf{IMDB}, base tables from the official IMDB (\url{https://datasets.imdbws.com}) are augmented with structured fields from the repository \texttt{jquigl/\allowbreak imdb-genres}. To support cross-table movie comparisons, the data is split by release year. In \textbf{MOLECULE}, we collected query templates from our agentic AI system built on a PubMed-derived molecule database (\url{https://pubmed.ncbi.nlm.nih.gov/}). These templates are used to simulate multi-step reasoning workflows (e.g., discovering candidate molecules by joining fragmented evidence). The dataset consists of Papers, Facts, and Molecules.
Benchmark and workload details are available in our technical report (\url{https://anonymous.4open.science/r/DASE-2C61/}).

We include two public benchmark datasets that allow us to examine specific aspects of \sys.
The first of these is \textbf{SemBench}~\cite{lao_sembench_2025}, which we use to explore the effectiveness of \sys{} in prefiltering for semantic queries.
Our second benchmark is \textbf{NFCorpus}~\cite{boteva2016full}, a 2,868-paper biomedical corpus with title embeddings, real intervention--condition and study--outcome links, and categorical attributes. Its depth-scaling workload ranges from two to eight join steps, providing diverse multi-step reasoning queries at different cardinalities.

\emph{Workloads.}
We developed synthetic workloads over our two datasets, following a taxonomy shown in Table~\ref{tab:workloads} and fully detailed in our technical report.  These cover all possible combinations of structured predicates (\(\mathcal{P}\)), multi-component scoring (\(\mathcal{S}\)), and cross-table joins (\(\mathcal{J}\)). In this framework, single-vector similarity search serves as the default configuration. The component \(\mathcal{S}\) represents the transition from simple embedding retrieval to a generalized scoring function \(f(s_1, \dots, s_n)\). This allows the system to aggregate multiple vector similarities, as required for complex discovery workflows.

The eight classes cover these requirements separately and in combination: W1 is the single-signal baseline; W2, W3, and W5 introduce filtering, multiple scores, and joins, respectively; and W4, W6, W7, and W8 combine them. The FSS/MSA expressions in Table~\ref{tab:workloads} illustrate their execution in \sys, rather than prescribing how other systems must execute these logical workloads.

Additional workloads are from the published SemBench and NFCorpus benchmark specifications.

\emph{Implementation strategies.}
\label{subsec:baselines}
In deep research tasks, we compare \sysb against five distinct execution strategies and DBMSs:

\begin{enumerate}[leftmargin=*]
    \item \textbf{PostgreSQL:} 
    The multi-step reasoning query is submitted directly to the PostgreSQL engine. The embedding columns are indexed using \texttt{pgvector}'s HNSW or IVFFlat structures. All execution decisions, including join order, index selection, and predicate pushdown, are delegated to the PostgreSQL optimizer.
    
    \item \textbf{Filter-First:} 
    Evaluates structural and exact-match predicates first (using the \texttt{WHERE} clause) to generate a candidate set of valid IDs. The system then performs random access to fetch the corresponding embeddings, computes the similarity scores on-the-fly, and maintains a Top-\(k\) heap to produce the result.

    \item \textbf{Score-First:} 
    An inverted strategy that prioritizes similarity search. It uses an \texttt{ORDER BY} clause to retrieve a continuous stream of records sorted by their joint similarity score. The system applies the structural predicates as a post-filter and halts the stream immediately once \(k\) valid entities are identified. 

    \item \textbf{Rerank:} 
    For queries involving multiple scoring signals, executes a retrieve-and-rerank pipeline. It streams candidates separately from each signal's index, applies predicates as post-filters, and dynamically evaluates the joint score to update the Top-\(k\) results until the exploration stopping criteria are met. For single-signal queries, it gracefully degrades to Score-First.
    
    \item \textbf{Milvus:} 
    Milvus \cite{wang_milvus_2021} is a state-of-the-art vector database. For single-signal queries (with or without structural predicates), we push down the filters directly into Milvus. Because Milvus lacks support for multi-signal scoring or relational joins, queries involving these operations are streamed to the application layer, which applies the Rerank strategy to finalize the Top-\(k\) results.
\end{enumerate}

\vspace{-4mm}

\subsection{Setup and Methodology}
\label{subsec:exp-setting}
\sysb is implemented in about 10{,}000 lines of Python and C, with part of the
development driven by Claude Code (Opus~4.6--4.8) and extensive manual code review. Experiments were conducted on PostgreSQL~18.3 with the DASE-modified pgvector
extension (v0.8.5). Text vectors use \texttt{Gemini-embedding-001} (1536
dimensions); MOLECULE additionally includes molecular descriptors and an ECFP
signal. We evaluated a set of multi-step reasoning queries involving structured
filters, multi-vector relevance scoring, and join-based conditions. All vector
columns use L2 HNSW indexes with \(m=16\), \(ef_{construction}=64\), and default
\(ef_{search}=40\).  
Each database instance runs on a dedicated set
of 16 physical AMD~EPYC~9575F cores (up to 5.0\,GHz) and 64\,GB of RAM. This
hardware supports AVX-512 instructions, which pgvector leverages
to accelerate vector distance computations. Evaluation was CPU-based; no GPUs
were used.

We report each method as a recall--throughput curve. For every method we sweep a search-effort parameter and record
recall against QPS: the HNSW \(ef_{search}\) and IVFFlat probe count for index
scans, the threshold-algorithm admission tolerance \(\varepsilon\) for DASE and
Rerank, the post-filter width for Milvus, and a per-query time budget for
best-effort strategies. Each operating point averages 5 runs after
discarding a warm-up run, and
shaded bands denote \(\pm 1\) standard deviation; per method we
retain the Pareto frontier over the resulting (recall, QPS) points. Recall is
measured against the released exhaustive top-\(20\) result for each planned query.
Index construction time is isolated and excluded from all QPS measurements. Table~\ref{tab:exp-setup} reports the
dataset-specific relation, \idx{}, and workload configurations; query-specific
thresholds and score weights are stored with the released workload instances.

For SemBench, \textbf{\sys} performs ranked embedding-based prefiltering, omitting LLM judgment, using counterfactual anchors for semantic filters and calibrated thresholds for semantic joins. \textbf{\sys + BigQuery} sends the same candidates to BigQuery semantic operators; its cost and latency therefore describe the composed pipeline. Table~\ref{tab:sembench} reports USD, the SemBench task metric, and end-to-end wall-clock latency including remote Gemini API time; \texttt{X} denotes unsupported configurations and scenario averages exclude them. The LOTUS, Palimpzest, ThalamusDB, and BigQuery columns reproduce SemBench's published \texttt{gemini-2.5-flash} configuration~\cite{lao_sembench_2025}; only the \sys{} and \sys{} + BigQuery columns are ours.

\subsection{Multi-Step Top-\(k\) Reasoning}

\emph{Deep Research.} 
We evaluate \sysb and the alternatives across the \textbf{IMDB} (Figure~\ref{fig:mrq-imdb}) and \textbf{MOLECULE} (Figure~\ref{fig:mrq-molecule}) datasets over our 8 workloads. Each panel plots the recall--throughput (QPS) Pareto frontier obtained by sweeping each method's search-effort as introduced in the setup section (Section~\ref{subsec:exp-setting}); up and to the right is better.
Milvus excels on the single-signal W1 workload, but its advantage reverses when queries require multi-score ranking. Filter-first becomes a sequential scan without predicates, rerank reduces to score-first with one signal, and score-first is infeasible for W7/W8 because it forms a full Cartesian product. On W3, \sysb outperforms round-robin rerank, indicating the value of adaptive stream selection; across W5--W8, \sysb and \idx{} sustain higher QPS at matched recall.
We separately study how performance scales with the number of coupled join positions in Section~\ref{sec:scaling_multi-step_k}.

\begin{figure*}[tb]
    \centering
    \begin{minipage}[c]{0.62\textwidth}
        \centering
        \includegraphics[width=0.9\linewidth]{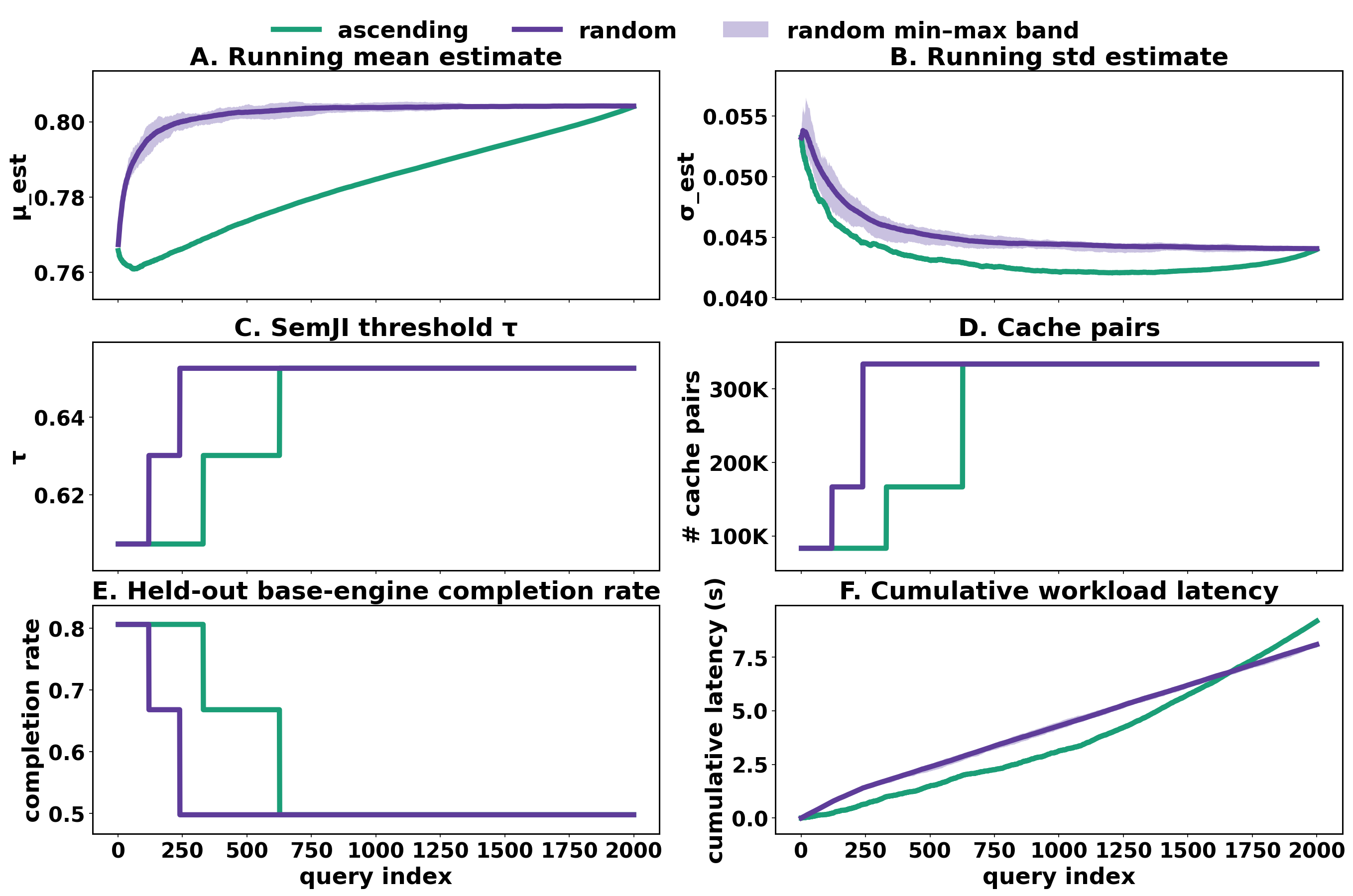}
        \vspace{-4mm}
        \caption{\idx evolution test.}
        \label{fig:semji_evolve}

        \vspace{3mm} 
        \centering
        \includegraphics[width=1\linewidth]{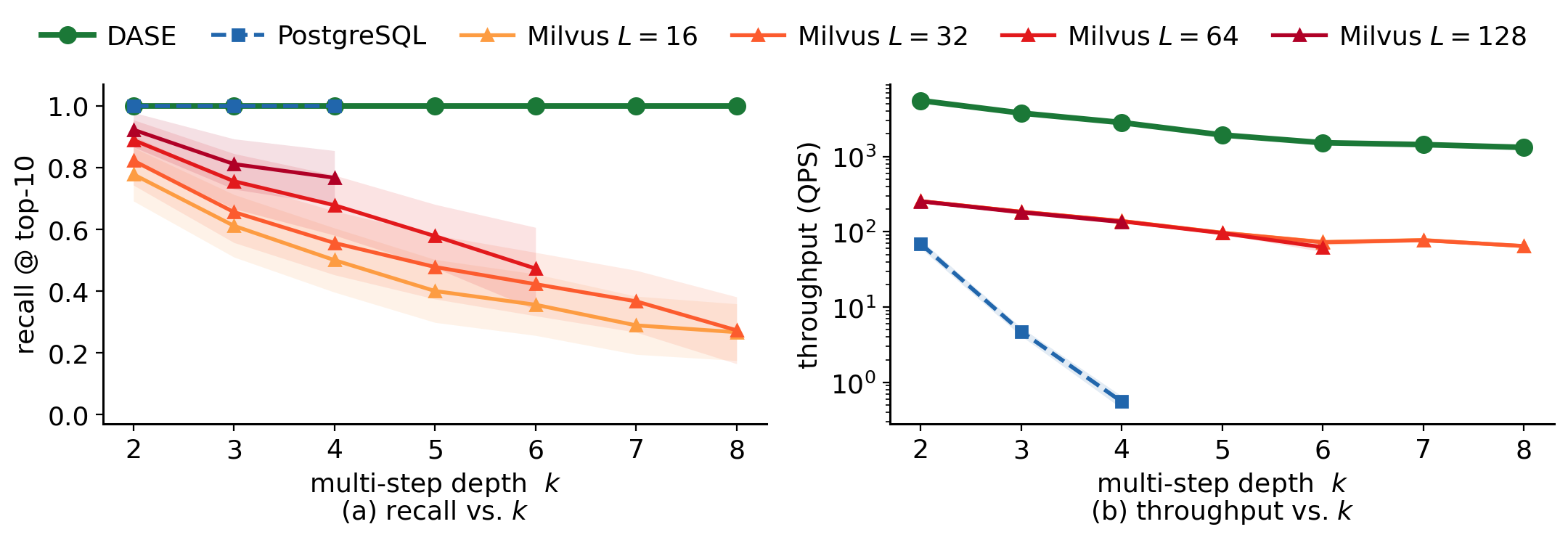}
        \vspace{-8mm}
        \caption{Number of Join Steps vs recall in NFCorpus}
        \label{fig:nfcorpus}
    \end{minipage}
    \hfill
    \begin{minipage}[c]{0.36\textwidth}
        \centering
        \normalsize 
        
        \captionof{table}{\idx Construction.}
        \label{tab:semji_construct}
        \vspace{-2mm}
        \begin{tabularx}{0.9\linewidth}{Xcc} 
            \toprule
            \textbf{Method} & \textbf{Recall} & \textbf{Time} \\
            \midrule
            Exact Construction & 1.00 & $\sim$81.2 h \\
            Iterative & 0.98 & 14.0 m \\
            Batched & 0.98 & 3.0 m \\
            Fully Vectorized& 0.98 & 1.6 m \\
            \bottomrule
        \end{tabularx}
        
        \bigskip 
        
        \captionof{table}{Filtered search results.}
        \label{tab:filter_results}
        \vspace{-4mm}
        \begin{tabularx}{0.9\linewidth}{lXcc} 
            \toprule
            \textbf{Method} & \textbf{Setting} & \textbf{Recall} & \textbf{QPS} \\
            \midrule
            baseline & - & 1.0  & 0.23 \\
            \midrule
            & 2-hop on & 0.905 & 9.6 \\
            Bitmap & 2-hop off & 0.884 & 10.5 \\
                   & adaptive & 0.903 & 10.1 \\
            \midrule
             & 4 bits/element & 0.817 & 9.1 \\
             Bloom & 12 b/element & 0.872 & 8.8 \\
            & 20 b/element & 0.883 & 8.2 \\
            \bottomrule
        \end{tabularx}

        \bigskip 

        \captionof{table}{Component ablation on \textsc{IMDB}. Each mechanism is disabled in isolation on the workload that exercises it; we report recall / throughput.}
        \label{tab:ablation}
        \vspace{-2mm}
        \resizebox{1.0\linewidth}{!}{
        \begin{tabular}{@{}llcc@{}}
            \toprule
            \textbf{Mechanism} & \textbf{Variant} & \textbf{Recall} & \textbf{QPS} \\
            \midrule
            \multirow{2}{*}{\idx{} (W5)}
              & \sys (with \idx{})      & \(0.95\) & \(92.0\) \\
              & \(-\) \idx{} (on-the-fly join) & \(0.71\) & \(5.3\) \\
            \midrule
            \multirow{2}{*}{Stream selection (W3)}
              & \sys (adaptive)        & \(0.98\) & \(57.8\) \\
              & \(-\) adaptive (round-robin) & \(0.98\) & \(29.8\) \\
            \bottomrule
        \end{tabular}
        }

    \end{minipage}
    
    \vspace{-2mm}
\end{figure*}

\emph{Semantic queries.} Table~\ref{tab:sembench} summarizes cost, quality, and latency on SemBench. \sys alone suffices for semantic-filter (\textbf{F}) tasks that reduce to a binary or coarse distinction, such as positive vs.\ negative reviews in \textit{Movie}: comparing embedding distance to counterfactual anchors returns valid answers at orders-of-magnitude lower cost and latency than any LLM-backed baseline.
For semantic-map (\textbf{M}) and LLM-semantic-join (\textbf{J}) tasks that exceed embedding-distance reasoning, \sys + BigQuery substantially raises quality (e.g., \textit{MMQA} avg.\ 0.11 \(\to\) 0.76). Prefiltering with \sys also lets BigQuery beat its own standalone numbers at a fraction of the cost: on \textit{E-Commerce}, \$0.54/38s at quality 0.80 versus BigQuery alone at \$2.42/45s and 0.67. The LLM sees only the prefilter's survivors, so it spends less and decides better.
The one failure mode is embedding quality itself: in \textit{Wildlife}, rare words like ``impala'' are poorly aligned, so neither configuration recovers fully.

Overall, \sys delivers low-cost answers when embedding distance suffices, and a high-recall prefilter that makes downstream LLM evaluation more efficient and accurate when it does not.

\vspace{-2mm}
\subsection{\idx Construction and Maintenance} 

We build \idx using HNSW-based ANN range queries with a radius threshold. Table~\ref{tab:semji_construct} 
shows an ablation from a naive iterative approach, through a batched 
lateral join, to a fully vectorized whole-table HNSW range query --- 
a single \texttt{JOIN LATERAL} that runs entirely in-database. This 
reduces construction time from 81.2 hours to 1.6 minutes, only losing
2\% from recall.

\paragraph{\idx{} evolution.}
We evaluate the capacity-based maintenance policy
(Section~\ref{sec:maintenance}) on the linkage-only MOLECULE W5 stream of
2{,}000 seed queries. A random warmup sample fixes
\(\tau_{init}=\mu-3\sigma=0.607\) before the workload is observed; overflowing
queries use the base-engine outer-band fallback while the controller grows the
index asynchronously.

On this stream, two capacity doublings grow \idx{} from
\(83\text{K}\) to \(333\text{K}\) pairs and reach
\(\tau_{converged}=0.6526\) (Figure~\ref{fig:semji_evolve}). The
final \(333\text{K}\) retained pairs are only \(0.011\%\) of the \(3.14\)B
fact--paper Cartesian product (6.7 links/fact and 5.3 links/paper on
average), so \idx{} remains highly compact relative to exact join enumeration.
The held-out
overflow rate falls from \(80.6\%\) to \(49.8\%\); the remaining overflows are
budget-limited sparse seeds. Ascending and random query orders reach the same
final radius (spread \(<0.001\)) with two expansions, differing only in when the
work is paid (a \(1.14\times\) cumulative-latency difference).

\subsection{Scaling Multi-step Reasoning Depth}
\label{sec:scaling_multi-step_k}

We isolate how each system scales vs{.} the number of coupled join positions~\(k\), on the multi-dimensional \textsc{NFCorpus} workload \(Q=(P,S,J)\). Here a \(k\)-step query fixes
per-position anchors \(a_1,\dots,a_k\) and asks for the top-\(n\) length-\(k\)
walks \((p_1,\dots,p_k)\) satisfying
\[
\min\ \tfrac{1}{k}\sum_{i=1}^{k}\mathrm{dist}(a_i,p_i)
\quad\text{s.t.} 
P(p_i)\ \forall i, 
\mathrm{dist}(p_i,p_{i+1})\le\tau\ \forall i,\quad
p_i\neq p_{i-2},
\]
i.e.\ a monotone score \(S\) (title-embedding distance) minimized subject to a
predicate \(P\) (selectivity \(\sigma{=}0.18\)) and a threshold
intervention--condition similarity join \(J\) (\(\tau{=}0.35\)) between consecutive
positions, without immediate backtracking. We sweep \(k{=}2\ldots8\) and report
recall against the exact top-\(n\) result and throughput (Figure~\ref{fig:nfcorpus};
\(95\%\) CI over \(90\) instances, \(3\) seeds \(\times\) \(30\)).

\textsc{PostgreSQL} hands the whole predicate-filtered \(k\)-way self-join to the SQL planner:
it uses the materialized edge relation, but produces a
combinatorial intermediate. \textsc{Milvus}
can apply per-position scalar filters, but cannot execute the
cross-position join \(J\) or rank the paths. It therefore retrieves the
per-position top-\(L\) nearest neighbors, applies \(P\), and reranks the surviving
length-\(k\) walks in the application layer (\(L\) is its accuracy knob). Given
the materialized \idx edge set, \sys recognizes the query as a \emph{minimum-cost
bounded-length walk} on the join graph and uses a
 Viterbi-like dynamic program with beam width \(n\) (the number of returned walks). It retains the top-\(n\) partial
paths per directed-edge state (remembering the preceding vertex to exclude
immediate backtracking); its worst-case time is
\(O\!\left(kn\sum_v \deg(v)^2\right)\), or \(O(kn|E|)\) for bounded-degree
graphs.

Both baselines fail as \(k\) grows (Figure~\ref{fig:nfcorpus}). \textsc{PostgreSQL}
is exact but its \(k\)-way join explodes, collapsing from \(70\) to \(0.55\) QPS
over \(k{=}2\ldots4\) and timing out for \(k{\ge}5\). \textsc{Milvus}'s qualifying
sets are rarely nearest neighbors, so its recall decays (\(L{=}16\):
\(0.78\!\to\!0.27\)) and its \(L^{k}\) enumeration turns the accurate settings
infeasible (\(L{=}128\) by \(k{=}6\)). \sys instead keeps full recall for every \(k\) while
sustaining \(5{,}540\) down to \(1{,}321\) QPS --- one-to-four
orders of magnitude above the planner and never infeasible. The gap \emph{widens}
with \(k\): the baselines time out or blow up while \sys stays flat, confirming
that join- and predicate-aware materialization pays off exactly in the multi-step
regime.

\vspace{-3mm}
\subsection{Filter Sets in the ANN Index} 

Table~\ref{tab:filter_results} evaluates predicate-aware HNSW search strategies. The exhaustive baseline returns the exact top-\(k\) answer (recall 1.0), but at low throughput (0.23 QPS). To accelerate filtering, we compare Bitmap and Bloom filter structures, finding that the Bitmap approach consistently delivers better recall and higher QPS across all configurations. Within the Bitmap approach, we further evaluate graph traversal methods. The 2-hop adaptive strategy yields the optimal trade-off: by dynamically exploring extended neighbors only when valid nodes are sparse, it achieves a roughly 40\(\times\) speedup (10.1 QPS) over the baseline while maintaining a highly competitive recall of 0.903. This adaptive mechanism balances the speed of the ``2-hop off'' method with the accuracy of the ``always-on'' method.

\vspace{-3mm}
\subsection{\sys{} Ablation Study}
\label{sec:ablation}

We isolate DASE's two central algorithmic contributions by disabling one at a time on the workload that exercises it in isolation, holding everything else fixed (Table \ref{tab:ablation}; \textsc{IMDB}, \(100\) queries, warm cache, same host).

\emph{\idx{} join materialization} (Table \ref{tab:ablation}, top). On the pure
semantic-join workload~W5 we keep DASE's seed streamer unchanged and switch
\emph{only} the partner check:
the full engine probes the materialized \textsc{\idx{}} index, while the ablated
variant resolves each candidate's partner on the fly with an ANN query against
the partner table. Removing \idx{} is strictly worse on \emph{both} axes --- recall
drops from \(0.95\) to \(0.71\) and throughput falls \(17\times\) (\(92.0\!\to\!5.3\)
QPS) --- because without the precomputed join the partner test must either
approximate (missing qualifying partners, hence the recall loss) or run exact
(far slower still). Materializing the join is what makes the join-bearing
workloads both accurate and fast.

\emph{Distribution-aware stream selection} (Table~\ref{tab:ablation}, bottom). On the
pure multi-scoring workload~W3 we replace DASE's threshold-algorithm stream
selector --- which
extends the stream with the largest expected gap reduction \(G_j\!=\!w_j/\mathrm{PDF}(x_j)\)
--- with vanilla round-robin extension. At the exact operating point (\(\varepsilon{=}0\),
recall~\(0.98\)) adaptive selection delivers \(1.9\times\) the throughput
(\(57.8\) vs \(29.8\) QPS): choosing the right stream closes the termination gap in
far fewer extension rounds. The gain is largest exactly here, in the
high-recall regime where many rounds are needed, and concentrates on queries
whose signals have heterogeneous score distributions.

\vspace{-2mm}
\subsection{Robustness to Correlated Embeddings}
\label{sec:pancaking-experiments}

The \emph{materialization radius} \(\tau_{cap}\) retains every endpoint pair within \(\mathrm{dist}(a,b)\leq\tau_{cap}\); \idx{} is the union of these per-endpoint neighborhoods. However, in a multi-predicate query it is not one ball around the results. A sample-derived cutoff can select the close-pair tail, but anisotropy can make a raw-distance cutoff unreliable.
We therefore investigated the distribution of distances for the IMDB and MOLECULE W7/W8 join endpoints. Their covariance participation ratios are only 75--93 and 112--166 of 1,536 dimensions (i.e., 5--11\% effective variance-carrying directions).

We then consider different methods for handling this anisotropy. For each corpus, we ran 200 queries over two independent 3K-by-3K reservoir samples (400 query--sample units per arm). During execution, we apply predicates before ranking and macro-average recall@20/precision@20 against each arm's exhaustive predicate-filtered top-20 oracle; these are candidate-retrieval, not final-answer, measures. We use a 95\% two-stage percentile bootstrap over endpoint samples and then queries. The raw-workload Recall/Precision intervals are 0.764--0.825/0.815--0.868 (IMDB) and 0.459--0.554/0.648--0.749 (MOLECULE); ABTT's are 0.982--0.998/0.999--1.000 and 0.954--0.981/0.954--0.981. Each arm calibrates \(\tau_{cap}\) in its own retrieval space; PCA uses 0.1 shrinkage and ABTT removes one leading component, both fit on each sample's 6K endpoints.

\begin{table}[htbp]
\centering
\caption{\label{tab:real-semji-robustness} Real predicate-filtered \idx{} quality. Each R/P entry is recall@20 / precision@20 relative to that arm's exhaustive oracle. The pairs column gives the materialized fraction for IMDB / MOLECULE; it is not a ratio.}
\vspace{-2mm}
\scriptsize
\begin{tabular}{lccc}
\toprule
\textbf{Arm} & \textbf{IMDB Rec./Prec.} & \textbf{MOLECULE Rec./Prec.} & \textbf{Pairs \% (I/M)} \\
\midrule
Raw workload cutoff & 0.795 / 0.842 & 0.506 / 0.700 & 0.60\% / 0.03\% \\
Raw \(\mu-3\sigma\) & 0.920 / 0.929 & 0.867 / 0.867 & 0.42\% / 0.62\% \\
PCA whitening & 0.910 / 0.920 & 0.868 / 0.868 & 0.44\% / 0.47\% \\
ABTT & \textbf{0.991 / 1.000} & \textbf{0.969 / 0.969} & 1.09\% / 0.93\% \\
Raw radius control & 0.954 / 0.962 & 0.881 / 0.881 & 1.09\% / 0.93\% \\
\bottomrule
\end{tabular}
\vspace{-3mm}
\end{table}

We observe from Table~\ref{tab:real-semji-robustness} that the raw workload cutoff strategy loses substantial coverage, especially on MOLECULE. \textbf{ABTT} has the highest within-arm coverage (0.991/1.000 on IMDB and 0.969/0.969 on MOLECULE). At the same materialized-pair fraction, the \textbf{raw radius control} remains 3.7 and 8.8 recall points lower, respectively; transformed-space calibration therefore improves which pairs are retained, not merely their number.


\section{Related Work}

\textbf{Semantic-operator systems.}
LOTUS~\cite{patel_semantic_2025} and Palimpzest~\cite{liu2025palimpzest} develop declarative operations whose predicates, joins, or transformations may require LLM-based evaluation. ThalamusDB~\cite{jo_thalamusdb_2024,trummer_implementing_2025} similarly applies approximate query processing to multimodal data. SemBench~\cite{lao_sembench_2025} compares these systems and others on semantic-filter, merge, and join tasks. DASE is complementary to these systems: it assumes that a semantic-query planner has supplied a retrieval plan and scoring function, then constructs and ranks a high-recall candidate set before final semantic judgment. Our \sys{} + BigQuery configuration instantiates this prefiltering role.
There are some operator-specific optimizations include BARGAIN~\cite{zeighami2025cut} for filtering and classification, and FDJ~\cite{zeighami2025join} for semantic joins.
\sys instead supports prefiltering across multiple semantic operators and integrates joins with top-\(k\) retrieval under monotone multi-signal scoring functions.

\textbf{Benchmarks for multi-source retrieval and reasoning.} \\ MMQA~\cite{wu2025mmqa} evaluates LLMs on multi-table retrieval, question answering, SQL generation, and identification of primary and foreign keys; its retrieval task ranks tables. STaRK~\cite{wu2024stark} benchmarks entity retrieval under combined textual and relational requirements, including multi-hop relations, and reports retrieval quality and latency. FDABench~\cite{wang2026fdabench} evaluates analytical agents that integrate structured and unstructured sources, measuring answer quality, tool use, and end-to-end cost and latency. These benchmarks assess retrieval or analytical workflows from natural-language requests. Our benchmark instead takes SQL queries and a multi-signal scoring function as given, and measures the recall--throughput trade-off for jointly scored evidence tuples. Its workloads combine structured constraints, exact joins, and thresholded embedding-similarity joins, with exhaustive top-\(k\) reference results defined by the fixed query and scoring semantics.

\textbf{Record linking.} Increasingly, record linking~\cite{christen_data_2012} tasks are being addressed using embeddings and LLMs~\cite{zeakis2025avenger}. \idx can be viewed as a materialized snapshot of links above a threshold.

\textbf{Hybrid DBMSs.}
Vector DBMSs such as Milvus~\cite{wang_milvus_2021} combine vector retrieval with scalar filtering, while filtered-ANN work such as ACORN~\cite{patel2024acorn} and Filtered-DiskANN~\cite{gollapudi_filtered-diskann_2023} improve \textbf{point-query vector search} under structured predicates. These methods are useful building blocks for DASE's filtered score streams, but their central access pattern is one query vector against one indexed collection. Our targets include M-to-N embedding-similarity joins between distinct relational rows, and exact joins that connect those rows into a pattern, with a monotone score that combines several lookup and join affinities. \idx{} materializes a sparse subset of those cross-table pairs so that the resulting pattern can be evaluated with relational operations rather than repeated point lookups.

\textbf{Ranked join processing.}
Threshold-style top-\(k\) algorithms~\cite{fagin_optimal_2003,kimelfeld_finding_2006,tziavelis_ranked_2024} combine sorted score streams with random access and stopping bounds. DASE adopts this execution principle: an embedding-similarity-join score belongs to a pair of rows that must be connected through the query pattern before it contributes to an output tuple. The materialized join relation, filtered score streams, and Multi-Scoring Aggregator provide the access paths and bound-aware coordination needed for this setting.

\textbf{Embedding geometry.}
Embedding whitening and anisotropy correction are well studied for NLP sentence and word representations~\cite{ethayarajh2019contextual,mu_allbutthetop_2018,su_whitening_2021,arora2017simple}. We use these techniques in a different role: when anisotropic or shared-context endpoint embeddings make a distribution-derived materialization radius unreliable, DASE separately calibrates a transformed embedding space and constructs the corresponding embedding-similarity join index there.

\section{Conclusions}

Multi-step reasoning queries frequently arise in agentic data engineering tasks such as scientific discovery. We introduced \idx{}, which facilitates efficient candidate exploration in top-\(k\) multi-step reasoning queries, along with ANN extensions to enable batched requests. We developed DASE, a query processing strategy that jointly optimizes attribute predicates and vector similarity scoring across joins.
We conducted an extensive evaluation on existing semantic benchmarks and a scientific-discovery workload that captures the structure of multi-step reasoning queries. Our results demonstrate the retrieval benefits and approximation trade-offs of \idx{}.
We also evaluated a robustness treatment for anisotropic endpoint embeddings, where a raw distribution-derived \idx{} materialization radius can miss useful links.

We hope that our formalization, methods, and benchmarks serve as a foundation for continued progress on query systems that bridge structured and unstructured data at scale. In ongoing work, we are focused on developing a high-level planner that uses costs to break agentic plans into stages with multi-step reasoning components.


\bibliographystyle{ACM-Reference-Format}
\bibliography{refs,references,references-mendeley}


\begin{thebibliography}{63}


\ifx \showCODEN    \undefined \def \showCODEN     #1{\unskip}     \fi
\ifx \showISBNx    \undefined \def \showISBNx     #1{\unskip}     \fi
\ifx \showISBNxiii \undefined \def \showISBNxiii  #1{\unskip}     \fi
\ifx \showISSN     \undefined \def \showISSN      #1{\unskip}     \fi
\ifx \showLCCN     \undefined \def \showLCCN      #1{\unskip}     \fi
\ifx \shownote     \undefined \def \shownote      #1{#1}          \fi
\ifx \showarticletitle \undefined \def \showarticletitle #1{#1}   \fi
\ifx \showURL      \undefined \def \showURL       {\relax}        \fi
\providecommand\bibfield[2]{#2}
\providecommand\bibinfo[2]{#2}
\providecommand\natexlab[1]{#1}
\providecommand\showeprint[2][]{arXiv:#2}

\bibitem[Anderson et~al\mbox{.}(2024)]%
        {anderson_design_2024}
\bibfield{author}{\bibinfo{person}{Eric Anderson}, \bibinfo{person}{Jonathan
  Fritz}, \bibinfo{person}{Austin Lee}, \bibinfo{person}{Bohou Li},
  \bibinfo{person}{Mark Lindblad}, \bibinfo{person}{Henry Lindeman},
  \bibinfo{person}{Alex Meyer}, \bibinfo{person}{Parth Parmar},
  \bibinfo{person}{Tanvi Ranade}, \bibinfo{person}{Mehul~A. Shah},
  \bibinfo{person}{Ben Sowell}, \bibinfo{person}{Dan~G. Tecuci},
  \bibinfo{person}{Vinayak Thapliyal}, {and} \bibinfo{person}{Matt Welsh}.}
  \bibinfo{year}{2024}\natexlab{}.
\newblock \showarticletitle{The {Design} of an {LLM}-powered {Unstructured}
  {Analytics} {System}}.
\newblock \bibinfo{journal}{\emph{ArXiv}}  \bibinfo{volume}{abs/2409.00847}
  (\bibinfo{year}{2024}).
\newblock
\urldef\tempurl%
\url{https://api.semanticscholar.org/CorpusID:272368251}
\showURL{%
\tempurl}


\bibitem[Arora et~al\mbox{.}(2017)]%
        {arora2017simple}
\bibfield{author}{\bibinfo{person}{Sanjeev Arora}, \bibinfo{person}{Yingyu
  Liang}, {and} \bibinfo{person}{Tengyu Ma}.} \bibinfo{year}{2017}\natexlab{}.
\newblock \showarticletitle{A simple but tough-to-beat baseline for sentence
  embeddings}. In \bibinfo{booktitle}{\emph{International conference on
  learning representations}}.
\newblock


\bibitem[Asai et~al\mbox{.}(2026)]%
        {asai2026synthesizing}
\bibfield{author}{\bibinfo{person}{Akari Asai}, \bibinfo{person}{Jacqueline
  He}, \bibinfo{person}{Rulin Shao}, \bibinfo{person}{Weijia Shi},
  \bibinfo{person}{Amanpreet Singh}, \bibinfo{person}{Joseph~Chee Chang},
  \bibinfo{person}{Kyle Lo}, \bibinfo{person}{Luca Soldaini},
  \bibinfo{person}{Sergey Feldman}, \bibinfo{person}{Mike D’arcy},
  {et~al\mbox{.}}} \bibinfo{year}{2026}\natexlab{}.
\newblock \showarticletitle{Synthesizing scientific literature with
  retrieval-augmented language models}.
\newblock \bibinfo{journal}{\emph{Nature}} \bibinfo{volume}{650},
  \bibinfo{number}{8103} (\bibinfo{year}{2026}), \bibinfo{pages}{857--863}.
\newblock


\bibitem[Bloom(1970)]%
        {bloom_spacetime_1970}
\bibfield{author}{\bibinfo{person}{Burton~H. Bloom}.}
  \bibinfo{year}{1970}\natexlab{}.
\newblock \showarticletitle{Space/{Time} {Trade}-offs in {Hash} {Coding} with
  {Allowable} {Errors}}.
\newblock \bibinfo{journal}{\emph{CACM}} \bibinfo{volume}{13},
  \bibinfo{number}{7} (\bibinfo{date}{July} \bibinfo{year}{1970}),
  \bibinfo{pages}{422--426}.
\newblock


\bibitem[Boteva et~al\mbox{.}(2016)]%
        {boteva2016full}
\bibfield{author}{\bibinfo{person}{Vera Boteva}, \bibinfo{person}{Demian
  Gholipour}, \bibinfo{person}{Artem Sokolov}, {and} \bibinfo{person}{Stefan
  Riezler}.} \bibinfo{year}{2016}\natexlab{}.
\newblock \showarticletitle{A full-text learning to rank dataset for medical
  information retrieval}. In \bibinfo{booktitle}{\emph{European Conference on
  Information Retrieval}}. Springer, \bibinfo{pages}{716--722}.
\newblock


\bibitem[Cao et~al\mbox{.}(2010)]%
        {cao_feedback-driven_2010}
\bibfield{author}{\bibinfo{person}{Huiping Cao}, \bibinfo{person}{Yan Qi},
  \bibinfo{person}{K~Selçuk Candan}, {and} \bibinfo{person}{Maria~Luisa
  Sapino}.} \bibinfo{year}{2010}\natexlab{}.
\newblock \showarticletitle{Feedback-driven result ranking and query refinement
  for exploring semi-structured data collections}. In
  \bibinfo{booktitle}{\emph{{EDBT}}}. \bibinfo{pages}{3--14}.
\newblock


\bibitem[Chen et~al\mbox{.}(2026)]%
        {chen2026mapreduce}
\bibfield{author}{\bibinfo{person}{Mingju Chen}, \bibinfo{person}{Guibin
  Zhang}, \bibinfo{person}{Heng Chang}, \bibinfo{person}{Yuchen Guo}, {and}
  \bibinfo{person}{Shiji Zhou}.} \bibinfo{year}{2026}\natexlab{}.
\newblock \showarticletitle{A-MapReduce: Executing Wide Search via Agentic
  MapReduce}.
\newblock \bibinfo{journal}{\emph{arXiv preprint arXiv:2602.01331}}
  (\bibinfo{year}{2026}).
\newblock


\bibitem[Chen et~al\mbox{.}(2016)]%
        {chen_learning_2016}
\bibfield{author}{\bibinfo{person}{Xu Chen}, \bibinfo{person}{Zheng Qin},
  \bibinfo{person}{Yongfeng Zhang}, {and} \bibinfo{person}{Tao Xu}.}
  \bibinfo{year}{2016}\natexlab{}.
\newblock \showarticletitle{Learning to rank features for recommendation over
  multiple categories}. In \bibinfo{booktitle}{\emph{Proceedings of the 39th
  {International} {ACM} {SIGIR} conference on {Research} and {Development} in
  {Information} {Retrieval}}}. \bibinfo{pages}{305--314}.
\newblock


\bibitem[Chithrananda et~al\mbox{.}(2020)]%
        {chithrananda_chemberta_2020}
\bibfield{author}{\bibinfo{person}{Seyone Chithrananda},
  \bibinfo{person}{Gabriel Grand}, {and} \bibinfo{person}{Bharath Ramsundar}.}
  \bibinfo{year}{2020}\natexlab{}.
\newblock \bibinfo{title}{{ChemBERTa}: {Large}-{Scale} {Self}-{Supervised}
  {Pretraining} for {Molecular} {Property} {Prediction}}.
\newblock
\href{https://doi.org/10.48550/arXiv.2010.09885}{doi:\nolinkurl{10.48550/arXiv.2010.09885}}
\newblock
\shownote{arXiv:2010.09885}.


\bibitem[Christen(2011)]%
        {christen_survey_2011}
\bibfield{author}{\bibinfo{person}{Peter Christen}.}
  \bibinfo{year}{2011}\natexlab{}.
\newblock \showarticletitle{A survey of indexing techniques for scalable record
  linkage and deduplication}.
\newblock \bibinfo{journal}{\emph{IEEE transactions on knowledge and data
  engineering}} \bibinfo{volume}{24}, \bibinfo{number}{9}
  (\bibinfo{year}{2011}), \bibinfo{pages}{1537--1555}.
\newblock


\bibitem[Christen(2012)]%
        {christen_data_2012}
\bibfield{author}{\bibinfo{person}{Peter Christen}.}
  \bibinfo{year}{2012}\natexlab{}.
\newblock \bibinfo{booktitle}{\emph{Data {Matching}: {Concepts} and
  {Techniques} for {Record} {Linkage}, {Entity} {Resolution}, and {Duplicate}
  {Detection}}}.
\newblock \bibinfo{publisher}{Springer}.
\newblock


\bibitem[Datar et~al\mbox{.}(2004)]%
        {datar2004locality}
\bibfield{author}{\bibinfo{person}{Mayur Datar}, \bibinfo{person}{Nicole
  Immorlica}, \bibinfo{person}{Piotr Indyk}, {and} \bibinfo{person}{Vahab~S
  Mirrokni}.} \bibinfo{year}{2004}\natexlab{}.
\newblock \showarticletitle{Locality-sensitive hashing scheme based on p-stable
  distributions}. In \bibinfo{booktitle}{\emph{Proceedings of the twentieth
  annual symposium on Computational geometry}}. \bibinfo{pages}{253--262}.
\newblock


\bibitem[Ethayarajh(2019)]%
        {ethayarajh2019contextual}
\bibfield{author}{\bibinfo{person}{Kawin Ethayarajh}.}
  \bibinfo{year}{2019}\natexlab{}.
\newblock \showarticletitle{How {Contextual} are {Contextualized} {Word}
  {Representations}? {Comparing} the {Geometry} of {BERT}, {ELMo}, and {GPT}-2
  {Embeddings}}. In \bibinfo{booktitle}{\emph{Proceedings of the 2019
  {Conference} on {Empirical} {Methods} in {Natural} {Language} {Processing}
  and the 9th {International} {Joint} {Conference} on {Natural} {Language}
  {Processing} ({EMNLP}-{IJCNLP})}}. \bibinfo{pages}{55--65}.
\newblock


\bibitem[Fagin et~al\mbox{.}(2003)]%
        {fagin_optimal_2003}
\bibfield{author}{\bibinfo{person}{Ronald Fagin}, \bibinfo{person}{Amnon
  Lotem}, {and} \bibinfo{person}{Moni Naor}.} \bibinfo{year}{2003}\natexlab{}.
\newblock \showarticletitle{Optimal aggregation algorithms for middleware}.
\newblock \bibinfo{journal}{\emph{Journal of computer and system sciences}}
  \bibinfo{volume}{66}, \bibinfo{number}{4} (\bibinfo{year}{2003}),
  \bibinfo{pages}{614--656}.
\newblock


\bibitem[Fernandes and Bernardino(2015)]%
        {fernandes2015bigquery}
\bibfield{author}{\bibinfo{person}{S{\'e}rgio Fernandes} {and}
  \bibinfo{person}{Jorge Bernardino}.} \bibinfo{year}{2015}\natexlab{}.
\newblock \showarticletitle{What is bigquery?}. In
  \bibinfo{booktitle}{\emph{Proceedings of the 19th International Database
  Engineering \& Applications Symposium}}. \bibinfo{pages}{202--203}.
\newblock


\bibitem[Gao et~al\mbox{.}(2023)]%
        {gao_retrieval-augmented_2023}
\bibfield{author}{\bibinfo{person}{Yunfan Gao}, \bibinfo{person}{Yun Xiong},
  \bibinfo{person}{Xinyu Gao}, \bibinfo{person}{Kangxiang Jia},
  \bibinfo{person}{Jinliu Pan}, \bibinfo{person}{Yuxi Bi}, \bibinfo{person}{Yi
  Dai}, \bibinfo{person}{Jiawei Sun}, \bibinfo{person}{Qianyu Guo},
  \bibinfo{person}{Meng Wang}, {and} \bibinfo{person}{Haofen Wang}.}
  \bibinfo{year}{2023}\natexlab{}.
\newblock \showarticletitle{Retrieval-{Augmented} {Generation} for {Large}
  {Language} {Models}: {A} {Survey}}.
\newblock \bibinfo{journal}{\emph{ArXiv}}  \bibinfo{volume}{abs/2312.10997}
  (\bibinfo{year}{2023}).
\newblock
\urldef\tempurl%
\url{https://api.semanticscholar.org/CorpusID:266359151}
\showURL{%
\tempurl}


\bibitem[Ghareeb et~al\mbox{.}(2026)]%
        {ghareeb2026multi}
\bibfield{author}{\bibinfo{person}{Ali~Essam Ghareeb},
  \bibinfo{person}{Benjamin Chang}, \bibinfo{person}{Ludovico Mitchener},
  \bibinfo{person}{Angela Yiu}, \bibinfo{person}{Caralyn~J Szostkiewicz},
  \bibinfo{person}{Dmytro Shved}, \bibinfo{person}{Gavin~J Gyimesi},
  \bibinfo{person}{Jon~M Laurent}, \bibinfo{person}{Samantha~M Wright},
  \bibinfo{person}{Muhammed~T Razzak}, {et~al\mbox{.}}}
  \bibinfo{year}{2026}\natexlab{}.
\newblock \showarticletitle{A multi-agent system for automating scientific
  discovery}.
\newblock \bibinfo{journal}{\emph{Nature}} (\bibinfo{year}{2026}),
  \bibinfo{pages}{1--3}.
\newblock


\bibitem[Gollapudi et~al\mbox{.}(2023)]%
        {gollapudi_filtered-diskann_2023}
\bibfield{author}{\bibinfo{person}{Siddharth Gollapudi}, \bibinfo{person}{Neel
  Karia}, \bibinfo{person}{Varun Sivashankar}, \bibinfo{person}{Ravishankar
  Krishnaswamy}, \bibinfo{person}{Nikit Begwani}, \bibinfo{person}{Swapnil
  Raz}, \bibinfo{person}{Yiyong Lin}, \bibinfo{person}{Yin Zhang},
  \bibinfo{person}{Neelam Mahapatro}, \bibinfo{person}{Premkumar Srinivasan},
  {and} \bibinfo{person}{{others}}.} \bibinfo{year}{2023}\natexlab{}.
\newblock \showarticletitle{Filtered-diskann: {Graph} algorithms for
  approximate nearest neighbor search with filters}. In
  \bibinfo{booktitle}{\emph{Proceedings of the {ACM} {Web} {Conference} 2023}}.
  \bibinfo{pages}{3406--3416}.
\newblock


\bibitem[Gottweis et~al\mbox{.}(2026)]%
        {gottweis2026accelerating}
\bibfield{author}{\bibinfo{person}{Juraj Gottweis}, \bibinfo{person}{Wei-Hung
  Weng}, \bibinfo{person}{Alexander Daryin}, \bibinfo{person}{Tao Tu},
  \bibinfo{person}{Petar Sirkovic}, \bibinfo{person}{Artiom Myaskovsky},
  \bibinfo{person}{Grzegorz Glowaty}, \bibinfo{person}{Felix Weissenberger},
  \bibinfo{person}{Alessio Orlandi}, \bibinfo{person}{Dan Popovici},
  {et~al\mbox{.}}} \bibinfo{year}{2026}\natexlab{}.
\newblock \showarticletitle{Accelerating scientific discovery with
  Co-Scientist}.
\newblock \bibinfo{journal}{\emph{Nature}} (\bibinfo{year}{2026}),
  \bibinfo{pages}{1--3}.
\newblock


\bibitem[Guo et~al\mbox{.}(2025)]%
        {guo_deepseek-r1_2025}
\bibfield{author}{\bibinfo{person}{Daya Guo}, \bibinfo{person}{Dejian Yang},
  \bibinfo{person}{Haowei Zhang}, \bibinfo{person}{Junxiao Song},
  \bibinfo{person}{Peiyi Wang}, \bibinfo{person}{Qihao Zhu},
  \bibinfo{person}{Runxin Xu}, \bibinfo{person}{Ruoyu Zhang},
  \bibinfo{person}{Shirong Ma}, \bibinfo{person}{Xiao Bi},
  \bibinfo{person}{Xiaokang Zhang}, \bibinfo{person}{Xingkai Yu},
  \bibinfo{person}{Yu Wu}, \bibinfo{person}{Z.~F. Wu}, \bibinfo{person}{Zhibin
  Gou}, \bibinfo{person}{Zhihong Shao}, \bibinfo{person}{Zhuoshu Li},
  \bibinfo{person}{Ziyi Gao}, \bibinfo{person}{Aixin Liu},
  \bibinfo{person}{Bing Xue}, \bibinfo{person}{Bingxuan Wang},
  \bibinfo{person}{Bochao Wu}, \bibinfo{person}{Bei Feng},
  \bibinfo{person}{Chengda Lu}, \bibinfo{person}{Chenggang Zhao},
  \bibinfo{person}{Chengqi Deng}, \bibinfo{person}{Chong Ruan},
  \bibinfo{person}{Damai Dai}, \bibinfo{person}{Deli Chen},
  \bibinfo{person}{Dongjie Ji}, \bibinfo{person}{Erhang Li},
  \bibinfo{person}{Fangyun Lin}, \bibinfo{person}{Fucong Dai},
  \bibinfo{person}{Fuli Luo}, \bibinfo{person}{Guangbo Hao},
  \bibinfo{person}{Guanting Chen}, \bibinfo{person}{Guowei Li},
  \bibinfo{person}{H. Zhang}, \bibinfo{person}{Hanwei Xu},
  \bibinfo{person}{Honghui Ding}, \bibinfo{person}{Huazuo Gao},
  \bibinfo{person}{Hui Qu}, \bibinfo{person}{Hui Li},
  \bibinfo{person}{Jianzhong Guo}, \bibinfo{person}{Jiashi Li},
  \bibinfo{person}{Jingchang Chen}, \bibinfo{person}{Jingyang Yuan},
  \bibinfo{person}{Jinhao Tu}, \bibinfo{person}{Junjie Qiu},
  \bibinfo{person}{Junlong Li}, \bibinfo{person}{J.~L. Cai},
  \bibinfo{person}{Jiaqi Ni}, \bibinfo{person}{Jian Liang},
  \bibinfo{person}{Jin Chen}, \bibinfo{person}{Kai Dong}, \bibinfo{person}{Kai
  Hu}, \bibinfo{person}{Kaichao You}, \bibinfo{person}{Kaige Gao},
  \bibinfo{person}{Kang Guan}, \bibinfo{person}{Kexin Huang},
  \bibinfo{person}{Kuai Yu}, \bibinfo{person}{Lean Wang},
  \bibinfo{person}{Lecong Zhang}, \bibinfo{person}{Liang Zhao},
  \bibinfo{person}{Litong Wang}, \bibinfo{person}{Liyue Zhang},
  \bibinfo{person}{Lei Xu}, \bibinfo{person}{Leyi Xia},
  \bibinfo{person}{Mingchuan Zhang}, \bibinfo{person}{Minghua Zhang},
  \bibinfo{person}{Minghui Tang}, \bibinfo{person}{Mingxu Zhou},
  \bibinfo{person}{Meng Li}, \bibinfo{person}{Miaojun Wang},
  \bibinfo{person}{Mingming Li}, \bibinfo{person}{Ning Tian},
  \bibinfo{person}{Panpan Huang}, \bibinfo{person}{Peng Zhang},
  \bibinfo{person}{Qiancheng Wang}, \bibinfo{person}{Qinyu Chen},
  \bibinfo{person}{Qiushi Du}, \bibinfo{person}{Ruiqi Ge},
  \bibinfo{person}{Ruisong Zhang}, \bibinfo{person}{Ruizhe Pan},
  \bibinfo{person}{Runji Wang}, \bibinfo{person}{R.~J. Chen},
  \bibinfo{person}{R.~L. Jin}, \bibinfo{person}{Ruyi Chen},
  \bibinfo{person}{Shanghao Lu}, \bibinfo{person}{Shangyan Zhou},
  \bibinfo{person}{Shanhuang Chen}, \bibinfo{person}{Shengfeng Ye},
  \bibinfo{person}{Shiyu Wang}, \bibinfo{person}{Shuiping Yu},
  \bibinfo{person}{Shunfeng Zhou}, \bibinfo{person}{Shuting Pan},
  \bibinfo{person}{S.~S. Li}, \bibinfo{person}{Shuang Zhou},
  \bibinfo{person}{Shaoqing Wu}, \bibinfo{person}{Tao Yun},
  \bibinfo{person}{Tian Pei}, \bibinfo{person}{Tianyu Sun}, \bibinfo{person}{T.
  Wang}, \bibinfo{person}{Wangding Zeng}, \bibinfo{person}{Wen Liu},
  \bibinfo{person}{Wenfeng Liang}, \bibinfo{person}{Wenjun Gao},
  \bibinfo{person}{Wenqin Yu}, \bibinfo{person}{Wentao Zhang},
  \bibinfo{person}{W.~L. Xiao}, \bibinfo{person}{Wei An},
  \bibinfo{person}{Xiaodong Liu}, \bibinfo{person}{Xiaohan Wang},
  \bibinfo{person}{Xiaokang Chen}, \bibinfo{person}{Xiaotao Nie},
  \bibinfo{person}{Xin Cheng}, \bibinfo{person}{Xin Liu}, \bibinfo{person}{Xin
  Xie}, \bibinfo{person}{Xingchao Liu}, \bibinfo{person}{Xinyu Yang},
  \bibinfo{person}{Xinyuan Li}, \bibinfo{person}{Xuecheng Su},
  \bibinfo{person}{Xuheng Lin}, \bibinfo{person}{X.~Q. Li},
  \bibinfo{person}{Xiangyue Jin}, \bibinfo{person}{Xiaojin Shen},
  \bibinfo{person}{Xiaosha Chen}, \bibinfo{person}{Xiaowen Sun},
  \bibinfo{person}{Xiaoxiang Wang}, \bibinfo{person}{Xinnan Song},
  \bibinfo{person}{Xinyi Zhou}, \bibinfo{person}{Xianzu Wang},
  \bibinfo{person}{Xinxia Shan}, \bibinfo{person}{Y.~K. Li},
  \bibinfo{person}{Y.~Q. Wang}, \bibinfo{person}{Y.~X. Wei},
  \bibinfo{person}{Yang Zhang}, \bibinfo{person}{Yanhong Xu},
  \bibinfo{person}{Yao Li}, \bibinfo{person}{Yao Zhao},
  \bibinfo{person}{Yaofeng Sun}, \bibinfo{person}{Yaohui Wang},
  \bibinfo{person}{Yi Yu}, \bibinfo{person}{Yichao Zhang},
  \bibinfo{person}{Yifan Shi}, \bibinfo{person}{Yiliang Xiong},
  \bibinfo{person}{Ying He}, \bibinfo{person}{Yishi Piao},
  \bibinfo{person}{Yisong Wang}, \bibinfo{person}{Yixuan Tan},
  \bibinfo{person}{Yiyang Ma}, \bibinfo{person}{Yiyuan Liu},
  \bibinfo{person}{Yongqiang Guo}, \bibinfo{person}{Yuan Ou},
  \bibinfo{person}{Yuduan Wang}, \bibinfo{person}{Yue Gong},
  \bibinfo{person}{Yuheng Zou}, \bibinfo{person}{Yujia He},
  \bibinfo{person}{Yunfan Xiong}, \bibinfo{person}{Yuxiang Luo},
  \bibinfo{person}{Yuxiang You}, \bibinfo{person}{Yuxuan Liu},
  \bibinfo{person}{Yuyang Zhou}, \bibinfo{person}{Y.~X. Zhu},
  \bibinfo{person}{Yanping Huang}, \bibinfo{person}{Yaohui Li},
  \bibinfo{person}{Yi Zheng}, \bibinfo{person}{Yuchen Zhu},
  \bibinfo{person}{Yunxian Ma}, \bibinfo{person}{Ying Tang},
  \bibinfo{person}{Yukun Zha}, \bibinfo{person}{Yuting Yan},
  \bibinfo{person}{Z.~Z. Ren}, \bibinfo{person}{Zehui Ren},
  \bibinfo{person}{Zhangli Sha}, \bibinfo{person}{Zhe Fu},
  \bibinfo{person}{Zhean Xu}, \bibinfo{person}{Zhenda Xie},
  \bibinfo{person}{Zhengyan Zhang}, \bibinfo{person}{Zhewen Hao},
  \bibinfo{person}{Zhicheng Ma}, \bibinfo{person}{Zhigang Yan},
  \bibinfo{person}{Zhiyu Wu}, \bibinfo{person}{Zihui Gu},
  \bibinfo{person}{Zijia Zhu}, \bibinfo{person}{Zijun Liu},
  \bibinfo{person}{Zilin Li}, \bibinfo{person}{Ziwei Xie},
  \bibinfo{person}{Ziyang Song}, \bibinfo{person}{Zizheng Pan},
  \bibinfo{person}{Zhen Huang}, \bibinfo{person}{Zhipeng Xu},
  \bibinfo{person}{Zhongyu Zhang}, {and} \bibinfo{person}{Zhen Zhang}.}
  \bibinfo{year}{2025}\natexlab{}.
\newblock \showarticletitle{{DeepSeek}-{R1} incentivizes reasoning in {LLMs}
  through reinforcement learning}.
\newblock \bibinfo{journal}{\emph{Nature}} \bibinfo{volume}{645},
  \bibinfo{number}{8081} (\bibinfo{date}{Sept.} \bibinfo{year}{2025}),
  \bibinfo{pages}{633--638}.
\newblock
\showISSN{0028-0836, 1476-4687}
\href{https://doi.org/10.1038/s41586-025-09422-z}{doi:\nolinkurl{10.1038/s41586-025-09422-z}}


\bibitem[Huang et~al\mbox{.}(2026)]%
        {huang2026wideseek}
\bibfield{author}{\bibinfo{person}{Ziyang Huang}, \bibinfo{person}{Haolin Ren},
  \bibinfo{person}{Xiaowei Yuan}, \bibinfo{person}{Jiawei Wang},
  \bibinfo{person}{Zhongtao Jiang}, \bibinfo{person}{Kun Xu},
  \bibinfo{person}{Shizhu He}, \bibinfo{person}{Jun Zhao}, {and}
  \bibinfo{person}{Kang Liu}.} \bibinfo{year}{2026}\natexlab{}.
\newblock \showarticletitle{WideSeek: Advancing Wide Research via Multi-Agent
  Scaling}.
\newblock \bibinfo{journal}{\emph{arXiv preprint arXiv:2602.02636}}
  (\bibinfo{year}{2026}).
\newblock


\bibitem[Jo and Trummer(2024)]%
        {jo_thalamusdb_2024}
\bibfield{author}{\bibinfo{person}{Saehan Jo} {and} \bibinfo{person}{Immanuel
  Trummer}.} \bibinfo{year}{2024}\natexlab{}.
\newblock \showarticletitle{{ThalamusDB}: {Approximate} {Query} {Processing} on
  {Multi}-{Modal} {Data}}.
\newblock \bibinfo{journal}{\emph{Proceedings of the ACM on Management of
  Data}} \bibinfo{volume}{2}, \bibinfo{number}{3} (\bibinfo{date}{May}
  \bibinfo{year}{2024}), \bibinfo{pages}{1--26}.
\newblock
\showISSN{2836-6573}
\href{https://doi.org/10.1145/3654989}{doi:\nolinkurl{10.1145/3654989}}


\bibitem[Kimelfeld and Sagiv(2006)]%
        {kimelfeld_finding_2006}
\bibfield{author}{\bibinfo{person}{Benny Kimelfeld} {and}
  \bibinfo{person}{Yehoshua Sagiv}.} \bibinfo{year}{2006}\natexlab{}.
\newblock \showarticletitle{Finding and approximating top-k answers in keyword
  proximity search}. In \bibinfo{booktitle}{\emph{{PODS}}}.
  \bibinfo{pages}{173--182}.
\newblock


\bibitem[Lan et~al\mbox{.}(2026)]%
        {lan2026table}
\bibfield{author}{\bibinfo{person}{Tian Lan}, \bibinfo{person}{Felix Henry},
  \bibinfo{person}{Bin Zhu}, \bibinfo{person}{Qianghuai Jia},
  \bibinfo{person}{Junyang Ren}, \bibinfo{person}{Qihang Pu},
  \bibinfo{person}{Haijun Li}, \bibinfo{person}{Longyue Wang},
  \bibinfo{person}{Zhao Xu}, {and} \bibinfo{person}{Weihua Luo}.}
  \bibinfo{year}{2026}\natexlab{}.
\newblock \showarticletitle{Table-as-Search: Formulate Long-Horizon Agentic
  Information Seeking as Table Completion}.
\newblock \bibinfo{journal}{\emph{arXiv preprint arXiv:2602.06724}}
  (\bibinfo{year}{2026}).
\newblock


\bibitem[Lan et~al\mbox{.}(2025)]%
        {lan2025deepwidesearch}
\bibfield{author}{\bibinfo{person}{Tian Lan}, \bibinfo{person}{Bin Zhu},
  \bibinfo{person}{Qianghuai Jia}, \bibinfo{person}{Junyang Ren},
  \bibinfo{person}{Haijun Li}, \bibinfo{person}{Longyue Wang},
  \bibinfo{person}{Zhao Xu}, \bibinfo{person}{Weihua Luo}, {and}
  \bibinfo{person}{Kaifu Zhang}.} \bibinfo{year}{2025}\natexlab{}.
\newblock \showarticletitle{Deepwidesearch: Benchmarking depth and width in
  agentic information seeking}.
\newblock \bibinfo{journal}{\emph{arXiv preprint arXiv:2510.20168}}
  (\bibinfo{year}{2025}).
\newblock


\bibitem[Lao et~al\mbox{.}(2025)]%
        {lao_sembench_2025}
\bibfield{author}{\bibinfo{person}{Jiale Lao}, \bibinfo{person}{Andreas
  Zimmerer}, \bibinfo{person}{Olga Ovcharenko}, \bibinfo{person}{Tianji Cong},
  \bibinfo{person}{Matthew Russo}, \bibinfo{person}{Gerardo Vitagliano},
  \bibinfo{person}{Michael Cochez}, \bibinfo{person}{Fatma Özcan},
  \bibinfo{person}{Gautam Gupta}, \bibinfo{person}{Thibaud Hottelier},
  \bibinfo{person}{H.~V. Jagadish}, \bibinfo{person}{Kris Kissel},
  \bibinfo{person}{Sebastian Schelter}, \bibinfo{person}{Andreas Kipf}, {and}
  \bibinfo{person}{Immanuel Trummer}.} \bibinfo{year}{2025}\natexlab{}.
\newblock \bibinfo{title}{{SemBench}: {A} {Benchmark} for {Semantic} {Query}
  {Processing} {Engines}}.
\newblock
\href{https://doi.org/10.48550/arXiv.2511.01716}{doi:\nolinkurl{10.48550/arXiv.2511.01716}}
\newblock
\shownote{arXiv:2511.01716}.


\bibitem[Li et~al\mbox{.}(2005)]%
        {li_ranksql_2005}
\bibfield{author}{\bibinfo{person}{Chengkai Li}, \bibinfo{person}{Kevin
  Chen-Chuan Chang}, \bibinfo{person}{Ihab~F. Ilyas}, {and}
  \bibinfo{person}{Sumin Song}.} \bibinfo{year}{2005}\natexlab{}.
\newblock \showarticletitle{{RankSQL}: {Query} {Algebra} and {Optimization} for
  {Relational} {Top}-k {Queries}}. In \bibinfo{booktitle}{\emph{{SIGMOD}}}.
  \bibinfo{pages}{131--142}.
\newblock


\bibitem[Li(2011)]%
        {li_learning_2011}
\bibfield{author}{\bibinfo{person}{Hang Li}.} \bibinfo{year}{2011}\natexlab{}.
\newblock \bibinfo{booktitle}{\emph{Learning to {Rank} for {Information}
  {Retrieval} and {Natural} {Language} {Processing}}}.
\newblock \bibinfo{publisher}{Morgan Claypool}.
\newblock


\bibitem[Li et~al\mbox{.}(2026)]%
        {li2026multi}
\bibfield{author}{\bibinfo{person}{Jisen Li}, \bibinfo{person}{Bingxuan Li},
  \bibinfo{person}{Nanyi Jiang}, \bibinfo{person}{Xuying Ning},
  \bibinfo{person}{Xiyao Wang}, \bibinfo{person}{Yifan Shen},
  \bibinfo{person}{Heng Wang}, \bibinfo{person}{Yuqing Jian},
  \bibinfo{person}{Xiaoxia Wu}, \bibinfo{person}{Ben Athiwaratkun},
  {et~al\mbox{.}}} \bibinfo{year}{2026}\natexlab{}.
\newblock \showarticletitle{Multi-Turn Agentic Scientific Literature Search via
  Workflow Induction}.
\newblock \bibinfo{journal}{\emph{arXiv preprint arXiv:2607.00597}}
  (\bibinfo{year}{2026}).
\newblock


\bibitem[Liu et~al\mbox{.}(2025b)]%
        {liu2025palimpzest}
\bibfield{author}{\bibinfo{person}{Chunwei Liu}, \bibinfo{person}{Matthew
  Russo}, \bibinfo{person}{Michael Cafarella}, \bibinfo{person}{Lei Cao},
  \bibinfo{person}{Peter~Baile Chen}, \bibinfo{person}{Zui Chen},
  \bibinfo{person}{Michael Franklin}, \bibinfo{person}{Tim Kraska},
  \bibinfo{person}{Samuel Madden}, \bibinfo{person}{Rana Shahout},
  {et~al\mbox{.}}} \bibinfo{year}{2025}\natexlab{b}.
\newblock \showarticletitle{Palimpzest: Optimizing ai-powered analytics with
  declarative query processing}. In \bibinfo{booktitle}{\emph{Proceedings of
  the Conference on Innovative Database Research (CIDR)}}. \bibinfo{pages}{2}.
\newblock


\bibitem[Liu et~al\mbox{.}(2025a)]%
        {liu2025supporting}
\bibfield{author}{\bibinfo{person}{Shu Liu}, \bibinfo{person}{Soujanya
  Ponnapalli}, \bibinfo{person}{Shreya Shankar}, \bibinfo{person}{Sepanta
  Zeighami}, \bibinfo{person}{Alan Zhu}, \bibinfo{person}{Shubham Agarwal},
  \bibinfo{person}{Ruiqi Chen}, \bibinfo{person}{Samion Suwito},
  \bibinfo{person}{Shuo Yuan}, \bibinfo{person}{Ion Stoica}, {et~al\mbox{.}}}
  \bibinfo{year}{2025}\natexlab{a}.
\newblock \showarticletitle{Supporting our ai overlords: Redesigning data
  systems to be agent-first}.
\newblock \bibinfo{journal}{\emph{arXiv preprint arXiv:2509.00997}}
  (\bibinfo{year}{2025}).
\newblock


\bibitem[Lu et~al\mbox{.}(2026)]%
        {lu2026depth}
\bibfield{author}{\bibinfo{person}{Duo Lu}, \bibinfo{person}{Helena Caminal},
  \bibinfo{person}{Manos Chatzakis}, \bibinfo{person}{Yannis Papakonstantinou},
  \bibinfo{person}{Yannis Chronis}, \bibinfo{person}{Vaibhav Jain}, {and}
  \bibinfo{person}{Fatma {\"O}zcan}.} \bibinfo{year}{2026}\natexlab{}.
\newblock \showarticletitle{An in-depth study of filter-agnostic vector search
  on a postgresql database system:[experiments \& analysis]}.
\newblock \bibinfo{journal}{\emph{Proceedings of the ACM on Management of
  Data}} \bibinfo{volume}{4}, \bibinfo{number}{3 (SIGMOD}
  (\bibinfo{year}{2026}), \bibinfo{pages}{1--26}.
\newblock


\bibitem[Mu and Viswanath(2018)]%
        {mu_allbutthetop_2018}
\bibfield{author}{\bibinfo{person}{Jiaqi Mu} {and} \bibinfo{person}{Pramod
  Viswanath}.} \bibinfo{year}{2018}\natexlab{}.
\newblock \showarticletitle{All-but-the-Top: Simple and Effective
  Postprocessing for Word Representations}. In
  \bibinfo{booktitle}{\emph{International {Conference} on {Learning}
  {Representations} ({ICLR})}}.
\newblock


\bibitem[Nogueira and Cho(2020)]%
        {nogueira_passage_2020}
\bibfield{author}{\bibinfo{person}{Rodrigo Nogueira} {and}
  \bibinfo{person}{Kyunghyun Cho}.} \bibinfo{year}{2020}\natexlab{}.
\newblock \bibinfo{title}{Passage {Re}-ranking with {BERT}}.
\newblock
\href{https://doi.org/10.48550/arXiv.1901.04085}{doi:\nolinkurl{10.48550/arXiv.1901.04085}}
\newblock
\shownote{arXiv:1901.04085}.


\bibitem[Patel et~al\mbox{.}(2024a)]%
        {patel_semantic_2024}
\bibfield{author}{\bibinfo{person}{Liana Patel}, \bibinfo{person}{Siddharth
  Jha}, \bibinfo{person}{Parth Asawa}, \bibinfo{person}{Melissa Pan},
  \bibinfo{person}{Carlos Guestrin}, {and} \bibinfo{person}{Matei Zaharia}.}
  \bibinfo{year}{2024}\natexlab{a}.
\newblock \showarticletitle{Semantic {Operators}: {A} {Declarative} {Model} for
  {Rich}, {AI}-based {Analytics} {Over} {Text} {Data}}.
\newblock
\urldef\tempurl%
\url{https://api.semanticscholar.org/CorpusID:271218837}
\showURL{%
\tempurl}


\bibitem[Patel et~al\mbox{.}(2025)]%
        {patel_semantic_2025}
\bibfield{author}{\bibinfo{person}{Liana Patel}, \bibinfo{person}{Siddharth
  Jha}, \bibinfo{person}{Melissa Pan}, \bibinfo{person}{Harshit Gupta},
  \bibinfo{person}{Parth Asawa}, \bibinfo{person}{Carlos Guestrin}, {and}
  \bibinfo{person}{Matei Zaharia}.} \bibinfo{year}{2025}\natexlab{}.
\newblock \showarticletitle{Semantic {Operators} and {Their} {Optimization}:
  {Enabling} {LLM}-{Based} {Data} {Processing} with {Accuracy} {Guarantees} in
  {LOTUS}}.
\newblock \bibinfo{journal}{\emph{Proceedings of the VLDB Endowment}}
  \bibinfo{volume}{18}, \bibinfo{number}{11} (\bibinfo{year}{2025}),
  \bibinfo{pages}{4171--4184}.
\newblock


\bibitem[Patel et~al\mbox{.}(2024b)]%
        {patel2024acorn}
\bibfield{author}{\bibinfo{person}{Liana Patel}, \bibinfo{person}{Peter Kraft},
  \bibinfo{person}{Carlos Guestrin}, {and} \bibinfo{person}{Matei Zaharia}.}
  \bibinfo{year}{2024}\natexlab{b}.
\newblock \showarticletitle{Acorn: Performant and predicate-agnostic search
  over vector embeddings and structured data}.
\newblock \bibinfo{journal}{\emph{Proceedings of the ACM on Management of
  Data}} \bibinfo{volume}{2}, \bibinfo{number}{3} (\bibinfo{year}{2024}),
  \bibinfo{pages}{1--27}.
\newblock


\bibitem[Peng et~al\mbox{.}(2024)]%
        {peng_graph_2024}
\bibfield{author}{\bibinfo{person}{Boci Peng}, \bibinfo{person}{Yun Zhu},
  \bibinfo{person}{Yongchao Liu}, \bibinfo{person}{Xiaohe Bo},
  \bibinfo{person}{Haizhou Shi}, \bibinfo{person}{Chuntao Hong},
  \bibinfo{person}{Yan Zhang}, {and} \bibinfo{person}{Siliang Tang}.}
  \bibinfo{year}{2024}\natexlab{}.
\newblock \bibinfo{title}{Graph {Retrieval}-{Augmented} {Generation}: {A}
  {Survey}}.
\newblock
\href{https://doi.org/10.48550/arXiv.2408.08921}{doi:\nolinkurl{10.48550/arXiv.2408.08921}}
\newblock
\shownote{arXiv:2408.08921}.


\bibitem[{Pinecone Systems}(2026)]%
        {pinecone}
\bibfield{author}{\bibinfo{person}{{Pinecone Systems}}.}
  \bibinfo{year}{2026}\natexlab{}.
\newblock \bibinfo{title}{Pinecone: The Vector Database for AI Applications}.
\newblock \bibinfo{howpublished}{\url{https://www.pinecone.io/}}.
\newblock
\newblock
\shownote{Accessed: 2026-04-15}.


\bibitem[Rogers and Hahn(2010)]%
        {rogers_extended-connectivity_2010}
\bibfield{author}{\bibinfo{person}{David Rogers} {and} \bibinfo{person}{Mathew
  Hahn}.} \bibinfo{year}{2010}\natexlab{}.
\newblock \showarticletitle{Extended-{Connectivity} {Fingerprints}}.
\newblock \bibinfo{journal}{\emph{Journal of Chemical Information and
  Modeling}} \bibinfo{volume}{50}, \bibinfo{number}{5} (\bibinfo{date}{May}
  \bibinfo{year}{2010}), \bibinfo{pages}{742--754}.
\newblock
\showISSN{1549-9596, 1549-960X}
\href{https://doi.org/10.1021/ci100050t}{doi:\nolinkurl{10.1021/ci100050t}}


\bibitem[Russo et~al\mbox{.}(2025)]%
        {russo_abacus_2025}
\bibfield{author}{\bibinfo{person}{Matthew Russo}, \bibinfo{person}{Sivaprasad
  Sudhir}, \bibinfo{person}{Gerardo Vitagliano}, \bibinfo{person}{Chunwei Liu},
  \bibinfo{person}{Tim Kraska}, \bibinfo{person}{Samuel Madden}, {and}
  \bibinfo{person}{Michael Cafarella}.} \bibinfo{year}{2025}\natexlab{}.
\newblock \showarticletitle{Abacus: {A} {Cost}-{Based} {Optimizer} for
  {Semantic} {Operator} {Systems}}.
\newblock \bibinfo{journal}{\emph{arXiv preprint arXiv:2505.14661}}
  (\bibinfo{year}{2025}).
\newblock


\bibitem[Saleh et~al\mbox{.}(2024)]%
        {saleh_sg-rag_2024}
\bibfield{author}{\bibinfo{person}{Ahmmad~OM Saleh}, \bibinfo{person}{Gokhan
  Tur}, {and} \bibinfo{person}{Yucel Saygin}.} \bibinfo{year}{2024}\natexlab{}.
\newblock \showarticletitle{{SG}-{RAG}: {Multi}-hop question answering with
  large language models through knowledge graphs}. In
  \bibinfo{booktitle}{\emph{Proceedings of the 7th {International} {Conference}
  on {Natural} {Language} and {Speech} {Processing} ({ICNLSP} 2024)}}.
  \bibinfo{pages}{439--448}.
\newblock


\bibitem[Shankar et~al\mbox{.}(2025)]%
        {shankar2025docetl}
\bibfield{author}{\bibinfo{person}{Shreya Shankar}, \bibinfo{person}{Tristan
  Chambers}, \bibinfo{person}{Tarak Shah}, \bibinfo{person}{Aditya~G
  Parameswaran}, {and} \bibinfo{person}{Eugene Wu}.}
  \bibinfo{year}{2025}\natexlab{}.
\newblock \showarticletitle{DocETL: Agentic Query Rewriting and Evaluation for
  Complex Document Processing}.
\newblock \bibinfo{journal}{\emph{Proceedings of the VLDB Endowment}}
  \bibinfo{volume}{18}, \bibinfo{number}{9} (\bibinfo{year}{2025}),
  \bibinfo{pages}{3035--3048}.
\newblock


\bibitem[Su et~al\mbox{.}(2021)]%
        {su_whitening_2021}
\bibfield{author}{\bibinfo{person}{Jianlin Su}, \bibinfo{person}{Jiarun Cao},
  \bibinfo{person}{Weijie Liu}, {and} \bibinfo{person}{Yangyiwen Ou}.}
  \bibinfo{year}{2021}\natexlab{}.
\newblock \showarticletitle{Whitening {Sentence} {Representations} for {Better}
  {Semantics} and {Faster} {Retrieval}}.
\newblock \bibinfo{journal}{\emph{arXiv preprint arXiv:2103.15316}}
  (\bibinfo{year}{2021}).
\newblock
\href{https://doi.org/10.48550/arXiv.2103.15316}{doi:\nolinkurl{10.48550/arXiv.2103.15316}}
\newblock
\shownote{arXiv:2103.15316}.


\bibitem[Trummer(2025)]%
        {trummer_implementing_2025}
\bibfield{author}{\bibinfo{person}{Immanuel Trummer}.}
  \bibinfo{year}{2025}\natexlab{}.
\newblock \bibinfo{title}{Implementing {Semantic} {Join} {Operators}
  {Efficiently}}.
\newblock
\href{https://doi.org/10.48550/arXiv.2510.08489}{doi:\nolinkurl{10.48550/arXiv.2510.08489}}
\newblock
\shownote{arXiv:2510.08489}.


\bibitem[Tziavelis et~al\mbox{.}(2024)]%
        {tziavelis_ranked_2024}
\bibfield{author}{\bibinfo{person}{Nikolaos Tziavelis},
  \bibinfo{person}{Wolfgang Gatterbauer}, {and} \bibinfo{person}{Mirek
  Riedewald}.} \bibinfo{year}{2024}\natexlab{}.
\newblock \showarticletitle{Ranked {Enumeration} for {Database} {Queries}}.
\newblock \bibinfo{journal}{\emph{ACM SIGMOD Record}} \bibinfo{volume}{53},
  \bibinfo{number}{3} (\bibinfo{date}{Nov.} \bibinfo{year}{2024}),
  \bibinfo{pages}{6--19}.
\newblock
\showISSN{0163-5808}
\href{https://doi.org/10.1145/3703922.3703924}{doi:\nolinkurl{10.1145/3703922.3703924}}


\bibitem[Vershynin(2018)]%
        {vershynin2018high}
\bibfield{author}{\bibinfo{person}{Roman Vershynin}.}
  \bibinfo{year}{2018}\natexlab{}.
\newblock \bibinfo{booktitle}{\emph{High-dimensional probability: An
  introduction with applications in data science}}. Vol.~\bibinfo{volume}{47}.
\newblock \bibinfo{publisher}{Cambridge university press}.
\newblock


\bibitem[Wang et~al\mbox{.}(2021)]%
        {wang_milvus_2021}
\bibfield{author}{\bibinfo{person}{Jianguo Wang}, \bibinfo{person}{Xiaomeng
  Yi}, \bibinfo{person}{Rentong Guo}, \bibinfo{person}{Hai Jin},
  \bibinfo{person}{Peng Xu}, \bibinfo{person}{Shengjun Li},
  \bibinfo{person}{Xiangyu Wang}, \bibinfo{person}{Xiangzhou Guo},
  \bibinfo{person}{Chengming Li}, \bibinfo{person}{Xiaohai Xu},
  \bibinfo{person}{Kun Yu}, \bibinfo{person}{Yuxing Yuan},
  \bibinfo{person}{Yinghao Zou}, \bibinfo{person}{Jiquan Long},
  \bibinfo{person}{Yudong Cai}, \bibinfo{person}{Zhenxiang Li},
  \bibinfo{person}{Zhifeng Zhang}, \bibinfo{person}{Yihua Mo},
  \bibinfo{person}{Jun Gu}, \bibinfo{person}{Ruiyi Jiang}, \bibinfo{person}{Yi
  Wei}, {and} \bibinfo{person}{Charles Xie}.} \bibinfo{year}{2021}\natexlab{}.
\newblock \showarticletitle{Milvus: {A} {Purpose}-{Built} {Vector} {Data}
  {Management} {System}}. In \bibinfo{booktitle}{\emph{Proceedings of the 2021
  {International} {Conference} on {Management} of {Data}}}.
  \bibinfo{publisher}{ACM}, \bibinfo{address}{Virtual Event China},
  \bibinfo{pages}{2614--2627}.
\newblock
\showISBNx{9781450383431}
\href{https://doi.org/10.1145/3448016.3457550}{doi:\nolinkurl{10.1145/3448016.3457550}}


\bibitem[Wang et~al\mbox{.}(2026)]%
        {wang2026fdabench}
\bibfield{author}{\bibinfo{person}{Ziting Wang}, \bibinfo{person}{Shize Zhang},
  \bibinfo{person}{Haitao Yuan}, \bibinfo{person}{Jinwei Zhu},
  \bibinfo{person}{Wei Dong}, {and} \bibinfo{person}{Gao Cong}.}
  \bibinfo{year}{2026}\natexlab{}.
\newblock \showarticletitle{{FDABench}: A Benchmark for Data Agents on
  Analytical Queries over Heterogeneous Data}. In
  \bibinfo{booktitle}{\emph{Proceedings of the 32nd ACM SIGKDD Conference on
  Knowledge Discovery and Data Mining V.2}}. \bibinfo{publisher}{ACM},
  \bibinfo{address}{New York, NY, USA}, \bibinfo{pages}{9985--9996}.
\newblock
\href{https://doi.org/10.1145/3770855.3817454}{doi:\nolinkurl{10.1145/3770855.3817454}}


\bibitem[{Weaviate}(2026)]%
        {weaviate}
\bibfield{author}{\bibinfo{person}{{Weaviate}}.}
  \bibinfo{year}{2026}\natexlab{}.
\newblock \bibinfo{title}{Weaviate: An Open-Source Cloud-Native Vector
  Database}.
\newblock \bibinfo{howpublished}{\url{https://github.com/weaviate/weaviate}}.
\newblock
\newblock
\shownote{Accessed: 2026-04-15}.


\bibitem[Wei et~al\mbox{.}(2022)]%
        {wei_chain--thought_2022}
\bibfield{author}{\bibinfo{person}{Jason Wei}, \bibinfo{person}{Xuezhi Wang},
  \bibinfo{person}{Dale Schuurmans}, \bibinfo{person}{Maarten Bosma},
  \bibinfo{person}{Fei Xia}, \bibinfo{person}{Ed Chi}, \bibinfo{person}{Quoc~V
  Le}, \bibinfo{person}{Denny Zhou}, {and} \bibinfo{person}{{others}}.}
  \bibinfo{year}{2022}\natexlab{}.
\newblock \showarticletitle{Chain-of-thought prompting elicits reasoning in
  large language models}.
\newblock \bibinfo{journal}{\emph{Advances in neural information processing
  systems}}  \bibinfo{volume}{35} (\bibinfo{year}{2022}),
  \bibinfo{pages}{24824--24837}.
\newblock


\bibitem[Weininger(1988)]%
        {weininger1988smiles}
\bibfield{author}{\bibinfo{person}{David Weininger}.}
  \bibinfo{year}{1988}\natexlab{}.
\newblock \showarticletitle{SMILES, a chemical language and information system.
  1. Introduction to methodology and encoding rules}.
\newblock \bibinfo{journal}{\emph{Journal of chemical information and computer
  sciences}} \bibinfo{volume}{28}, \bibinfo{number}{1} (\bibinfo{year}{1988}),
  \bibinfo{pages}{31--36}.
\newblock


\bibitem[Wong et~al\mbox{.}(2025)]%
        {widesearch}
\bibfield{author}{\bibinfo{person}{Ryan Wong}, \bibinfo{person}{Jiawei Wang},
  \bibinfo{person}{Junjie Zhao}, \bibinfo{person}{Li Chen},
  \bibinfo{person}{Yan Gao}, \bibinfo{person}{Long Zhang},
  \bibinfo{person}{Xuan Zhou}, \bibinfo{person}{Zuo Wang}, \bibinfo{person}{Kai
  Xiang}, \bibinfo{person}{Ge Zhang}, \bibinfo{person}{Wenhao Huang},
  \bibinfo{person}{Yang Wang}, {and} \bibinfo{person}{Ke Wang}.}
  \bibinfo{year}{2025}\natexlab{}.
\newblock \bibinfo{title}{WideSearch: Benchmarking Agentic Broad Info-Seeking}.
\newblock
\showeprint[arxiv]{2508.07999}~[cs.CL]
\urldef\tempurl%
\url{https://arxiv.org/abs/2508.07999}
\showURL{%
\tempurl}


\bibitem[Wu et~al\mbox{.}(2025)]%
        {wu2025mmqa}
\bibfield{author}{\bibinfo{person}{Jian Wu}, \bibinfo{person}{Linyi Yang},
  \bibinfo{person}{Dongyuan Li}, \bibinfo{person}{Yuliang Ji},
  \bibinfo{person}{Manabu Okumura}, {and} \bibinfo{person}{Yue Zhang}.}
  \bibinfo{year}{2025}\natexlab{}.
\newblock \showarticletitle{MMQA: Evaluating LLMs with multi-table multi-hop
  complex questions}. In \bibinfo{booktitle}{\emph{International Conference on
  Learning Representations}}, Vol.~\bibinfo{volume}{2025}.
  \bibinfo{pages}{48626--48643}.
\newblock


\bibitem[Wu et~al\mbox{.}(2024)]%
        {wu2024stark}
\bibfield{author}{\bibinfo{person}{Shirley Wu}, \bibinfo{person}{Shiyu Zhao},
  \bibinfo{person}{Michihiro Yasunaga}, \bibinfo{person}{Kexin Huang},
  \bibinfo{person}{Kaidi Cao}, \bibinfo{person}{Qian Huang},
  \bibinfo{person}{Vassilis~N. Ioannidis}, \bibinfo{person}{Karthik Subbian},
  \bibinfo{person}{James Zou}, {and} \bibinfo{person}{Jure Leskovec}.}
  \bibinfo{year}{2024}\natexlab{}.
\newblock \showarticletitle{{STaRK}: Benchmarking {LLM} Retrieval on Textual
  and Relational Knowledge Bases}. In \bibinfo{booktitle}{\emph{Advances in
  Neural Information Processing Systems}},
  \bibfield{editor}{\bibinfo{person}{A.~Globerson},
  \bibinfo{person}{L.~Mackey}, \bibinfo{person}{D.~Belgrave},
  \bibinfo{person}{A.~Fan}, \bibinfo{person}{U.~Paquet},
  \bibinfo{person}{J.~Tomczak}, {and} \bibinfo{person}{C.~Zhang}} (Eds.),
  Vol.~\bibinfo{volume}{37}. \bibinfo{publisher}{Curran Associates, Inc.},
  \bibinfo{pages}{127129--127153}.
\newblock
\href{https://doi.org/10.52202/079017-4037}{doi:\nolinkurl{10.52202/079017-4037}}


\bibitem[Xu et~al\mbox{.}(2026)]%
        {xu2026wideseekr1exploringwidthscaling}
\bibfield{author}{\bibinfo{person}{Zelai Xu}, \bibinfo{person}{Zhexuan Xu},
  \bibinfo{person}{Ruize Zhang}, \bibinfo{person}{Chunyang Zhu},
  \bibinfo{person}{Shi Yu}, \bibinfo{person}{Weilin Liu},
  \bibinfo{person}{Quanlu Zhang}, \bibinfo{person}{Wenbo Ding},
  \bibinfo{person}{Chao Yu}, {and} \bibinfo{person}{Yu Wang}.}
  \bibinfo{year}{2026}\natexlab{}.
\newblock \bibinfo{title}{WideSeek-R1: Exploring Width Scaling for Broad
  Information Seeking via Multi-Agent Reinforcement Learning}.
\newblock
\showeprint[arxiv]{2602.04634}~[cs.AI]
\urldef\tempurl%
\url{https://arxiv.org/abs/2602.04634}
\showURL{%
\tempurl}


\bibitem[Yan et~al\mbox{.}(2013)]%
        {yan_actively_2013}
\bibfield{author}{\bibinfo{person}{Zhepeng Yan}, \bibinfo{person}{Nan Zheng},
  \bibinfo{person}{Zachary Ives}, \bibinfo{person}{Partha Talukdar}, {and}
  \bibinfo{person}{Cong Yu}.} \bibinfo{year}{2013}\natexlab{}.
\newblock \showarticletitle{Actively {Soliciting} {Feedback} for {Query}
  {Answers} in {Keyword} {Search}-{Based} {Data} {Integration}}.
\newblock \bibinfo{journal}{\emph{PVLDB}} (\bibinfo{year}{2013}).
\newblock


\bibitem[Yang et~al\mbox{.}(2023)]%
        {yang_gpt4tools_2023}
\bibfield{author}{\bibinfo{person}{Rui Yang}, \bibinfo{person}{Lin Song},
  \bibinfo{person}{Yanwei Li}, \bibinfo{person}{Sijie Zhao},
  \bibinfo{person}{Yixiao Ge}, \bibinfo{person}{Xiu Li}, {and}
  \bibinfo{person}{Ying Shan}.} \bibinfo{year}{2023}\natexlab{}.
\newblock \showarticletitle{Gpt4tools: {Teaching} large language model to use
  tools via self-instruction}.
\newblock \bibinfo{journal}{\emph{Advances in Neural Information Processing
  Systems}}  \bibinfo{volume}{36} (\bibinfo{year}{2023}),
  \bibinfo{pages}{71995--72007}.
\newblock


\bibitem[Yao et~al\mbox{.}(2022)]%
        {yao_react_2022}
\bibfield{author}{\bibinfo{person}{Shunyu Yao}, \bibinfo{person}{Jeffrey Zhao},
  \bibinfo{person}{Dian Yu}, \bibinfo{person}{Nan Du}, \bibinfo{person}{Izhak
  Shafran}, \bibinfo{person}{Karthik Narasimhan}, {and} \bibinfo{person}{Yuan
  Cao}.} \bibinfo{year}{2022}\natexlab{}.
\newblock \showarticletitle{{ReAct}: {Synergizing} {Reasoning} and {Acting} in
  {Language} {Models}}.
\newblock \bibinfo{journal}{\emph{International Conference on Learning
  Representations}}  \bibinfo{volume}{abs/2210.03629} (\bibinfo{year}{2022}).
\newblock


\bibitem[Zeakis et~al\mbox{.}(2025)]%
        {zeakis2025avenger}
\bibfield{author}{\bibinfo{person}{Alexandros Zeakis}, \bibinfo{person}{George
  Papadakis}, \bibinfo{person}{Dimitrios Skoutas}, {and}
  \bibinfo{person}{Manolis Koubarakis}.} \bibinfo{year}{2025}\natexlab{}.
\newblock \showarticletitle{AvengER: Ensembling and Fine-Tuning LLMs for SELECT
  Prompts in Entity Resolution}. In \bibinfo{booktitle}{\emph{European Semantic
  Web Conference}}. Springer, \bibinfo{pages}{301--320}.
\newblock


\bibitem[Zeighami et~al\mbox{.}(2025)]%
        {zeighami2025cut}
\bibfield{author}{\bibinfo{person}{Sepanta Zeighami}, \bibinfo{person}{Shreya
  Shankar}, {and} \bibinfo{person}{Aditya Parameswaran}.}
  \bibinfo{year}{2025}\natexlab{}.
\newblock \showarticletitle{Cut costs, not accuracy: Llm-powered data
  processing with guarantees}.
\newblock \bibinfo{journal}{\emph{Proceedings of the ACM on Management of
  Data}} \bibinfo{volume}{3}, \bibinfo{number}{6} (\bibinfo{year}{2025}),
  \bibinfo{pages}{1--26}.
\newblock


\bibitem[Zeighami et~al\mbox{.}(2026)]%
        {zeighami2025join}
\bibfield{author}{\bibinfo{person}{Sepanta Zeighami}, \bibinfo{person}{Shreya
  Shankar}, {and} \bibinfo{person}{Aditya Parameswaran}.}
  \bibinfo{year}{2026}\natexlab{}.
\newblock \showarticletitle{Featurized-Decomposition Join: Low-Cost Semantic
  Joins with Guarantees}.
\newblock \bibinfo{journal}{\emph{Proc. VLDB Endow.}} \bibinfo{volume}{19},
  \bibinfo{number}{11} (\bibinfo{date}{Sept.} \bibinfo{year}{2026}),
  \bibinfo{pages}{3048–3061}.
\newblock
\showISSN{2150-8097}
\href{https://doi.org/10.14778/3836663.3836672}{doi:\nolinkurl{10.14778/3836663.3836672}}


\bibitem[Zheng et~al\mbox{.}(2025)]%
        {zheng2025deepresearcher}
\bibfield{author}{\bibinfo{person}{Yuxiang Zheng}, \bibinfo{person}{Dayuan Fu},
  \bibinfo{person}{Xiangkun Hu}, \bibinfo{person}{Xiaojie Cai},
  \bibinfo{person}{Lyumanshan Ye}, \bibinfo{person}{Pengrui Lu}, {and}
  \bibinfo{person}{Pengfei Liu}.} \bibinfo{year}{2025}\natexlab{}.
\newblock \showarticletitle{Deepresearcher: Scaling deep research via
  reinforcement learning in real-world environments}. In
  \bibinfo{booktitle}{\emph{Proceedings of the 2025 Conference on Empirical
  Methods in Natural Language Processing}}. \bibinfo{pages}{414--431}.
\newblock


\end{thebibliography}

\appendix

\begingroup
\emergencystretch=1em
\section{Benchmark Dataset}
\label{sec:benchmark}

\subsection{IMDB Benchmark}
\label{sec:bench-imdb}

\subsubsection{Dataset Description}
The IMDB dataset is designed to simulate a movie discovery and recommendation environment. It allows for querying films based on a combination of categorical metadata (e.g., genres, ratings, release years) and rich textual profiles. To set up a scenario facilitating cross-era movie discovery (e.g., finding modern successors to classic cinema), we partition the data into two relational tables, $T_1$ and $T_2$, representing different time periods. Each record encapsulates a comprehensive movie profile, including a synthesized crew description and high-dimensional embeddings for its plot and title.

\subsubsection{Data Source and Construction}
To construct this benchmark, we integrated official metadata with unstructured content:
\begin{itemize}
    \item \textbf{Data Sourcing:} We extracted structured attributes (title, rating, crew identifiers) from the official IMDB non-commercial datasets. To ensure a high-quality corpus, we limited the scope to non-adult feature films released between 1950 and 2020. This was augmented with detailed plot descriptions sourced from the \texttt{jquigl/imdb-genres} collection.
    \item \textbf{Profile Construction:} We synthesized an \texttt{actor\_director} textual field for each movie by joining the crew identifiers with the name basics table, resolving top-billed cast members and directors into human-readable strings. The final consolidated table, consisting of 85,615 cleaned records, was merged on the composite key of movie title and release year.
    \item \textbf{Vectorization / Partition:} We generated 1536-dimensional embeddings for the \texttt{title}, \texttt{plot}, and \texttt{actor\_director} fields using the Gemini-embedding-001 model. Finally, the dataset was split by the median release year (2007) into table \texttt{imdb\_t1} (42,378 records, pre-2007) and table \texttt{imdb\_t2} (43,237 records, 2007--2020) to provide physical targets for relational semantic joins.
\end{itemize}

\begin{figure}[t]
  \centering
  \includegraphics[width=\linewidth]{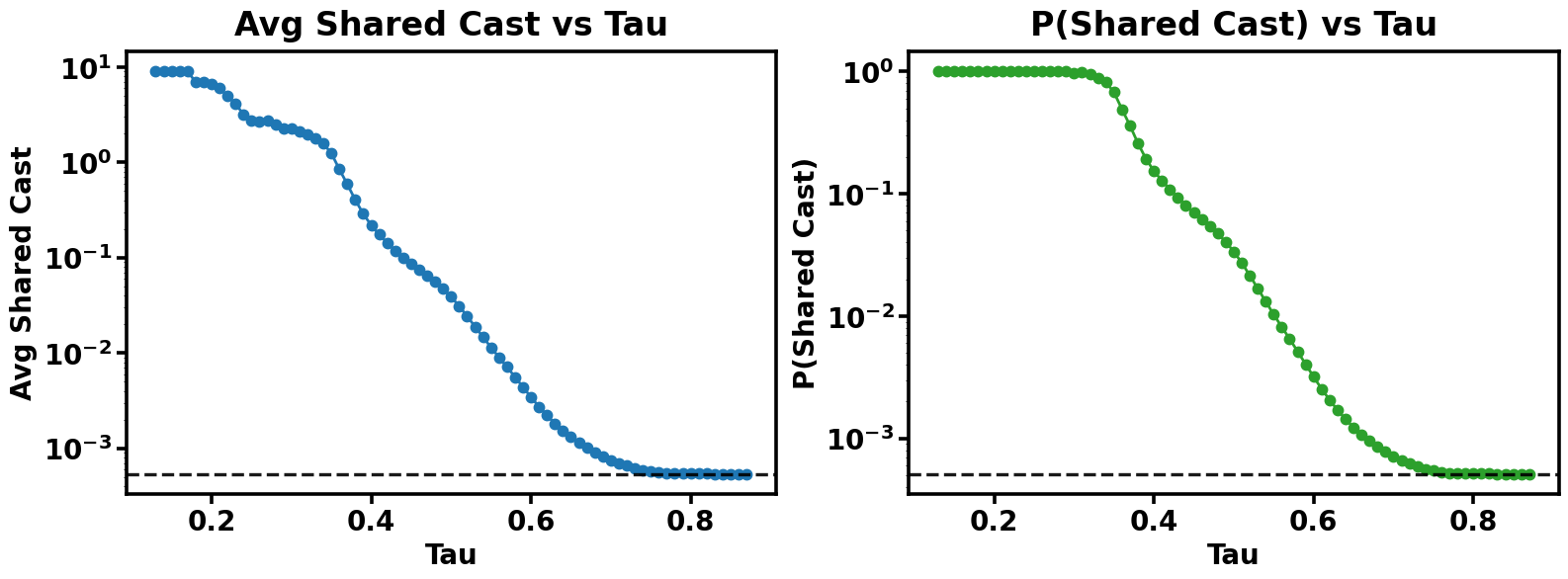}
  \caption{Validation of the semantic join proxy: shared cast metrics versus embedding distance $\tau$. Dashed lines indicate global baselines.}
  \label{fig:shared_cast_vs_tau}
\end{figure}

\subsubsection{Workload Construction}
\label{sec:imdb-workload-construction}

To systematically evaluate the execution strategies, we generate a synthetic workload suite that instantiates the three core dimensions of our multi-step reasoning query model: structured predicates ($\mathcal{P}$), multi-vector scoring ($\mathcal{S}$), and semantic join conditions ($\mathcal{J}$).

\textbf{Predicates and Scoring ($\mathcal{P}$ and $\mathcal{S}$).} 
We instantiate hard logical constraints ($\mathcal{P}$) using the exact relational metadata, such as bounding the release years or applying exact-match filters on categorical genres. The multi-dimensional ranking objective ($\mathcal{S}$) is formulated by aggregating the semantic similarities between user-provided query anchors and the movies' high-dimensional \texttt{plot} and \texttt{title} embeddings. 

\textbf{Semantic Joins ($\mathcal{J}$).} 
To evaluate inter-entity synthesis (workloads W5--W8), we simulate a cross-era discovery scenario that links classic films in \texttt{imdb\_t1} with their modern counterparts in \texttt{imdb\_t2}. This linkage is fundamentally driven by a semantic join over the \texttt{actor\_director} embeddings. 

To validate that the geometric distance ($\tau$) in this specific embedding space serves as a robust proxy for real-world semantic relatedness, we profiled the pairwise distances across the dataset. As illustrated in Figure~\ref{fig:shared_cast_vs_tau}, we measured both the average number of shared cast members (left) and the probability of two movies sharing at least one cast member (right) as a function of the distance threshold $\tau$. Note that the y-axes are scaled logarithmically, and the horizontal dashed lines denote the global expected baselines across the unconstrained Cartesian product. 

The empirical results exhibit a pronounced correlation: at tight distance thresholds (small $\tau$), both the expected density and the probability of shared cast members increase exponentially, vastly outperforming the global average. This observation perfectly aligns with our geometric premise that meaningful semantic affinities manifest strictly as extreme tail events. Consequently, this justifies our methodology of utilizing the \texttt{actor\_director} embedding distance as the underlying relational proxy to construct and evaluate ground-truth semantic join pairs.

\paragraph{\textbf{W1: Find movies by a plot description (Semantic Retrieval).}} 
To establish a performance baseline, we synthesize a query set of 100 seed phrases simulating authentic user search behaviors. We generate these search intents using OpenAI GPT-4o ($\text{temperature}=0.8$) with the following exact prompt: \textit{``Generate exactly [N] short phrases that someone might type when searching for movies. Examples: 'thriller with twist', 'romantic comedy in Paris'. Output one phrase per line, no numbering, no other text.''} Each generated phrase is then vectorized into 1536-dimensional embeddings using the Gemini-embedding-001 model. To ensure uniform coverage across the partitioned corpus, the generator randomly assigns each query to one of the four possible target spaces: either the \texttt{plot} or \texttt{title} embeddings within \texttt{imdb\_t1} or \texttt{imdb\_t2}. Each query is parameterized with $K=20$. This workload represents the simplest form of single-signal semantic retrieval, providing the ground-truth baseline for the system's raw vector search throughput without structural constraints.

\begin{lstlisting}[language=SQL, caption={Example W1 Query}, label={lst:w1-sql}]
-- Searching for "a thriller with unexpected twist and mystery"
SELECT tconst, title, plot
FROM imdb_t1
ORDER BY plot_emb <-> '[0.0336, -0.0163, ...]' 
LIMIT 20;
\end{lstlisting}

\paragraph{\textbf{W2: Find movies by a plot description, within genre/era (Filtered Retrieval).}} 
Building upon W1, workload W2 introduces exact structured predicates ($\mathcal{P}$) to evaluate filtered semantic search. To construct this workload, we keep the same query-text prompt setting as W1, generate 100 intents, and vectorize each into a 1536-dimensional embedding. We then augment each query with 1 to 3 relational filters applied strictly to target-table metadata. The predicate space is bounded by three explicit rules: (1) temporal constraints applying $\ge$ or $\le$ operators on \texttt{year} within $[1980, 2020]$; (2) quality thresholds applying the $\ge$ operator on \texttt{rating} within $[1.0, 10.0]$; and (3) categorical constraints applying the IN operator over random subsets of 12 predefined cinematic \texttt{genre}s. Paired with a uniformly selected target semantic space (either \texttt{plot\_emb} or \texttt{title\_emb} within \texttt{imdb\_t1} or \texttt{imdb\_t2}) and parameterized with $K=20$, this workload benchmarks hybrid execution with relational pruning plus dense vector scans.

\begin{lstlisting}[language=SQL, caption={Example W2 Query}, label={lst:w2-sql}]
-- Q: "a thriller with unexpected twist and mystery" 
-- Filtered by specific categorical genres
SELECT tconst, title, genre, plot
FROM imdb_t1
WHERE genre IN ('Action', 'Fantasy', 'Thriller')
ORDER BY plot_emb <-> '[-0.007, -0.028, ...]' 
LIMIT 20;
\end{lstlisting}

\paragraph{\textbf{W3: Find movies matching a multi-aspect semantic profile (Multi-scoring retrieval).}} 
To evaluate multi-signal aggregation ($\text{Multi-}\mathcal{S}$) over a single entity without structural predicates ($\mathcal{P} = \emptyset$), workload W3 synthesizes user intents spanning multiple semantic dimensions. We generate 100 queries uniformly targeting either \texttt{imdb\_t1} or \texttt{imdb\_t2}. Query text generation uses a 3-slot prompt template (plot description \(|\) title style \(|\) actor/director vibe), then maps each slot to \texttt{plot\_emb}, \texttt{title\_emb}, and \texttt{actor\_director\_emb}, respectively. Each component phrase is embedded with Gemini-embedding-001 (1536-d). Ranking uses a late-binding weighted sum over these $L_2$ vector distances, with per-query weights normalized to 1.0 and $K=20$.

\begin{lstlisting}[language=SQL, caption={Example W3 Query}, label={lst:w3-sql}]
-- Q: Find movies in imdb_t1 matching a multi-aspect profile:
-- famous thriller actors AND unexpected twist plot motif
SELECT tconst, title, year
FROM imdb_t1
ORDER BY 
  0.60 * (actor_director_emb <-> '[-0.033, -0.031, ...]') 
  -- demographic cast intent mapping (L2 distance)
+ 0.40 * (plot_emb <-> '[-0.007, -0.028, ...]') 
  -- narrative plot intent mapping (L2 distance)
LIMIT 20;
\end{lstlisting}

\paragraph{\textbf{W4: Find movies matching a multi-aspect semantic profile within strict relational filters (Filtered multi-scoring).}} 
Building upon W3, workload W4 introduces exact structural predicates ($\mathcal{P}$) to evaluate filtered multi-signal aggregation ($\text{Multi-}\mathcal{S}$) over a single entity without cross-table joins ($\mathcal{J} = \emptyset$). We synthesize 1000 queries uniformly targeting either \texttt{imdb\_t1} or \texttt{imdb\_t2}. Query text generation follows the same three-slot semantics as W3, using a tab-separated triple format to improve parsing robustness at larger scale. The three text slots are mapped to \texttt{plot\_emb}, \texttt{title\_emb}, and \texttt{actor\_director\_emb}, and each is embedded with Gemini-embedding-001 (1536-d). 

To instantiate the structured predicates, each query is dynamically augmented with 1 to 2 relational filter groups applied strictly to the target table's metadata. The predicate generation space is strictly bounded by three explicit logical rules: (1) temporal constraints applying the $\ge$ operator on \texttt{year} sampled within $[1983, 2008]$, or applying both $\ge$ and $\le$ operators bounded by intervals $[1983, 2003]$ and $[1995, 2018]$; (2) quality thresholds applying the $\ge$ operator on \texttt{rating} sampled within $[3.5, 7.5]$, or both $\ge$ and $\le$ operators bounded by $[3.5, 6.5]$ and $[5.0, 9.0]$; and (3) categorical constraints applying the IN operator over random subsets of 1 to 4 predefined cinematic \texttt{genre}s from a pool of 12 options. Logically, the query formulates a top-$K$ ($K=20$) retrieval where the final ranking is determined by a late-binding, weighted sum of the $L_2$ distances across 2 to 3 corresponding continuous vector spaces. Each component signal is assigned a randomized interpolation weight summing to 1.0. This workload benchmarks the capability of the database engine to efficiently execute hybrid queries combining diverse relational filters with dense multi-stream vector scans.

\begin{lstlisting}[language=SQL, caption={Example W4 Query}, label={lst:w4-sql}]
-- Q: Find movies in imdb_t2 matching a multi-aspect profile:
-- "enchanting romance film" (title), "magical romance movie stars" (cast),
-- and "romantic fantasy involving mythical realms" (plot)
SELECT tconst, title, year, genre
FROM imdb_t2
WHERE year >= 2005 
  AND genre IN ('Adventure', 'Mystery')
ORDER BY 
  0.440 * (plot_emb <-> '[-0.028, -0.012, ...]') 
  -- narrative plot intent (L2 distance)
+ 0.426 * (actor_director_emb <-> '[-0.039, -0.034, ...]') 
  -- demographic cast intent (L2 distance)
+ 0.134 * (title_emb <-> '[-0.058, -0.016, ...]') 
  -- stylistic title intent (L2 distance)
LIMIT 20;
\end{lstlisting}

\paragraph{\textbf{W5: Pair up similar-cast films that fit a plot or title theme (Semantic join).}} 
Transitioning to Inter-Entity Synthesis ($\mathcal{J}$), workload W5 evaluates semantic joins without structural predicates ($\mathcal{P} = \emptyset$). At a high level, this query pattern represents a cross-era discovery task: it finds pairs of classic and modern movies that share a highly identical demographic cast or crew, and ranks these linked pairs globally based on how well one of the movies matches a specific narrative or stylistic search intent. To construct this workload, we synthesize 100 movie search intents using OpenAI GPT-4o ($\text{temperature}=0.8$) with the following exact prompt: \textit{``Generate exactly [N] short phrases that someone might type when searching for movies. Examples: 'thriller with twist', 'romantic comedy in Paris'. Output one phrase per line, no numbering, no other text.''} Each generated phrase is independently vectorized into a 1536-dimensional embedding using the Gemini-embedding-001 model. 

Logically, the query formulates a distance-threshold $\epsilon$-join on the \texttt{actor\_director\_emb} vectors between the \texttt{imdb\_t1} and \texttt{imdb\_t2} tables. To comprehensively evaluate the engine across varying linkage densities, the parameter $\tau$ is strictly sampled from predefined discrete semantic similarity tiers ($0.40, 0.45, 0.50, 0.55, 0.60$). These distinct values map directly to physical real-world relationships: a tight threshold ($\tau = 0.40$) captures highly identical cast combinations yielding an intermediate join selectivity of $0.01\%$, a medium threshold ($\tau = 0.50$) captures similar associations yielding $0.23\%$ selectivity, while a relaxed threshold ($\tau = 0.60$) captures loosely related connections yielding an exponentially larger $5.13\%$ intermediate join selectivity. The semantic scoring signal ($\mathcal{S}$) is uniformly assigned to target either \texttt{plot\_emb} or \texttt{title\_emb} on either the \texttt{imdb\_t1} or \texttt{imdb\_t2} table. Crucially, to prevent a single highly-ranked anchor film from dominating the result set with multiple join partners, the valid joined pairs are explicitly deduplicated by the primary key (\texttt{tconst}) of the scored table. The pairs are then globally ranked by the $L_2$ distance between the assigned target embedding and the query vector, retaining the overall top $K=20$ distinct pairs.

\begin{lstlisting}[language=SQL, caption={Example W5 Query}, label={lst:w5-sql}]
-- Q: Find classic-modern movie pairs with similar cast/crew (L2 <= 0.40),
-- ensure distinct classic movies, globally ranked by how well t1 fits a plot intent
SELECT t1_tconst, t1_title, t2_title
FROM (
  SELECT DISTINCT ON (t1.tconst) 
         t1.tconst AS t1_tconst, 
         t1.title AS t1_title, 
         t2.title AS t2_title,
         t1.plot_emb <-> '[-0.007, -0.028, ...]' AS score
  FROM imdb_t1 t1
  JOIN imdb_t2 t2
    ON (t1.actor_director_emb <-> t2.actor_director_emb) <= 0.40
    -- cast similarity join (sampled from discrete similarity tiers)
  ORDER BY t1.tconst, score
) distinct_pairs
ORDER BY score
LIMIT 20;
\end{lstlisting}

\paragraph{\textbf{W6: Pair up similar-cast films that fit a plot theme, within genre/era on each side (Filtered join).}} 
Building upon W5, workload W6 introduces exact structural predicates ($\mathcal{P}$) on both sides of the semantic join to evaluate filtered inter-entity synthesis. At a high level, this query pattern represents a constrained cross-era discovery task: it finds pairs of classic and modern movies that share a highly identical demographic cast or crew, satisfying rigorous structural constraints on both sides (restricting release eras or demanding specific minimum ratings), and ranks these linked pairs globally based on how well the primary movie matches a specific narrative search intent. To construct this workload, we synthesize 100 movie plot motifs using OpenAI GPT-4o ($\text{temperature}=0.8$) with the following exact prompt: \textit{``Generate exactly [N] short phrases describing a specific movie theme, setting, or plot motif. Examples: 'a detective solving a murder in a small town', 'spaceship crew encountering an alien anomaly'. Output one phrase per line, no numbering, no other text.''} Each generated phrase is independently vectorized into a 1536-dimensional embedding using the Gemini-embedding-001 model to serve as the ranking intent on \texttt{plot\_emb}. 

To instantiate the structured predicates, each query is dynamically augmented with relational filters applied strictly to the metadata of both \texttt{imdb\_t1} and \texttt{imdb\_t2}. The predicate generation space is strictly bounded by explicit logical rules: temporal constraints applying the $\ge$ or $\le$ operators on \texttt{year} sampled within the interval $[1980, 2020]$; quality thresholds applying the $\ge$ or $\le$ operators on \texttt{rating} sampled within $[1.0, 10.0]$; and categorical constraints applying the IN operator over random subsets of 12 predefined cinematic \texttt{genre}s. The join threshold tiers, deduplication policy, and sparse-to-dense linkage settings follow W5. W6 adds only dual-sided relational filtering before final ranking, and then ranks surviving pairs by the $L_2$ distance between \texttt{imdb\_t1.plot\_emb} and the query vector (top $K=20$).

\begin{lstlisting}[language=SQL, caption={Example W6 Query}, label={lst:w6-sql}]
-- Q: Find thriller plot movies in imdb_t1 linked to imdb_t2 movies with 
-- similar cast/crew (L2 <= 0.45), satisfying specific year and rating filters
SELECT t1_tconst, t1_title, t2_title
FROM (
  SELECT DISTINCT ON (t1.tconst) 
         t1.tconst AS t1_tconst, 
         t1.title AS t1_title, 
         t2.title AS t2_title,
         t1.plot_emb <-> '[-0.007, -0.027, ...]' AS score
  FROM imdb_t1 t1
  JOIN imdb_t2 t2
    ON (t1.actor_director_emb <-> t2.actor_director_emb) <= 0.45
    -- cast similarity join (sampled from discrete similarity tiers)
  WHERE t1.year >= 1985 
    AND t2.rating >= 5.7
  ORDER BY t1.tconst, score
) distinct_pairs
ORDER BY score
LIMIT 20;
\end{lstlisting}

\paragraph{\textbf{W7: Pair up similar-cast films across two plot themes (Multi-scoring join).}} 
Building upon W5, workload W7 introduces complex multi-signal aggregation ($\text{Multi-}\mathcal{S}$) into the inter-entity synthesis pipeline ($\mathcal{J}$) without structural predicates ($\mathcal{P} = \emptyset$). At a high level, this query pattern represents a dual-intent cross-era discovery task: it finds tuples of classic and modern movies that share a highly identical demographic cast or crew, and globally ranks these pairs based on how well the primary movie matches a first narrative motif, how well the secondary movie matches a second narrative motif, and the absolute tightness of their shared cast similarity. To construct this workload, we synthesize 1000 search intents per table using OpenAI GPT-4o ($\text{temperature}=1.0$) with the following exact prompt: \textit{``Generate exactly [N] short, diverse phrases (3-8 words each) that someone might type when searching for movies. They should cover a wide range of genres, moods, settings, and themes. Avoid repeating themes already covered.\textbackslash nOutput one phrase per line, no numbering, no other text.''} Each phrase is independently vectorized into a 1536-dimensional embedding using the Gemini-embedding-001 model to target either the \texttt{imdb\_t1} \texttt{plot\_emb} or \texttt{imdb\_t2} \texttt{plot\_emb}. The primary and secondary query vectors are paired dynamically across specific correlation tiers (anti, low, medium, high) to rigorously control query hardness.

The join threshold tiers and sparse-to-dense linkage settings follow W5. The key change in W7 is the ranking objective: globally valid pairs are scored by a late-binding weighted sum of three $L_2$ distances, combining \texttt{imdb\_\allowbreak t1.\allowbreak plot\_\allowbreak emb} to the first intent, \texttt{imdb\_\allowbreak t2.\allowbreak plot\_\allowbreak emb} to the second intent, and the cross-table \texttt{actor\_\allowbreak director\_\allowbreak emb} distance. Each weight vector is sampled and normalized to 1.0, and the top $K=20$ pairs are retained.

\begin{lstlisting}[language=SQL, caption={Example W7 Query}, label={lst:w7-sql}]
-- Q: Find classic-modern movie tuples with similar cast/crew (L2 <= 0.50),
-- ranked globally by a 3-way multi-scoring objective
SELECT t1.tconst AS t1_tconst, 
       t2.tconst AS t2_tconst,
       ( 0.20 * (t1.plot_emb <-> '[-0.009, 0.021, ...]') 
         -- primary plot intent (L2 distance)
       + 0.20 * (t2.plot_emb <-> '[-0.019, -0.008, ...]') 
         -- secondary plot intent (L2 distance)
       + 0.60 * (t1.actor_director_emb <-> t2.actor_director_emb) 
         -- actual cast similarity (L2 distance)
       ) AS score
FROM imdb_t1 t1
JOIN imdb_t2 t2
  ON (t1.actor_director_emb <-> t2.actor_director_emb) <= 0.50
  -- cast similarity join (sampled from discrete similarity tiers)
ORDER BY score
LIMIT 20;
\end{lstlisting}

\paragraph{\textbf{W8: Pair up similar-cast films across two plot themes, within genre/era on each side (Filtered multi-scoring join).}} 
Building upon W7, workload W8 represents the most comprehensive query pattern, introducing exact structural predicates ($\mathcal{P}$) on both sides of the multi-signal aggregation ($\text{Multi-}\mathcal{S}$) and inter-entity synthesis pipeline ($\mathcal{J}$). At a high level, this query pattern represents a constrained, dual-intent cross-era discovery task: it finds tuples of classic and modern movies that share a highly identical demographic cast or crew, satisfying rigorous structural constraints on both sides (restricting release eras or demanding specific minimum ratings), and globally ranks these linked pairs based on how well the primary movie matches a first narrative motif, how well the secondary movie matches a second narrative motif, and the absolute tightness of their shared cast similarity. To construct this workload, we synthesize 100 pairs of distinct movie concepts using OpenAI GPT-4o ($\text{temperature}=0.8$) with the following exact prompt: \textit{``Generate exactly [N] pairs of movie concepts. Each line should contain two distinct concepts separated by a pipe character '|'. For example: 'A gritty sci-fi thriller about time travel | A lighthearted comedy about a robot learning to love'. Output only the pipe-separated pairs, one pair per line.''} Each text field in the generated pair is independently vectorized into a 1536-dimensional embedding using the Gemini-embedding-001 model, with the first concept targeting the \texttt{imdb\_t1} \texttt{plot\_emb} and the second targeting the \texttt{imdb\_t2} \texttt{plot\_emb}.

To instantiate the structured predicates, each query is dynamically augmented with relational filters applied strictly to the metadata of both \texttt{imdb\_t1} and \texttt{imdb\_t2}. The predicate generation space is strictly bounded by explicit logical rules: temporal constraints applying the $\ge$ or $\le$ operators on \texttt{year} sampled within the interval $[1980, 2020]$; quality thresholds applying the $\ge$ or $\le$ operators on \texttt{rating} sampled within $[1.0, 10.0]$; and categorical constraints applying the IN operator over random subsets of 12 predefined cinematic \texttt{genre}s. W8 follows W5 for join tiers, and follows W7 for three-signal late-binding ranking; its only additional step is applying dual-sided relational filters before ranking. The final result keeps the top $K=20$ distinct pairs.

\begin{lstlisting}[language=SQL, caption={Example W8 Query}, label={lst:w8-sql}]
-- Q: Find "Sci-fi movies with robots" in imdb_t1 (year >= 1985) linked to 
-- "Exploration documentaries of outer space" in imdb_t2 (rating >= 5.7) 
-- with similar cast/crew (L2 <= 0.50), ranked globally by a 3-way objective
SELECT t1.tconst AS t1_tconst, 
       t1.title AS t1_title, 
       t2.title AS t2_title,
       ( 0.25 * (t1.plot_emb <-> '[-0.009, 0.021, ...]') 
         -- primary plot intent (L2 distance)
       + 0.25 * (t2.plot_emb <-> '[-0.027, -0.013, ...]') 
         -- secondary plot intent (L2 distance)
       + 0.50 * (t1.actor_director_emb <-> t2.actor_director_emb) 
         -- actual cast similarity (L2 distance)
       ) AS score
FROM imdb_t1 t1
JOIN imdb_t2 t2
  ON (t1.actor_director_emb <-> t2.actor_director_emb) <= 0.50
  -- cast similarity join (sampled from discrete similarity tiers)
WHERE t1.year >= 1985 
  AND t2.rating >= 5.7
ORDER BY score
LIMIT 20;
\end{lstlisting}

\subsection{MOLECULE Benchmark}
\label{sec:bench-molecule}

\subsubsection{Dataset Description}
Unlike IMDB, which is positioned as a fabricated discovery benchmark, MOLECULE provides real hybrid workloads extracted from an operational Biofoundry project dedicated to academic knowledge integration and retrieval. The underlying database naturally stores heterogeneous entities—including research papers, researchers, organizations, and biochemical concepts—augmented with semantic tags, structured attributes, and multiple field-level embeddings. To expose this complex, evidence-centric biomedical reasoning in a relational benchmark, we project the relevant entity interactions into a bipartite schema: a fact-side relation (\texttt{molecule\_\allowbreak fact}) and a paper-side relation (\texttt{molecule\_\allowbreak paper}). This layout successfully captures the target user behavior of our deep research agent system: proposing molecule-related concepts, retrieving supporting literature, applying metadata constraints, and ranking candidates under multi-signal criteria.

\subsubsection{Data Source and Construction}
To construct this benchmark, we combine two sources: (i) molecule-centric fact/paper data extracted from the broader heterogeneous knowledge graph and (ii) deep research agent query traces. The construction pipeline is:
\begin{itemize}
    \item \textbf{Agent-Trace Grounding:} We collect historical deep research agent queries and cluster them into eight workload intents (W1--W8). For each intent class \(x\), we retain a set of representative trace templates \(T_x\), where each template includes the query skeleton and the fields that must be instantiated (\texttt{query\_text}, predicates, join tier, weight vector, or combinations thereof).
    \item \textbf{Data Sourcing:} We collect molecule facts and publication records with complete text fields and normalized molecule identifiers from the Biofoundry integration engine, then retain only rows with valid metadata required by the template instantiation fields.
    \item \textbf{Profile Construction:} On the fact side, each row stores a fact statement plus structured molecule descriptors; on the paper side, each row stores publication metadata and abstract-level content. This yields a physically separated but semantically linkable two-relation layout aligned with agent reasoning hops.
    \item \textbf{Vectorization / Structural Features:} We encode textual fields with Gemini-embedding-001 into 1536-dimensional vectors and represent molecular structure with 1024-bit ECFP fingerprints. These representations are exactly the fields consumed by template-driven augmentation from \(T_x\).
\end{itemize}

\subsubsection{Workload Construction}
\label{sec:molecule-workload-construction}

To systematically evaluate execution strategies, we construct eight workloads that cover the full taxonomy of our query model: structured predicates ($\mathcal{P}$), multi-signal scoring ($\mathcal{S}$), and semantic joins ($\mathcal{J}$). For each workload \(x\), we follow an agent-grounded augmentation recipe: (1) select representative templates \(T_x\) from deep research agent traces, (2) map template slots to physical fields, (3) generate or sample slot values, and (4) materialize augmented queries as workload JSON.

\textbf{Predicates and Scoring ($\mathcal{P}$ and $\mathcal{S}$).}
For fact-side filtering, predicate atoms are sampled from categorical fields (\texttt{fact\_\allowbreak type}, \texttt{claim\_\allowbreak polarity}, \texttt{mol\_\allowbreak source\_\allowbreak origin}, \texttt{mol\_\allowbreak toxicity\_\allowbreak flag}, \texttt{mol\_\allowbreak moa}, \texttt{mol\_\allowbreak target\_\allowbreak type}, \texttt{evidence\_\allowbreak type}, \texttt{organism}), numeric fields (\texttt{mol\_\allowbreak mw}, \texttt{mol\_\allowbreak logp}, \texttt{mol\_\allowbreak tpsa}, \texttt{mol\_\allowbreak num\_\allowbreak atoms}, \texttt{mol\_\allowbreak num\_\allowbreak rings}, \texttt{mol\_\allowbreak num\_\allowbreak hbd}, \texttt{mol\_\allowbreak num\_\allowbreak hba}), and array containment on \texttt{mol\_\allowbreak therapeutic\_\allowbreak area}. Multi-signal scoring combines fact-text semantic distance and molecule-structure Jaccard distance, with per-query randomized weights normalized to sum to 1.

\textbf{Semantic Joins ($\mathcal{J}$).}
For inter-entity synthesis (W5--W8), we link fact rows to paper rows using a distance-threshold join over textual evidence vectors. The threshold $\tau$ is sampled from discrete tiers $\{0.45, 0.50, 0.55, 0.65\}$, labeled \textit{very\_similar}, \textit{similar}, \textit{related}, and \textit{loose}. These tiers induce exponentially different candidate densities (from thousands to millions of pairs), which we intentionally preserve to stress both sparse and dense join regimes.

To avoid synthetic key joins, all cross-table linkage is parameterized geometrically by $\tau$; smaller tiers produce strict semantic evidence alignment, and larger tiers expand recall with sharply higher pair counts. This discrete-tier design mirrors realistic retrieval behavior where semantic relatedness appears as tail events in embedding space.

\paragraph{\textbf{W1: Find facts on a research topic (Semantic retrieval).}} 
For W1, we first extract a template set \(T_1\) from agent traces whose intent is single-hop fact retrieval without filters or joins. We then instantiate the \texttt{query\_text} slot using OpenAI GPT-4o ($\text{temperature}=0.9$) with the exact prompt: \textit{``Generate exactly [N] diverse natural-language queries a biomedical researcher might type when searching a database of facts about drug molecules and proteins. Cover varied topics: drug mechanism of action, toxicity, pharmacokinetics, structure-activity relationships, clinical outcomes, natural products, therapeutic areas (oncology / neurology / infectious / metabolic / cardio / psychiatry / immunology), organisms tested (human / mouse / rat / cell\_line / bacteria), in vitro vs in vivo evidence, and synthesis routes. Vary phrasing style: keyword style, question style, long-sentence style. Keep each under 15 words, realistic, no numbering, one phrase per line.''} Instantiated queries are deduplicated, embedded by Gemini-embedding-001 (1536-d), and bound to \texttt{molecule\_fact.fact\_emb} with $K=20$. This yields pure semantic retrieval with $\mathcal{P}=\emptyset$ and $\mathcal{J}=\emptyset$.

\begin{lstlisting}[language=SQL, caption={Example W1 Query}, label={lst:mol-w1-sql}]
-- Q: "inhibits beta-amyloid aggregation"
SELECT id, organism, property_type, fact_description
FROM molecule_fact
ORDER BY fact_emb <-> '[-0.0125, 0.0341, ...]' 
LIMIT 20;
\end{lstlisting}

\paragraph{\textbf{W2: Find facts on a research topic, within organism/property scope (Filtered retrieval).}} 
For W2, we extract templates \(T_2\) from agent traces where semantic retrieval is explicitly conditioned by metadata constraints. We instantiate \texttt{query\_\allowbreak text} by referring to the W1 generation setting, then augment each template with sampled predicate slots (1--3 predicates) over \texttt{molecule\_\allowbreak fact}. The predicate slot space is explicitly bounded: categorical IN predicates over \texttt{fact\_\allowbreak type}, \texttt{claim\_\allowbreak polarity}, \texttt{mol\_\allowbreak source\_\allowbreak origin}, \texttt{mol\_\allowbreak toxicity\_\allowbreak flag}, \texttt{mol\_\allowbreak moa}, \texttt{mol\_\allowbreak target\_\allowbreak type}, \texttt{evidence\_\allowbreak type}, \texttt{organism}; numeric threshold predicates ($\ge,\le$) over \texttt{mol\_\allowbreak mw}, \texttt{mol\_\allowbreak logp}, \texttt{mol\_\allowbreak tpsa}, \texttt{mol\_\allowbreak num\_\allowbreak atoms}, \texttt{mol\_\allowbreak num\_\allowbreak rings}, \texttt{mol\_\allowbreak num\_\allowbreak hbd}, \texttt{mol\_\allowbreak num\_\allowbreak hba}; and array containment (\texttt{@>}) on \texttt{mol\_\allowbreak therapeutic\_\allowbreak area}. We retain augmented queries by selectivity buckets: \textit{tight} $[0.01,0.10)$ at 30\%, \textit{medium} $[0.10,0.30)$ at 40\%, and \textit{loose} $[0.30,0.70)$ at 30\%.

\begin{lstlisting}[language=SQL, caption={Example W2 Query}, label={lst:mol-w2-sql}]
-- Q: "drug interaction mechanisms with cytochrome P450 enzymes"
-- medium-selectivity bundle: numeric + categorical + array containment
SELECT id, organism, property_type, fact_description
FROM molecule_fact
WHERE mol_tpsa >= 115.54
  AND organism IN ('human', 'mouse')
  AND mol_therapeutic_area @> ARRAY['oncology']
ORDER BY fact_emb <-> '[-0.0102, -0.0001, ...]' 
LIMIT 20;
\end{lstlisting}

\paragraph{\textbf{W3: Find facts on a topic about a reference molecule (Multi-scoring retrieval).}} 
For W3, we extract templates \(T_3\) from traces where the agent jointly reasons over textual evidence and molecule structure in one retrieval hop. We instantiate two template slots: (i) \texttt{query\_text} (referring to the W1 generation setting) mapped to \texttt{molecule\_fact.fact\_emb} and (ii) a reference molecule slot mapped to \texttt{molecule\_fact.mol\_ecfp} as a 1024-bit ECFP anchor. We then instantiate per-query weight slots and normalize to 1.0, producing a late-binding weighted sum across L2 text distance and Jaccard structure distance.

\begin{lstlisting}[language=SQL, caption={Example W3 Query}, label={lst:mol-w3-sql}]
-- Q: "drug interaction mechanisms..." related to a specific reference molecule
SELECT id, organism, property_type, fact_description
FROM molecule_fact
ORDER BY 
  0.60 * (fact_emb <-> '[-0.0101, -0.0001, ...]') 
  -- biological topic intent (semantic distance)
+ 0.40 * (mol_ecfp <%> '[0, 0, 0, 1, ...]') 
  -- structural chemical similarity (Jaccard distance)
LIMIT 20;
\end{lstlisting}

\paragraph{\textbf{W4: Find facts on a topic about a reference molecule, within organism/property scope (Filtered multi-scoring).}} 
For W4, we extract templates \(T_4\) from traces that combine dual-signal scoring with explicit fact-side filtering. We instantiate text, structure, and weight slots exactly as in W3, then instantiate 1--3 predicate slots by referring to the W2 predicate setting and retain queries under the same selectivity-bucket policy (\textit{tight} 30\%, \textit{medium} 40\%, \textit{loose} 30\%). This yields a template-grounded filtered multi-scoring workload on a single relation.

\begin{lstlisting}[language=SQL, caption={Example W4 Query}, label={lst:mol-w4-sql}]
-- Q: "drug interaction mechanisms..." related to a specific reference molecule
-- filtered multi-scoring: same two signals as W3 + predicates following W2
SELECT id, organism, property_type, fact_description
FROM molecule_fact
WHERE mol_tpsa >= 115.54
  AND evidence_type IN ('in_vitro', 'in_vivo')
  AND mol_target_type IN ('enzyme', 'receptor')
ORDER BY 
  0.60 * (fact_emb <-> '[-0.0101, -0.0001, ...]') 
  -- biological topic intent (semantic distance)
+ 0.40 * (mol_ecfp <%> '[0, 0, 0, 1, ...]') 
  -- structural chemical similarity (Jaccard distance)
LIMIT 20;
\end{lstlisting}

\paragraph{\textbf{W5: Find paper-supported facts about a reference molecule (Semantic join).}} 
For W5, we extract templates \(T_5\) from traces where the agent retrieves fact candidates and then links them to supporting literature. We instantiate a structure-anchor slot (sampled valid \texttt{mol\_id}, 1024-bit ECFP) and a join-threshold slot for \texttt{molecule\_\allowbreak fact.\allowbreak fact\_\allowbreak emb} $\leftrightarrow$ \texttt{molecule\_\allowbreak paper.\allowbreak abstract\_\allowbreak emb}. The join-threshold slot is sampled from discrete tiers: \texttt{very\_similar} ($0.45$), \texttt{similar} ($0.50$), \texttt{related} ($0.55$), \texttt{loose} ($0.65$), and \texttt{unrelated} ($0.75$), preserving the observed sparse-to-dense evidence-linking regimes in agent behavior.

\begin{lstlisting}[language=SQL, caption={Example W5 Query}, label={lst:mol-w5-sql}]
-- Q: Find facts matching a specific molecular structure, 
-- then join papers providing semantic textual evidence (L2 <= 0.45)
SELECT fact_id, paper_id
FROM (
  SELECT DISTINCT ON (f.id)
         f.id AS fact_id,
         p.pmid AS paper_id,
         (f.fact_emb <-> p.abstract_emb) AS join_dist
  FROM (
    SELECT id, fact_emb
    FROM molecule_fact
    ORDER BY mol_ecfp <%> '[0, 0, 0, 0, ...]'  -- fact-side structure intent
    LIMIT 20
  ) AS f
  JOIN molecule_paper AS p
    ON (f.fact_emb <-> p.abstract_emb) <= 0.45   -- very_similar tier
  ORDER BY f.id, join_dist
) AS dedup
ORDER BY join_dist
LIMIT 20;
\end{lstlisting}

\paragraph{\textbf{W6: Find paper-supported facts about a reference molecule, within fact/paper filters (Filtered join).}} 
For W6, we extract templates \(T_6\) from traces where evidence-linking is constrained by both fact-side and paper-side metadata. We instantiate W5 structure-anchor and join slots, then instantiate dual-sided predicate slots using an oversampled candidate pool: fact-side predicates by referring to the W2 predicate setting and paper-side predicates over \texttt{year}, \texttt{journal}, \texttt{pub\_type}, and \texttt{mesh\_terms}. Candidate templates are retained via soft selectivity acceptance (knee at 0.10), producing a broad left/right selectivity grid aligned with agent-side constrained retrieval behavior.

\begin{lstlisting}[language=SQL, caption={Example W6 Query}, label={lst:mol-w6-sql}]
-- Q: Facts for a reference molecule (natural/microbial origin),
-- joined with US Gov supported papers providing semantic evidence
SELECT fact_id, paper_id
FROM (
  SELECT DISTINCT ON (f.id)
         f.id AS fact_id,
         p.pmid AS paper_id,
         (f.fact_emb <-> p.abstract_emb) AS join_dist
  FROM (
    SELECT id, fact_emb
    FROM molecule_fact
    WHERE mol_source_origin IN ('natural_product', 'microbial')
      AND mol_toxicity_flag IN ('non_toxic')
    ORDER BY mol_ecfp <%> '[0, 0, 0, 0, ...]'
    LIMIT 20
  ) AS f
  JOIN molecule_paper AS p
    ON (f.fact_emb <-> p.abstract_emb) <= 0.45
   AND p.pub_type @> ARRAY['Research Support, U.S. Gov''t']
   AND p.mesh_terms @> ARRAY['Humans']
  ORDER BY f.id, join_dist
) AS dedup
ORDER BY join_dist
LIMIT 20;
\end{lstlisting}

\paragraph{\textbf{W7: Find paper-supported facts combining a reference molecule, a topic, and tight fact-paper alignment (Multi-scoring join).}} 
For W7, we extract templates \(T_7\) from traces where the agent performs cross-table tuple retrieval with three simultaneous ranking signals. We instantiate a molecule-anchor slot, a topic-text slot, a join-tier slot, and a three-weight slot. Topic texts are generated by OpenAI GPT-4o ($\text{temperature}=0.8$) from four exact prompt templates (\texttt{mol\_\allowbreak specific}, \texttt{area\_\allowbreak specific}, \texttt{target\_\allowbreak topic}, \texttt{unrelated}) and embedded by Gemini-embedding-001 onto \texttt{molecule\_\allowbreak paper.\allowbreak abstract\_\allowbreak emb}. Join tiers are sampled from $\{0.45,0.50,0.55,0.65\}$ and weight slots are normalized to 1.0, yielding a template-grounded late-binding multi-signal join workload for global tuple pairs.

\begin{lstlisting}[language=SQL, caption={Example W7 Query}, label={lst:mol-w7-sql}]
-- Q: Find fact-paper tuples matching a molecular structure and topic,
-- ranked globally by a 3-way objective
SELECT f.id AS fact_id,
       p.pmid AS paper_id,
       ( 0.35 * (f.mol_ecfp <%> '[0, 1, 0, ...]')      -- fact molecule intent
       + 0.33 * (p.abstract_emb <-> '[-0.009, 0.021, ...]') -- paper topic intent
       + 0.32 * (f.fact_emb <-> p.abstract_emb) ) AS score  -- join penalty
FROM molecule_fact AS f
JOIN molecule_paper AS p
  ON (f.fact_emb <-> p.abstract_emb) <= 0.50  -- similar tier
ORDER BY score
LIMIT 20;
\end{lstlisting}

\paragraph{\textbf{W8: Find paper-supported facts combining molecule/\allowbreak topic/\allowbreak fact-paper alignment, within fact/\allowbreak paper filters (Filtered multi-scoring join).}} 
For W8, we extract templates \(T_8\) from traces representing full deep-research behavior: constrained cross-entity tuple linking with multi-signal ranking. We instantiate W7 slots (molecule anchor, topic embedding, join tier, normalized three-weight vector) and W6 slots (dual-sided predicate bundle) together. Operationally, we pair an oversampled W6 predicate pool with W7 query templates, then select a final subset that preserves diversity over filter hardness, join density tier, and topic-to-molecule correlation tier. This yields the comprehensive agent-grounded tuple workload where $\mathcal{P}$, $\mathcal{S}$, and $\mathcal{J}$ are jointly active.

\begin{lstlisting}[language=SQL, caption={Example W8 Query}, label={lst:mol-w8-sql}]
-- Q: Find tuples of Catalyst facts + NIH-funded papers, ranked 
-- globally by molecule intent + topic intent + alignment penalty
SELECT f.id AS fact_id,
       p.pmid AS paper_id,
       ( 0.35 * (f.mol_ecfp <%> '[0, 1, 0, ...]')
       + 0.33 * (p.abstract_emb <-> '[-0.006, 0.012, ...]')
       + 0.32 * (f.fact_emb <-> p.abstract_emb) ) AS score
FROM molecule_fact AS f
JOIN molecule_paper AS p
  ON (f.fact_emb <-> p.abstract_emb) <= 0.55   -- related tier
WHERE f.mol_moa IN ('catalyst')
  AND f.evidence_type IN ('in_vivo', 'clinical')
  AND p.pub_type @> ARRAY['Research Support, N.I.H., Extramural']
  AND p.year >= 2005
ORDER BY score
LIMIT 20;
\end{lstlisting}
\endgroup
\begingroup
\emergencystretch=2em
\section{Integration with Semantic Operator Systems}
\label{sec:integrate_sem_op}

As established in Sec.~\ref{sec:multistep_query}, \sys{} is designed to assist \emph{semantic operator systems} \cite{russo_abacus_2025, trummer_implementing_2025,patel_semantic_2024} by acting as a cost-effective prefiltering stage that prunes input data before any LLM invocation. These systems typically implement a suite of LLM-backed \emph{semantic operators}, including \textsc{sem\_filter}, \textsc{sem\_classify}, \textsc{sem\_rank}, \textsc{sem\_join}, and \textsc{sem\_map}. \sys{} functions as a plug-in prefilter for these operators, leveraging multi-facet vector embeddings to identify the most promising candidate tuples for downstream LLM evaluation. Detailed SemBench performance results are provided in Table~\ref{tab:sembench}.

\subsection{The Prefilter Pipeline}
\label{sec:int-pipeline}

In this architecture, \sys{} serves as a pluggable prefilter component. The overall workflow typically operates in two stages: \sys{} first recommends a high-recall candidate set, which a full-featured semantic operator engine subsequently evaluates and refines using targeted LLM calls. Alternatively, in scenarios where the downstream semantic operator is omitted, the workflow bypasses LLM evaluation entirely and simply accepts all of \sys{}'s multi-dimensional recommendations directly as the final output. This ``\sys{} alone'' configuration (denoted as \textit{DASE} in Table~\ref{tab:sembench}) remains highly competitive whenever the embedding distance effectively separates relevant tuples from the background distribution.

\subsection{Per-Operator Strategies}
\label{sec:int-strategies}

The integration logic is specifically tailored to the semantics of each operator. Crucially, \sys{} treats every natural-language query as a \emph{multi-dimensional evaluation task}: it decomposes the user's semantic intent into a statistical proxy comprising weighted aggregations over multiple, potentially heterogeneous embedding streams (e.g., visual, textual, and structural). At the execution layer, these operators are natively compiled into the multi-scoring architecture introduced in Sec.~\ref{sec:execution}, where independent Filtered-Score Streamers (FSS) are synchronized and pruned by the Multi-Scoring Aggregator (MSA).

\subsubsection{\textsc{sem\_filter}}
\label{sec:int-filter}

A \textsc{sem\_filter} evaluates a natural-language Boolean predicate $\ell$ on an input relation $T$. For a query such as ``finding a blue shirt with positive reviews,'' the semantics are not restricted to a single modality. \sys{} captures this by modeling the filter as a joint multi-scoring task across all relevant feature spaces $\mathcal{F}$ (e.g., image embeddings for visual traits and text embeddings for sentiment).

For each space $f \in \mathcal{F}$, \sys{} derives a specific set of positive anchors $P_f$ and antithetical anchors $N_f$. The MSA then synchronizes these independent FSS streams to compute a cross-dimensional contrastive margin on the fly:
\[
  s_i \;=\; \sum_{f \in \mathcal{F}} w_f \left(
    \tfrac{1}{|P_f|}\sum_{p\in P_f}\!\cos(\mathbf{e}_p, \mathbf{e}_i^{(f)})
    \;-\;
    \tfrac{1}{|N_f|}\sum_{n\in N_f}\!\cos(\mathbf{e}_n, \mathbf{e}_i^{(f)})
  \right)
\]
where $\mathbf{e}_i^{(f)}$ is the embedding of tuple $i$ in space $f$, and $w_f$ balances the contribution of each modality. This mechanism natively aggregates signals---such as visual blueness and textual sentiment---to identify top candidates. Because \sys{} acts as a high-recall prefilter, avoiding false negatives is paramount. We employ two recommendation rules that utilize intentionally relaxed boundaries:
\begin{itemize}
\item \emph{Margin-based uncertainty band.} For total labeling tasks (e.g., \emph{Wildlife}~Q1), \sys{} applies thresholds to the joint margin $s_i$. To aggressively protect recall, it forwards a conservatively wide uncertain band (typically $\alpha \approx 0.2$--$0.5$) for LLM evaluation, while accepting the cheap sign of $s_i$ for confidently extreme rows.
\item \emph{Top-$K'$ shortlist.} For \texttt{LIMIT}~$K$ queries (e.g., \emph{Movie}~Q1), the MSA seamlessly streams the $K'$ tuples with the highest aggregate $s_i$. The downstream engine's \texttt{AI.IF} then evaluates this multi-dimensional shortlist, short-circuiting at the $K$-th valid survivor. This heuristic reduces the query latency for \emph{Movie}~Q1 from 26.3\,s to 0.5\,s.
\end{itemize}

\subsubsection{\textsc{sem\_classify}}
\label{sec:int-classify}

A \textsc{sem\_classify} operator assigns each tuple to one of $K$ enumerated classes $\mathcal{C}=\{c_1,\dots,c_K\}$. Whereas \textsc{sem\_filter} contrasts a single positive concept against its antithetical paraphrases, \textsc{sem\_classify} generalises this to a $K$-way contrast: each class becomes its own anchor, and the per-row decision reduces to identifying which anchor wins. For instance, in \emph{Cars}~Q10 the engine must classify automotive complaints into 24 distinct problem categories, where a complaint may comprise both a textual description and an image of the defect, requiring multi-modal reasoning.

\sys{} natively maps this multi-class evaluation onto a parallel multi-scoring retrieval task. It embeds each class anchor prompt $c_k$ once per facet to obtain $\mathbf{e}_{c_k}^{(f)}$, initialising an independent FSS stream per category. The MSA then synchronises these $K$ streams across the required feature spaces $\mathcal{F}$ (e.g., visual embeddings of the damaged car part and text embeddings of the complaint description), aggregating per-facet cosine similarities into a per-class affinity $s_i^{(k)}$, from which a proxy label $\hat{k}_i$ and confidence margin $m_i$ follow directly:
\begin{gather*}
  s_i^{(k)} \;=\; \sum_{f \in \mathcal{F}} w_f \cos\!\bigl(\mathbf{e}_{c_k}^{(f)},\, \mathbf{e}_i^{(f)}\bigr), \ k = 1,\dots,K, \\[4pt]
  \hat{k}_i \;=\; \operatorname*{arg\,max}_{k}\, s_i^{(k)},
  \
  m_i \;=\; s_i^{(\hat{k}_i)} - \max_{k \neq \hat{k}_i}\, s_i^{(k)}.
\end{gather*}
All three quantities are computed without any LLM invocation. \sys{} then partitions the result set on $m_i$: the bottom-$\alpha$ tuples (smallest margin) are forwarded to the downstream engine for rigorous refinement via \texttt{AI.CLASSIFY}, while the engine safely accepts the cheap argmax $\hat{k}_i$ on the remaining confident rows. The final output unions the LLM-resolved tail with the \sys{}-asserted confident bulk, achieving macro $F_1$ parity with an exhaustive all-LLM baseline at roughly half the inference cost (Table~\ref{tab:sembench}).

\subsubsection{\textsc{sem\_rank}}
\label{sec:int-rank}

A \textsc{sem\_rank} operator orders or scores tuples according to a defined rubric. \sys{} compiles this requirement into two distinct multi-scoring execution paths based on the query objective:
\begin{itemize}
\item \emph{Top-$K$ retrieval.} When the query strictly demands the highest-ranking candidates, the execution naturally reduces to the \textsc{sem\_filter}\,+\,\texttt{LIMIT} paradigm of Sec.~\ref{sec:int-filter}. \sys{} leverages the MSA to synchronise contrastive margin streams across all targeted feature spaces $\mathcal{F}$, organically short-circuiting the retrieval once $K$ confidently top-ranked tuples survive downstream \texttt{AI.IF} validation.
\item \emph{Rubric scoring.} When a strict numeric label is required for \emph{every} row (e.g., \emph{Movie}~Q9, which scores from 1 to 5 how much a reviewer liked a movie), the task is structurally a \textsc{sem\_classify} over an \emph{ordinal} class set: each rubric grade $g \in \{1,\dots,G\}$ plays the role of a class anchor $c_g$, and the per-grade affinity $s_i^{(g)}$, proxy label $\hat{g}_i = \arg\max_g s_i^{(g)}$, and confidence margin $m_i$ all carry over verbatim from the formulation in Sec.~\ref{sec:int-classify}. The only operator-specific change lies in the cascade policy. Because global rank-correlation metrics like Spearman integrate over the entire score distribution, a single mis-ranked tuple can disproportionately degrade the global metric, so \sys{} cannot afford the aggressive accept-rate used for nominal classification. The bottom-$\alpha$ tail (smallest $m_i$) is therefore widened substantially ($\alpha \approx 0.7$) and refined via \texttt{AI.SCORE} rather than \texttt{AI.CLASSIFY}; the proxy argmax $\hat{g}_i$ is accepted only on the confident head, and the two streams are concatenated into the final ranking.
\end{itemize}

\subsubsection{\textsc{sem\_join}}
\label{sec:int-join}

A \textsc{sem\_join} evaluates a match predicate over pairs $(t_i,t_j)\!\in\!T_1\!\times\!T_2$, such as verifying if two items share the same brand and category (\emph{Ecomm}~Q7). Unlike the per-row operators above, the candidate space here is the Cartesian product $|T_1|\!\cdot\!|T_2|$, so prefiltering is not merely an optimisation but a feasibility requirement. \sys{} evaluates the match proxy directly against the \idx{}, which strictly decouples continuous multi-facet similarities in \texttt{\idx{}(embed)} from discrete semantic truths in \texttt{\idx{}(AI.IF)}.

During execution, \sys{} first probes \texttt{\idx{}(AI.IF)} to resolve previously verified pairs at zero LLM cost. For unresolved pairs, the MSA translates the NL match criteria into a multi-scoring problem, synchronising the traversal across multiple \texttt{\idx{}(embed)} streams (e.g., product titles \emph{and} images) to compute a joint similarity score.
\begin{itemize}
\item \emph{Two-sided uncertainty band.} For joins like \emph{Ecomm}~Q7, the MSA organically identifies pairs with the highest aggregate scores ($S_{ij}\!\geq\!\tau_{\mathrm{hi}}$) as confident matches. Pairs in the intermediate range ($\tau_{\mathrm{lo}}{<}S_{ij}{<}\tau_{\mathrm{hi}}$)---where combined visual and textual cues are strong but exact identity is ambiguous---are forwarded to the LLM. The MSA traversal short-circuits once the upper bound of the joint score drops below $\tau_{\mathrm{lo}}$, implicitly pruning the massive tail of certain negatives. If the materialised caches exhaust before reaching $\tau_{\mathrm{lo}}$, \sys{} gracefully resumes the stream by dynamically evaluating distances via the base vector indices as discussed in Sec.~\ref{sec:maintenance}.
\item \emph{Top-$K$ poles.} For queries seeking extreme pairs (e.g., \emph{Movie}~Q5), the MSA fetches the $K$ highest- and lowest-scoring pairs by directly aggregating the underlying facet streams.
\end{itemize}
Crucially, as the downstream LLM resolves uncertain cases via \texttt{AI.IF}, the deterministic results are written back exclusively to \texttt{\idx{}(AI.IF)}, effectively amortising the cost of multi-faceted reasoning for subsequent queries.

\subsubsection{\textsc{sem\_map}}
\label{sec:int-map}

A \textsc{sem\_map} call applies a per-row natural-language transformation that returns a free-form string, e.g., brand extraction (\emph{Ecomm}~Q3), colour labelling (\emph{Ecomm}~Q12), or genre clustering (\emph{MMQA}~Q4). Because the output domain is open-vocabulary, neither the contrastive margin of \textsc{sem\_filter} nor the $K$-way argmax of \textsc{sem\_classify} applies---there is no fixed anchor set to contrast against. Instead, \sys{} exploits a complementary form of semantic redundancy: tuples whose embeddings are near-duplicates across the query-relevant feature spaces $\mathcal{F}$ are statistically guaranteed to map to the same target string, so a single LLM call on one representative safely labels its entire neighbourhood.

To compose this joint similarity, the MSA initialises a parallel FSS per facet $f \in \mathcal{F}$ and aggregates the per-facet cosine streams into a unified affinity
$\sigma_{ij} = \sum_{f \in \mathcal{F}} w_f\,\cos(\mathbf{e}_i^{(f)},\mathbf{e}_j^{(f)})$
on the fly. \sys{} then clusters the input tuples by agglomerative clustering on this aggregate affinity (complete linkage, distance threshold $\tau_d{=}0.10$, i.e., $\sigma_{ij}\!\geq\!0.90$) and recommends only the medoid nearest each cluster centroid. The downstream engine runs \texttt{AI.GENERATE} on this condensed $K$-tuple shortlist (e.g., $\sim$90 medoids out of 500 rows on \emph{Ecomm}~Q3), and the returned label is broadcast to the entire cluster via a SQL join on the synthetic cluster ID. Because clusters form in the joint multi-facet space rather than over any single embedding axis, the medoid is robust to single-modality drift---visually similar but textually distinct items remain in separate clusters. Quality degrades only when $\tau_d$ is lax enough to fuse genuinely distinct entities, which on \emph{Ecomm} costs less than $0.01$ ARI while reducing LLM invocations by an order of magnitude.

\endgroup

\end{document}